\documentclass{jpp}

\usepackage[toc, xindy, acronym]{glossaries}
\usepackage{natbib}

\usepackage[title]{appendix}

\usepackage[setpagesize=false]{hyperref} 
\hypersetup{
    colorlinks,
    linkcolor={blue!80!black},
    citecolor={blue!80!black},
    urlcolor={blue!80!black}
}

\usepackage{zref-savepos}
\usepackage{cleveref}
\crefformat{section}{\S#2#1#3} 
\crefformat{subsection}{\S#2#1#3}
\crefformat{subsubsection}{\S#2#1#3}
\crefformat{figure}{figure~#2#1#3}
\crefformat{equation}{(#2#1#3)}
\crefrangeformat{equation}{(#3#1#4)--(#5#2#6)}
\crefrangeformat{figure}{figures #3#1#4--#5#2#6}

\usepackage[makeroom]{cancel}
\usepackage{scrextend} 
\usepackage{bold-extra} 
\usepackage{color} 
\usepackage{fancyhdr}
\usepackage{ulem}
\usepackage{xcolor}
\usepackage{relsize}
\usepackage{textcomp}
\usepackage{lipsum}
\usepackage{setspace}
\usepackage{empheq}

\usepackage{float} 

\usepackage{epstopdf, epsfig}

\usepackage{pdfpages} 
\usepackage{graphicx} 
\usepackage{graphics} 

\usepackage{caption}
\usepackage{tabu} 
\usepackage{tabulary} 

\usepackage{framed} 
\usepackage{listings} 

\usepackage{amsmath} 
\usepackage{amssymb} 
\usepackage{mathrsfs} 

\usepackage{algorithm} 
\usepackage{algpseudocode} 

\usepackage{tensor}
\usepackage{upgreek}
\usepackage{physics}

\usepackage{tikz}
\usepackage{circuitikz}
\usetikzlibrary{calc}
\usetikzlibrary{patterns}
\usetikzlibrary{arrows}
\usetikzlibrary{shapes}
\usetikzlibrary{plotmarks}
\usetikzlibrary{decorations.markings}

\usepackage{pgfplots}
\usepgfplotslibrary{polar}
\usepgflibrary{shapes.geometric}
\usepackage[]{pstricks}

\tikzset{
	set arrow inside/.code={\pgfqkeys{/tikz/arrow inside}{#1}},
	set arrow inside={end/.initial=>, opt/.initial=},
	/pgf/decoration/Mark/.style={
		mark/.expanded=at position #1 with
		{
			\noexpand\arrow[\pgfkeysvalueof{/tikz/arrow inside/opt}]{\pgfkeysvalueof{/tikz/arrow inside/end}}
		}
	},
	arrow inside/.style 2 args={
		set arrow inside={#1},
		postaction={
			decorate,decoration={
				markings,Mark/.list={#2}
			}
		}
	},
}

\newcommand{\gradd}{{\boldsymbol{\nabla}}}

\renewcommand{\vec}[1]{\boldsymbol{#1}}
\newcommand{\uvec}[1]{\hat{\boldsymbol{#1}}}

\newcommand{\df}{\delta \! f}

\newcommand{\dBpar}{\delta \! B_\parallel}

\newcommand{\rmd}{\mathrm{d}}
\newcommand{\sgn}{\mathrm{sgn}}
\renewcommand{\Re}{\mathrm{Re}}
\renewcommand{\Im}{\mathrm{Im}}
\newcommand{\s}{s}

\newcommand{\bea}{\begin{eqnarray}}
\newcommand{\eea}{\end{eqnarray}}
\newcommand{\beq}{\begin{equation}}
\newcommand{\eeq}{\end{equation}}

\newcommand{\kperp}{k_\perp}

\newcommand{\kpar}{k_\parallel}

\newcommand{\vperp}{v_\perp}
\newcommand{\vpar}{v_\parallel}

\newcommand{\vths}{v_{{\rm th}\s}}
\newcommand{\vthi}{v_{{\rm th}i}}
\newcommand{\vthe}{v_{{\rm th}e}}

\newcommand{\upar}{u_\parallel}

\renewcommand{\tt}{\tilde t}

\newcounter{NoTableEntry}
\renewcommand*{\theNoTableEntry}{NTE-\the\value{NoTableEntry}}

\newcommand{\customlabel}[2]{%
   \protected@write \@auxout {}{\string \newlabel {#1}{{#2}{\thepage}{#2}{#1}{}} }%
   \hypertarget{#1}{#2}
}

\makeatletter
\def\blfootnote{\xdef\@thefnmark{}\@footnotetext}
\makeatother

\newcommand{\I}{I}
\newcommand{\J}{J}
\newcommand{\Z}{Z}

\newcommand{\Ical}{\mathcal{I}}
\newcommand{\Jcal}{\mathcal{J}}

\newcommand{\M}{\mathcal{M}}
\newcommand{\K}{\mathcal{K}}
\renewcommand{\L}{\mathcal{L}}
\newcommand{\Q}{Q}
\newcommand{\Lmat}{\mathbf{L}}

\newcommand{\CB}{{C_\sigma}}
\newcommand{\CInvL}{{C_\rho}}

\newcommand{\Iab}{\I_{a,b}}
\newcommand{\Jab}{\J_{a,b}}

\newcommand{\Iabp}{\I_{a,b}^{+}}

\newcommand{\sqrtp}[1]{\sqrt[+]{#1}}
\newcommand{\sqrtm}[1]{\sqrt[-]{#1}}
\newcommand{\sqrtbr}[1]{\sqrt[\branchsymb]{#1}}

\newcommand{\hsk}{{h_\s}_{\vec{k}}}
\newcommand{\hskhat}{{\hat{h}}_{\s \vec{k}}}
		
\newcommand{\chik}{{\chi}_{\vec{k}}}
\newcommand{\chikhat}{{\hat{\chi}}_{\vec{k}}}

\newcommand{\phik}{{\phi}_{\vec{k}}}
\newcommand{\phikhat}{{\hat{\phi}}_{\vec{k}}}

\newcommand{\Apark}{A_{\parallel \vec{k}}}

\newcommand{\dBpark}{{\delta \! B}_{\parallel \vec{k}}}

\newcommand{\gsk}{{g}_{\s \vec{k}}}

\newcommand{\vthr}{v_{{\rm th}r}}

\newcommand{\xplus}{u_+}
\newcommand{\xminus}{u_-}
\newcommand{\xpm}{u_\pm}
\newcommand{\xmp}{u_\mp}

\newcommand{\zetaplus}{\zeta_+}
\newcommand{\zetaminus}{\zeta_-}
\newcommand{\zetapm}{\zeta_\pm}

\newcommand{\branchsymb}{\lambda}
\newcommand{\mainD}{\mathcal{D}}

\newcommand{\ptf}[1]{\frac{\partial {#1}}{\partial t}}

\newcommand{\pyf}[1]{\frac{\partial {#1}}{\partial y}}
\newcommand{\pzf}[1]{\frac{\partial {#1}}{\partial z}}

\title[]{An analytical form of the dispersion function for local linear gyrokinetics in a curved magnetic field}
\author[P. G. Ivanov and T. Adkins] 
{
P.~G.~Ivanov$^{1, 2}$\thanks{Email: plamen.ivanov@physics.ox.ac.uk}
and
T.~Adkins$^{1,2,3}$
}

\affiliation{
$^1$Rudolf Peierls Centre for Theoretical Physics, University of Oxford,\\ 
Oxford, OX1 3PU, UK
\\[\affilskip]
$^2$Culham Centre for Fusion Energy, United Kingdom Atomic Energy Authority,\\ Abingdon, OX14 3DB, UK
\\[\affilskip]
$^3$Merton College, Oxford, OX1 4JD, UK

}

\begin{document}

\maketitle

\begin{abstract}
Starting from the equations of collisionless linear gyrokinetics for magnetised plasmas with an imposed inhomogeneous magnetic field, we present the first known analytical, closed-form solution for the resulting velocity-space integrals in the presence of resonances due to both parallel streaming and constant magnetic drifts. These integrals are written in terms of the well-known plasma dispersion function \citep{faddeeva54, fried61}, rendering the subsequent expressions simpler to treat analytically and more efficient to compute numerically. We demonstrate that our results converge to the well-known ones in the straight-magnetic-field and two-dimensional limits, and show good agreement with the numerical solver by \citet{gurcan14}. By way of example, we calculate the exact dispersion relation for a simple electrostatic, ion-temperature-gradient-driven instability, and compare it with approximate kinetic and fluid models.
\end{abstract}

\section{Introduction}
\label{sec:introduction}
The investigation of the linear-stability properties of magnetically confined plasmas is crucial for the design of magnetic-confinement-fusion devices. The heat and particle losses in these devices are dominated by turbulent fluctuations, which are themselves excited by linear instabilities driven by the gradients of the plasma equilibrium \citep{rudakov61, pogutse68, coppi67, guzdar83, hugill83, liewer85, waltz88, wootton90, cowley91, kotschenreuther95, xanthopoulos2007, ongena2016}. In most cases, the strong toroidal magnetic field constrains the plasma fluctuations to have typical temporal scales that are slow compared to the frequency of the Larmor motion of the particles, and to be anisotropic in space: length scales along the magnetic field are comparable to the size of the device, while ones perpendicular to it are comparable to the Larmor radii of the particles. Therefore, the plasma dynamics can often be treated using the gyrokinetic formalism \citep{frieman82, sugama96, howes06, abel13, catto2019}. 

When solving the linear gyrokinetic equation, one inevitably encounters resonant velocity-space integrals that need to be evaluated, analytically or numerically, in order to obtain the dispersion relation for the linear modes present within the system. The most basic of these resonances results from the parallel (to the magnetic field) streaming of particles, first discussed by \citet{landau46}. However, in the presence of an inhomogeneous equilibrium magnetic field, one is presented with a qualitatively different type of resonance due to the magnetic drifts of the particles. Evaluating these resonant integrals analytically, in the presence of both parallel streaming and magnetic drifts that are constant along the magnetic field, and without further approximations \citep[such as those used in, e.g.,][]{terry82,kim94} has remained an open research question, despite some progress being made numerically \citep{gurcan14,gultekin18,gultekin20,parisi20}. On the other hand, it is well-known that there are instabilities that exist only in the presence of curved magnetic fields, e.g., the toroidal ion-temperature-gradient (ITG) instability \citep{pogutse68, guzdar83, waltz88, kotschenreuther95}. Often, such instabilities are the dominant ones in toroidal plasmas. Thus, the exact inclusion of the magnetic-drift resonance in the analytical theory of linear gyrokinetics is expected to lead to qualitative changes in the behaviour of the resulting dispersion relation and to allow for a more complete treatment of the linear-stability properties of strongly magnetised plasmas.

In this work, we present closed forms for the aforementioned resonant integrals. These are written in terms of the plasma dispersion function \citep{faddeeva54, fried61}. They allow us to find a closed expression for the drift-kinetic dispersion relation, or an absolutely convergent series for the gyrokinetic one via Taylor expansions. The inclusion of magnetic drifts in the linear gyrokinetic problem introduces two distinct changes: (i) quantitatively, in that it significantly modifies the growth rates and frequencies of linear solutions; and (ii) qualitatively, by introducing a multivalued dispersion function. The latter has important consequences for the form of the dispersion relation and its solution, some of which have already been described in the literature \citep{kuroda98, sugama99}. 

The rest of the paper is organised as follows. We begin by summarising how the gyrokinetic dispersion relation, and the resonant velocity-space integrals of which it is comprised, emerge from the Fourier-Laplace transform of the linear gyrokinetic equations in \cref{sec:gyrokinetic_linear_theory}. Then, in \cref{sec:previous_solutions}, we discuss already-known solutions for these integrals and the asymptotic limits in which they apply. The main result of this work is presented in \cref{sec:the_generalised_plasma_dispersion_function}, where we derive the exact solution to one particular resonant integral --- the `generalised plasma dispersion function' --- to which all others will be related. In \cref{sec:asymptotic_known}, we show both analytically and numerically that the generalised plasma dispersion function asymptotes to the known solutions in the cases of zero magnetic curvature and of two-dimensional perturbations, while \cref{sec:numerical_comparison} demonstrates that our expressions are in agreement with the numerical solver published by \citet{gurcan14}. Section \cref{sec:analytic_continuation} discusses the analytic continuation of these functions and the subsequent solution to the inverse-Laplace-transform problem by which we obtain the solution to the linear gyrokinetic system. In \cref{sec:from_DK_to_GK}, we show how the results obtained in \cref{sec:the_generalised_plasma_dispersion_function} can be generalised to the gyrokinetic case via absolutely convergent Taylor expansions. In \cref{sec:itg}, we give an example calculation for the electrostatic ITG instability and compare it with known kinetic and fluid limits. Finally, our results are summarised and possible extensions discussed in \cref{sec:summary}.

\section{Collisionless gyrokinetic linear theory}
\label{sec:gyrokinetic_linear_theory}

In this section, we demonstrate how the resonant kinetic integrals that are the main focus of this paper emerge naturally from considerations of linear, collisionless local gyrokinetic theory with constant geometric coefficients [see the discussion following \cref{eq:perpendicular_amperes_law}]. Readers already familiar with gyrokinetic theory may wish to skip ahead to \cref{sec:low_beta_drift_kinetic_limit}, working backwards where further clarification is required. 

\subsection{Gyrokinetics}
\label{sec:gyrokinetics}
As is often the case in the study of magnetically confined plasmas, we shall assume that the fluctuations within our plasma obey the standard gyrokinetic ordering (see, e.g., \citealt{abel13} or \citealt{catto2019}); that is, for fluctuations with a characteristic frequency $\omega$ and wavenumbers $k_\parallel$ and $k_\perp$ parallel and perpendicular to the equilibrium magnetic field direction $\vec{b}_0 = \vec{B}_0/B_0$, we have 
\begin{equation}
	\frac{\omega}{\Omega_\s} \sim \frac{\nu_{\s \s'}}{\Omega_\s} \sim \frac{k_\parallel}{k_\perp} \sim \frac{q_\s \phi}{T_{0\s}} \sim \frac{\dBpar}{B_0} \sim \frac{\delta \! \vec{B}_\perp}{B_0} \sim \frac{\rho_\s}{L} \ll 1,
	\label{eq:gyrokinetic_ordering}
\end{equation}
where $\Omega_\s = q_\s B_0/m_\s c$ is the cyclotron frequency of species $\s$ with charge $q_\s$, equilibrium density and temperature $n_{0s}$ and $T_{0s}$, respectively, mass $m_\s$ and thermal speed \mbox{$\vths = \sqrt{2T_{0\s}/m_\s}$, $\nu_{\s \s'}$} is the typical collision frequency, $\rho_\s = \vths/|\Omega_\s|$ is the thermal Larmor radius, $\dBpar$ and $\delta \! \vec{B}_\perp$ are the fluctuations of the magnetic field parallel and perpendicular to the equilibrium direction, respectively, and $L$ is a typical equilibrium length scale. It is assumed that all equilibrium quantities evolve on the (long) transport timescale $\tau_E^{-1} \sim (\rho_\s/L)^3 \Omega_\s$, and so will be considered static throughout the remainder of this paper. 

Under the ordering \eqref{eq:gyrokinetic_ordering}, the perturbed distribution function $\df_\s$ consists of the Boltzmann and gyrokinetic parts:
\begin{equation}
	\df_\s(\vec{r},\vec{v},t) = - \frac{q_\s \phi(\vec{r},t)}{T_{0\s}} f_{0\s}(x,\vec{v}) + h_\s(\vec{R}_\s, v_\perp, \vpar,t),
	\label{eq:perturbed_distribution_function}
\end{equation}
where $\vec{R}_\s = \vec{r} - \vec{b}_0 \times \vec{v}_\perp/\Omega_\s$ is the guiding-centre position, and $h_\s$ evolves according to the gyrokinetic equation
\begin{equation}
	\frac{\partial}{\partial t} \left( h_\s - \frac{q_\s \left< \chi \right>_{\vec{R}_\s}}{T_{0\s}} f_{0\s} \right) + \left(\vpar \vec{b}_0 + \vec{v}_{d \s} \right) \bcdot \gradd h_\s + \vec{v}_\chi \bcdot \gradd_\perp \left( h_\s + f_{0\s} \right) = \left( \frac{\partial h_\s}{\partial t} \right)_c.
	\label{eq:gyrokinetic_equation}
\end{equation}
In the above, and throughout this paper, $\left< ... \right>_{\vec{R}_\s}$ denotes the standard gyroaverage at constant \(\vec{R}_\s\). Here, $\chi = \phi - \vec{v}\bcdot \vec{A}/c$ is the gyrokinetic potential ($\phi$ and $\vec{A}$ are the scalar and vector potential, respectively, under the Coulomb gauge \(\grad \bcdot \vec{A} = 0\)) that gives rise to the drift velocity
\begin{equation}
	\vec{v}_\chi =\frac{c}{B_0}\vec{b}_0 \times \frac{\partial \left< \chi \right>_{\vec{R}_\s}}{\partial \vec{R}_\s},
	\label{eq:drift_velocity}
\end{equation}
which includes the $\vec{E} \times \vec{B}$ drift, the parallel streaming along perturbed field lines, and the $\gradd \! B$ drift associated with the perturbed magnetic field. This gives rise to nonlinearities (with which we will not be concerned in this paper), as well as the familiar gyrokinetic drive associated with the equilibrium distribution $f_{0s}$, viz.,
\begin{align}
	\vec{v}_\chi \bcdot \gradd_\perp f_{0s} =  - \frac{c}{B_0} \left( \vec{b}_0 \times \frac{\partial \left< \chi \right>_{\vec{R}_\s}}{\partial \vec{R}_\s} \right) \bcdot \gradd x  \left[\frac{1}{L_{n_s}} + \frac{\eta_\s}{L_{n_s}} \left( \frac{v^2}{\vths^2} - \frac{3}{2} \right) \right] f_{0s},
	\label{eq:equilibrium_injection}
\end{align}
where
\begin{align}
	L_{n_\s}^{-1} = - \frac{1}{n_{0\s}} \frac{\partial n_{0\s}}{\partial x}, \quad L_{T_\s}^{-1} = - \frac{1}{T_{0\s}} \frac{\partial T_{0\s}}{\partial x}, \quad \eta_\s = \frac{L_{n_\s}}{L_{T_\s}},
	\label{eq:equilibrium_gradients}
\end{align}
are the characteristic length scales associated with the radial equilibrium gradients of both density and temperature, respectively, $\eta_\s$ is their ratio, and \(x\) is the direction of the equilibrium gradients. The magnetic drifts associated with the equilibrium field are 
\begin{equation}
	\vec{v}_{d\s} = \frac{\vec{b}_0}{\Omega_\s} \times \left[ \vpar^2 \vec{b}_0\bcdot \grad\vec{b}_0 + \frac{1}{2}\vperp^2 \grad\log B_0 \right].
	\label{eq:magnetic_drifts}
\end{equation}
The last term on the right-hand side of \eqref{eq:gyrokinetic_equation} is the (linearised) collision operator, which we henceforth neglect given that we are interested in studying collisionless dynamics. The electromagnetic fields appearing in the gyrokinetic equation \eqref{eq:gyrokinetic_equation} are determined by the quasineutrality condition
\begin{equation}
	0 = \sum_{\s} q_\s \delta n_\s = \sum_\s q_\s \left[ -\frac{q_\s \phi}{T_{0\s}}  n_{0\s}    + \int \rmd^3 \vec{v}  \left< h_\s \right>_{\vec{r}}\right],
	\label{eq:quasineutrality}
\end{equation}
where $\left< ... \right>_{\vec{r}}$ denotes the gyroaverage at constant \(\vec{r}\), and by the parallel and perpendicular parts of Amp\`ere's law, which are, respectively,
\begin{align}
	\gradd_\perp^2 A_\parallel &= - \frac{4\pi}{c} \sum_\s q_\s \int \rmd^3 \vec{v} \: \vpar \left< h_\s \right>_{\vec{r}},
	\label{eq:parallel_amperes_law} \\
	\gradd_\perp^2 \dBpar& = - \frac{4\pi}{c} \vec{b}_0 \bcdot \left[ \gradd_\perp \times \sum_\s q_\s \int \rmd^3 \vec{v} \left< \vec{v}_\perp h_\s \right>_{\vec{r}} \right].
	\label{eq:perpendicular_amperes_law}
\end{align}
Together, \eqref{eq:gyrokinetic_equation} and \eqref{eq:quasineutrality}--\cref{eq:perpendicular_amperes_law} form a closed system of equations that, in principle, allows us to determine $h_\s$ and thus the evolution of the fluctuations in our plasma. In this work, we solve the linear part of this system in the `local' limit \citep{beer95}: we assume that the gradients of all equilibrium quantities are constant --- including the geometric coefficients \(\vec{b}_0\bcdot\grad\vec{b}_0\) and \(\grad\log B_0\) that appear in the magnetic drifts --- and choose orthonormal coordinates \((x, y, z)\), in which \(\uvec{z}=\vec{b}_0\) is the direction of the magnetic field, \(\uvec{x}\) is, as above, the direction of the equilibrium gradients (cf. the radial direction in toroidal geometry), and \(\uvec{y} \equiv \vec{b}_0 \times \uvec{x}\) is the binormal direction (cf. the poloidal direction in toroidal geometry). One can think of this geometry as that of a \(Z\)-pinch \citep[see][]{ricci06, ivanov20, ivanov22, adkins22} due to the assumption of constant magnetic curvature and lack of magnetic shear, which we have implicitly assumed. Under these assumptions, the system of equations \eqref{eq:gyrokinetic_equation}, \eqref{eq:quasineutrality}--\cref{eq:perpendicular_amperes_law} is homogeneous in space, allowing us to impose periodic boundary conditions in all three spatial dimensions. 

In the next section, we consider the time evolution of a single Fourier mode and obtain the resulting gyrokinetic dispersion relation.

\subsection{Linear gyrokinetic problem}
\label{sec:gyrokinetic_linear_eigenvalue_problem}
Neglecting the nonlinear term and introducing the spatial Fourier decomposition:
\begin{align}
	h_\s(\vec{R}_\s, v_\perp, v_\parallel, t) = \sum_{\vec{k}} \hsk(\vperp, \vpar, t) e^{i\vec{k}\bcdot \vec{R}_{\s}}, \quad \chi(\vec{r},t) = \sum_{\vec{k}}\chik(t)  e^{i\vec{k} \bcdot \vec{r} },
	\label{eq:fourier_space}
\end{align}
with $\vec{k} = \vec{k_\perp} + k_\parallel \vec{b}_0 $, the Fourier modes \(\hsk\) and \(\chik\) can be shown to satisfy
\begin{align}
	\frac{\partial}{\partial t} \left( \hsk - \frac{q_{\s} \left<\chik \right>_{\vec{R}_{\s}}}{T_{0\s}}f_{0\s} \right) + i k_\parallel v_\parallel \hsk + i \omega_{D\s} \hsk - i \omega_{*\s}^T \frac{q_{\s} \left<\chik \right>_{\vec{R}_{\s}}}{T_{0{\s}}}f_{0{\s}} = 0, 
	\label{eq:gk_fourier_space}
\end{align}
where we have defined the drift frequencies associated with the equilibrium gradients of species $\s$ [cf. \eqref{eq:equilibrium_injection}]:
\begin{align}
	\omega_{*\s}^T = \omega_{*\s} \left[ 1 + \eta_\s \left( \frac{v^2}{\vths^2} - \frac{3}{2} \right) \right], \quad \omega_{*\s} = - \frac{k_y c T_{0\s}}{q_\s B_0 L_{n_\s}},
	\label{eq:drift_frequency}
\end{align}
and with the equilibrium magnetic field curvature and gradient, respectively [cf. \eqref{eq:magnetic_drifts}]:
\begin{align}
	\omega_{D\s} = \frac{2\vpar^2}{\vths^2} \omega_{\kappa s} + \frac{\vperp^2}{\vths^2} \omega_{\gradd \! B \s},
	\label{eq:magnetic_drift}
\end{align}
where
\begin{align}
	\omega_{\kappa \s} = \frac{\vths^2 }{2\Omega_s} \vec{k}_\perp \bcdot \left[\vec{b}_0 \times (\vec{b}_0 \bcdot \gradd )\vec{b}_0 \right], \quad \omega_{\gradd \! B\s} = \frac{\vths^2}{2\Omega_s} \vec{k}_\perp \bcdot \left(\vec{b}_0 \times \gradd \log B_0 \right).
	\label{eq:magnetic_drift_frequencies}
\end{align}
Starting from the perpendicular force balance of the gyrokinetic equilibrium [see equation (128) in \citealt{abel13}], it is straightforward to show that the difference between these two drifts is given by 
\begin{align}
	 \omega_{\kappa \s} - \omega_{\gradd \! B \s} =  \frac{\vths^2}{2\Omega_\s} \vec{k_\perp} \bcdot \left( \vec{b}_0 \times \gradd x \right) \left.\frac{\partial}{\partial x}\right|_{B_0} \sum_{\s'} \frac{\beta_{\s'}}{2} ,
	 \label{eq:magnetic_drifts_difference} 
\end{align}
where $\beta_\s = 8 \pi n_{0\s} T_{0\s}/B_0^2$ is the plasma beta of species $\s$.
Lastly, the gyroaveraged Fourier-transformed gyrokinetic potential is 
\begin{align}
	\left< \chik \right>_{\vec{R}_s} = J_0(b_\s) \left(\phik - \frac{v_\parallel \Apark}{c} \right) +  \frac{2 J_1 (b_\s)}{b_\s} \frac{T_{0\s}}{q_\s} \frac{v_\perp^2 }{\vths^2}\frac{\dBpark}{B_0},
	\label{eq:fourier_transformed_gyrokinetic_potential}
\end{align}
while the field equations \eqref{eq:quasineutrality}--\eqref{eq:perpendicular_amperes_law} can be written as (see, e.g., \citealt{howes06})
\begin{align}
	\sum_\s \frac{q_\s^2 n_{0\s}}{T_{0\s}} \phik & =  \sum_\s q_\s   \int \rmd^3 \vec{v} \: J_0(b_\s) \tilde{h}_\s, \label{eq:quasineutrality_fourier} \\
	k_\perp^2 \Apark & = \frac{4\pi}{c} \sum_\s q_\s \int \rmd^3 \vec{v} \: v_\parallel J_0(b_\s)\tilde{h}_s , \label{eq:parallel_amperes_fourier} \\
	\frac{\dBpark}{B_0} &= - \frac{1}{2} \sum_\s \frac{\beta_\s}{n_{0\s}} \int \rmd^3 \vec{v} \: \frac{v_\perp^2}{\vths^2} \frac{2 J_1 (b_\s)}{b_\s} \tilde{h}_s,
	\label{eq:perpendicular_amperes_law_fourier}
\end{align}
where $b_\s = k_\perp v_\perp/\Omega_\s$, and $J_0$, $J_1$ are the Bessel functions of the first kind (\citealt{abramowitz72}) that capture finite-Larmor-radius effects. It will prove convenient to combine \(\phik\), \(\Apark\), and \(\dBpark\) into a single vector \(\vec{\chi}_{\vec{k}}\) given by
\begin{align}
	\vec{\chi}_{\vec{k}} = \left(
	\frac{q_r \phik}{T_{0r}}, \:
	\frac{k_\parallel }{|k_\parallel| } \frac{\Apark}{\rho_r B_0}, \:
	\frac{\dBpark}{B_0}
	\right)^T.
	\label{eq:vector_of_fields}
\end{align}
Here, and in what follows, we normalise the electromagnetic fields using an arbitrary reference mass $m_r$, density $n_{0r}$, thermal velocity $\vthr$, temperature $T_{0r}$, and gyroradius~$\rho_r$.

\begin{figure}
	\centering
	
	\scalebox{0.9}{\begin{tikzpicture}[scale=1, thick, every node/.style={scale=1.2}]
			\newcommand{\arrowIn}{
				\tikz \draw[-latex] (-1pt,0) -- (1pt,0);
			}
		
            \def\xaxislength{10}
            \def\yaxislength{7.5}
            \def\xoffset{0.25}
            \def\yoffset{2}
            \def\zoffset{0.1}
            
            \draw[-latex] (\xoffset - \xaxislength/3,\yoffset,\zoffset) -- (\xoffset,\yoffset,\zoffset) -- (\xoffset + \xaxislength/2,\yoffset,\zoffset);
            \draw (\xoffset + \xaxislength/2,\yoffset,\zoffset) node[anchor=north] {$\Re(p)$};
            
            \draw[-latex] (\xoffset,\yoffset - \yaxislength/2,\zoffset) -- (\xoffset,\yoffset,\zoffset) -- (\xoffset,\yoffset + \yaxislength/2,\zoffset);
            \draw (\xoffset,\yoffset +\yaxislength/2 ,\zoffset) node[anchor=east] {$\Im(p)$};
            
            \def\sigmaoffset{\xaxislength/4}
            \draw (\xoffset + \sigmaoffset,\yoffset,\zoffset) node[anchor= north west] {$\sigma$};
            \draw[solid] (\xoffset + \sigmaoffset,\yoffset - \yaxislength/2,\zoffset) -- (\xoffset + \sigmaoffset,\yoffset + \yaxislength/2,\zoffset);
            \draw (\xoffset + \sigmaoffset,\yoffset - \yaxislength/4,\zoffset) node[anchor = center, rotate = 90] {\arrowIn}; 
            \draw (\xoffset + \sigmaoffset,\yoffset + \yaxislength/4,\zoffset) node[anchor = center, rotate = 90] {\arrowIn};
            
            \fill [gray, opacity=0.15] (\xoffset + \sigmaoffset,\yoffset + \yaxislength/2 ,\zoffset) -- (\xoffset + \xaxislength/2,\yoffset + \yaxislength/2 ,\zoffset) -- (\xoffset + \xaxislength/2,\yoffset - \yaxislength/2 ,\zoffset) -- (\xoffset + \sigmaoffset,\yoffset - \yaxislength/2 ,\zoffset) -- cycle;
            
            \draw (\xoffset + \sigmaoffset/2,\yoffset + \yaxislength/3 ,\zoffset) node[anchor = center] {$\times$};
            \draw (\xoffset - \sigmaoffset/2,\yoffset + \yaxislength/5 ,\zoffset) node[anchor = center] {$\times$};
            \draw (\xoffset + \sigmaoffset/4,\yoffset - \yaxislength/8 ,\zoffset) node[anchor = center] {$\times$};
            \draw (\xoffset + 2*\sigmaoffset/4,\yoffset - 7*\yaxislength/16,\zoffset) node[anchor = center] {$\times$};
            
            \draw[decorate, decoration = zigzag]  (\xoffset - \xaxislength/3,\yoffset - 9*\yaxislength/32,\zoffset) -- (\xoffset,\yoffset - 9*\yaxislength/32,\zoffset);
            
            \draw (\xoffset + \sigmaoffset,\yoffset + \yaxislength/6,\zoffset) node[anchor = north west
            ] {$\CB$};

        \end{tikzpicture}}
	
	 \caption{The complex $p$ plane, with $\Re(p)$ and $\Im(p)$ shown on the horizontal and vertical axes, respectively. The contour of integration for the inverse Laplace transform $\CB$ is is a vertical straight line at $\Re(p) = \sigma$, to the right of which (i.e, in the shaded grey region) the functions $\hskhat$ and $\chikhat$ are guaranteed to be analytic. Singularities, such as poles (indicated by crosses) or branch cuts (indicated by the zigzag line), could exist at $\Re(p) < \sigma$.}
	 \label{fig:inverse_laplace_transform}
\end{figure}
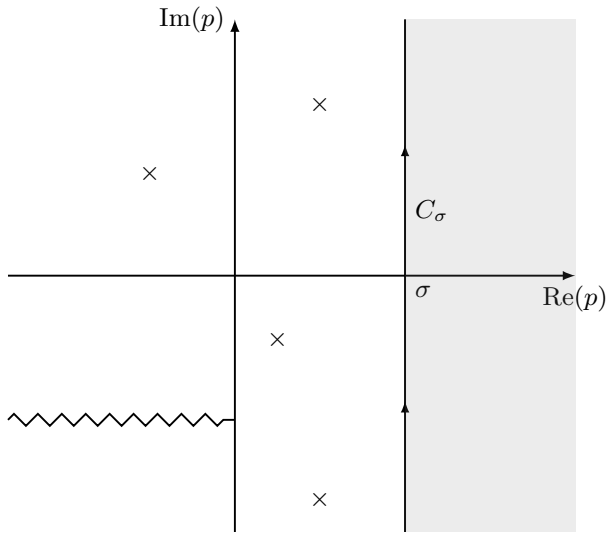

Following \cite{landau46}, we consider an initial-value problem and introduce the Laplace transformations 
\begin{align}
	\hskhat(\vperp, \vpar, p) = \int_0^{\infty} \rmd t \: e^{-pt} \hsk(\vperp, \vpar, t), \quad \hat{\vec{\chi}}_{\vec{k}}(p) = \int_0^{\infty} \rmd t \: e^{-pt} \vec{\chi}_{\vec{k}}(t).
	\label{eq:laplace_transform}
\end{align}
Assuming there exist positive real \(m\) and \(M\) such that
\begin{align}
	\left|\hsk(\vperp, \vpar, t)\right|, \: \left| \vec{\chi}_{\vec{k}}(t) \right| \leqslant Me^{m t},
	\label{eq:sigma_definition}
\end{align}
for all \(t > 0\), and picking any real \(\sigma\) with \(\sigma > m\), the integrals in \cref{eq:laplace_transform} converge and the transformed distributions \(\hskhat\) and fields \(\hat{\vec{\chi}}_{\vec{k}}\) are analytic for all complex values of $p$ with $\Re(p) \geqslant \sigma$. The inverse transformations are given by
\begin{align}
	 \hsk(\vperp, \vpar, t) = \frac{1}{2 \pi i} \int_{\CB} \rmd p \: e^{pt} 	\hskhat(\vperp, \vpar, p), \quad \vec{\chi}_{\vec{k}}(t) = \frac{1}{2 \pi i} \int_{\CB} \rmd p \: e^{pt} \hat{\vec{\chi}}_{\vec{k}}(p),
	 \label{eq:inverse_laplace_transform}
\end{align}
where the contour of integration $\CB$ is along a straight line parallel to the imaginary axis and intersecting the real axis at $\Re(p) = \sigma$, as in \cref{fig:inverse_laplace_transform} (this is the so-called Bromwich contour).  

Performing the Laplace transform as in \cref{eq:laplace_transform}, \cref{eq:gk_fourier_space} straightforwardly becomes
\begin{align}
	\hskhat = \frac{p + i \omega_{*\s}^T}{p + i k_\parallel \vpar + i \omega_{D\s}} \frac{q_{\s} \left<\chikhat \right>_{\vec{R}_{\s}}}{T_{0\s}}f_{0\s} + \frac{{\gsk}}{p + i k_\parallel \vpar + i \omega_{D\s}},
	\label{eq:gk_fourier_laplace}
\end{align}
where $\gsk$ is the initial condition:
\begin{align}
	\gsk(\vperp, \vpar) = \hsk(\vperp, \vpar, {t=0}) - \frac{q_{\s} \left<\chikhat ({t=0}) \right>_{\vec{R}_{\s}}}{T_{0\s}}f_{0\s}.
	\label{eq:gsk_definition}
\end{align}
Then, normalising the characteristic frequencies to the parallel-streaming rate\footnote{Note that normalising to $|\kpar| \vths$ rather than $\kpar \vths$ means that the condition for analyticity $ \Re(p) \geqslant \sigma > 0$ implies \(\Im(\zeta_\s) > 0\), regardless of the sign of $\kpar$. \label{footnote:kpar} }
\begin{align}
	\zeta_\s = \frac{ip}{|k_\parallel| \vths}, \quad \zeta_{*\s} = \frac{\omega_{*s}}{|k_\parallel| \vths}, \quad \zeta_{\kappa \s} = \frac{\omega_{\kappa s}}{|k_\parallel| \vths}, \quad \zeta_{\gradd \! B \s} = \frac{\omega_{\gradd \! B \s}}{|k_\parallel| \vths},
	\label{eq:normalised_frequencies}
\end{align}
and defining the dimensionless velocity variables
\begin{align}
	u = \frac{k_\parallel }{|k_\parallel| } \frac{v_\parallel}{\vths}, \quad \mu = \frac{v_\perp^2}{\vths^2},
	\label{eq:normalised_velocity variables}
\end{align}
we substitute \eqref{eq:gk_fourier_laplace} into the Laplace transforms of the field equations \eqref{eq:quasineutrality_fourier}--\eqref{eq:perpendicular_amperes_law_fourier} to obtain the linear eigenvalue problem
\begin{align}
	\mathbf{L} \hat{\vec{\chi}}_{\vec{k}}
		+ \vec{G} = 0 ,
	\label{eq:eigenvalue_problem}
\end{align}
in which $\mathbf{L}$ is the linear coefficient matrix and $\vec{G}$ is the vector of the initial conditions of the fields. The components of $\mathbf{L}$ are given by 
\begin{align}
	\text{L}_{\phi \phi} & = - \sum_\s \frac{q_\s^2 n_{0\s} T_{0r}}{q_r^2 n_{0r} T_{0\s}} \left\{ 1+ \left[\zeta_\s - \zeta_{*\s} + \eta_\s \zeta_{*\s} \left( \partial_a + \partial_b + \frac{3}{2} \right) \right] \left.\Ical_{a,b}^{(\s)} \right|_{a=b=1}\right\} ,\label{eq:L_phiphi} \\
	\text{L}_{\phi A} & =  2\sum_\s \frac{q_\s^2 n_{0\s} \vths T_{0r}}{q_r^2 n_{0r} \vthr T_{0\s}} \left[\zeta_\s - \zeta_{*\s} + \eta_\s \zeta_{*\s} \left( \partial_a + \partial_b + \frac{3}{2} \right) \right] \left.\Jcal_{a,b}^{(\s)}\right|_{a=b=1}, \label{eq:L_phia} \\
	\text{L}_{\phi B} & = \sum_\s \frac{q_\s n_{0\s}}{q_r n_{0r}} \left[\zeta_\s - \zeta_{*\s} + \eta_\s \zeta_{*\s} \left( \partial_a + \partial_b + \frac{3}{2} \right) \right] \partial_b \left.\K_{a,b}^{(\s)}\right|_{a=b=1}, \label{eq:L_phib} \\
	\text{L}_{A \phi} & = - \sum_\s \frac{q_\s^2 n_{0\s} \vths T_{0r}}{q_r^2 n_{0r} \vthr T_{0\s}} \left[\zeta_\s - \zeta_{*\s} + \eta_\s \zeta_{*\s} \left( \partial_a + \partial_b + \frac{3}{2} \right) \right] \left.\Jcal_{a,b}^{(\s)}\right|_{a=b=1} ,\label{eq:L_aphi}\\
	\text{L}_{AA} & = - \frac{B_0^2 ( k_\perp \rho_r)^2}{8 \pi n_{0r} T_{0r}} - 2 \sum_\s \frac{q_\s^2 n_{0\s} m_r}{q_r^2 n_{0r} m_\s} \left[\zeta_\s - \zeta_{*\s} + \eta_\s \zeta_{*\s} \left( \partial_a + \partial_b + \frac{3}{2} \right) \right] \partial_a \left.\Ical_{a,b}^{(\s)}\right|_{a=b=1}, \label{eq:L_aa} \\
	\text{L}_{AB} & = \sum_s \frac{q_\s n_{0\s} \vths}{q_r n_{0r} \vthr} \left[\zeta_\s - \zeta_{*\s} + \eta_\s \zeta_{*\s} \left( \partial_a + \partial_b + \frac{3}{2} \right) \right] \partial_b \left.\L_{a,b}^{(\s)}\right|_{a=b=1}, \label{eq:L_ab} \\
	\text{L}_{B\phi} & = - \sum_\s  \frac{\beta_\s}{2} \frac{q_\s T_{0r}}{q_r T_{0s}} \left[\zeta_\s - \zeta_{*\s} + \eta_\s \zeta_{*\s} \left( \partial_a + \partial_b + \frac{3}{2} \right) \right] \partial_b \left.\K_{a,b}^{(\s)}\right|_{a=b=1}, \label{eq:L_bphi} \\
	\text{L}_{B A} & = \sum_\s  \beta_\s \frac{q_\s T_{0r} \vths}{q_r T_{0s} \vthr} \left[\zeta_\s - \zeta_{*\s} + \eta_\s \zeta_{*\s} \left( \partial_a + \partial_b + \frac{3}{2} \right) \right] \partial_b \left.\L_{a,b}^{(\s)} \right|_{a=b=1},\label{eq:L_ba} \\
	\text{L}_{BB} & = - 1 +  \sum_\s \frac{\beta_\s}{2} \left[\zeta_\s - \zeta_{*\s} + \eta_\s \zeta_{*\s} \left( \partial_a + \partial_b + \frac{3}{2} \right) \right] \partial_b^2 \left.\M_{a,b}^{(\s)}\right|_{a=b=1} ,\label{eq:L_bb}
\end{align}
where we have defined the following integrals
\begin{align}
	\Ical_{a,b}^{(\s)} & = \frac{1}{\sqrt{\pi}} \int_{-\infty}^\infty \rmd u \int_0^\infty \rmd \mu \: \frac{ e^{-a u^2 - b \mu}}{u - \zeta_\s + \left(2u^2 \zeta_{\kappa \s} + \mu  \zeta_{Bs} \right)} J_0^2(b_\s),  \label{eq:iab_flr} \\
	\Jcal_{a,b}^{(\s)} & = \frac{1}{\sqrt{\pi}} \int_{-\infty}^\infty \rmd u \int_0^\infty \rmd \mu \: \frac{ u e^{-a u^2 - b \mu}}{u - \zeta_\s + \left(2u^2 \zeta_{\kappa \s} + \mu  \zeta_{Bs} \right)} J_0^2(b_\s),  \label{eq:jab_flr} \\
	\K_{a,b}^{(\s)} & = \frac{1}{\sqrt{\pi}} \int_{-\infty}^\infty \rmd u \int_0^\infty \rmd \mu \: \frac{ e^{-a u^2 - b \mu}}{u - \zeta_\s + \left(2u^2 \zeta_{\kappa \s} + \mu  \zeta_{Bs} \right)} \frac{2 J_0 (b_\s) J_1 (b_\s)}{b_\s},  \label{eq:kab_flr} \\
	\L_{a,b}^{(\s)} & = \frac{1}{\sqrt{\pi}} \int_{-\infty}^\infty \rmd u \int_0^\infty \rmd \mu \: \frac{ u e^{-a u^2 - b \mu}}{u - \zeta_\s + \left(2u^2 \zeta_{\kappa \s} + \mu  \zeta_{Bs} \right)} \frac{2 J_0 (b_\s) J_1 (b_\s)}{b_\s},  \label{eq:lab_flr} \\
	\M_{a,b}^{(\s)} & = \frac{1}{\sqrt{\pi}} \int_{-\infty}^\infty \rmd u \int_0^\infty \rmd \mu \: \frac{  e^{-a u^2 - b \mu}}{u - \zeta_\s + \left(2u^2 \zeta_{\kappa \s} + \mu  \zeta_{Bs} \right)} \left[\frac{2 J_1 (b_\s)}{b_\s} \right]^2. \label{eq:mab_flr} 
\end{align}
Here, and throughout the remainder of this paper, the parameters $a$ and $b$ are assumed to be both real and positive, ensuring integral convergence. Finally, the components of $\vec{G} = (G_\phi, G_A, G_B)^T$ are given by 
\begin{align}
	G_\phi & = \sum_\s \frac{q_\s n_{0\s}}{q_r n_{0r}} \frac{1}{n_{0\s}} \int \rmd^3 \vec{v} \:  \frac{\gsk}{p + i \kpar \vpar + i \omega_{D\s}}J_0 (b_\s),\label{eq:g_phi} \\
	G_A & = \sum_s \frac{q_\s n_{0\s} \vths}{q_r n_{0r} \vthr } \frac{1}{n_{0\s}}\int \rmd^3 \vec{v} \: \frac{\vpar}{\vths} \frac{\gsk}{p + i \kpar \vpar + i \omega_{D\s}}J_0 (b_\s)  \label{eq:g_A}, \\
	G_B & = - \sum_\s \frac{\beta_\s}{2}\frac{1}{n_{0\s}}\int \rmd^3 \vec{v} \: \frac{\vperp^2}{\vths^2}  \frac{\gsk}{p + i \kpar \vpar + i \omega_{D\s}}\frac{2 J_1(b_\s)}{b_\s}.\label{eq:g_B}
\end{align}

The eigenvalue problem \eqref{eq:eigenvalue_problem} can be inverted in order to solve for the fields in the usual way, viz., 
\begin{align}
	\hat{\vec{\chi}}_{\vec{k}}(p)  = \frac{(\text{adj}\:\mathbf{L})	\vec{G}	}{\det\mathbf{L}},
	\label{eq:solving_for_fields}
\end{align}
where $\text{adj}\:\mathbf{L}$ and $\det\mathbf{L}$ are the adjugate matrix and determinant of the linear matrix $\mathbf{L}$, respectively. The time-dependent fields are then determined by the inverse Laplace transform of \cref{eq:solving_for_fields}. As discussed above, the integrals in \cref{eq:inverse_laplace_transform} are, before analytic continuation, defined for $\Re(p) \geqslant \sigma > 0$. For these values of \(p\), \(\Im(\zeta_\s) > 0\), and so the integrals in \crefrange{eq:iab_flr}{eq:mab_flr} converge and are analytic functions of \(p\). Note that the equation
\begin{equation}
	\label{eq:dispersion_relation}
	D(p)\equiv\det\mathbf{L}(p) = 0
\end{equation}
is commonly known as the `dispersion relation', while we shall refer to \(D\) itself as the `dispersion function'.

\subsection{Drift-kinetic limit}
\label{sec:low_beta_drift_kinetic_limit}
To evaluate the integrals \crefrange{eq:iab_flr}{eq:mab_flr}, we specialise to the drift-kinetic limit, in which the perpendicular wavenumbers of the perturbations are assumed small in comparison to the species' gyroradii, viz., 
\begin{align}
	b_\s \sim \kperp \rho_s \ll 1.
	\label{eq:drift_kinetic_limit}
\end{align}
In this limit, the Bessel functions can be expanded as 
\begin{align}
	J_0(b_\s) = 1 + O(b_\s^2), \quad 2J_1(b_\s)/b_\s = 1 + O(b_\s^2),
	\label{eq:bessel_function_expansion}
\end{align}
meaning that, to leading order in $b_\s$, the contributions of the Bessel functions to the integrals \eqref{eq:iab_flr}-\eqref{eq:mab_flr} are equal to one, and we may write 
\begin{align}
	\I_{a,b}^{(\s)} = \Ical_{a,b}^{(\s)} = \K_{a,b}^{(\s)} = \M_{a,b}^{(\s)} , \quad \J_{a,b}^{(\s)} = \Jcal_{a,b}^{(\s)} = \L_{a,b}^{(\s)},
	\label{eq:integral_equalities}
\end{align}
where $\I_{a,b}^{(\s)}$ and $\J_{a,b}^{(\s)}$ are given by 
\begin{align}
	\I_{a,b}^{(\s)}(\zeta_{\s}, \zeta_{\kappa \s}, \zeta_{Bs}) & = \frac{1}{\sqrt{\pi}} \int_{-\infty}^\infty \rmd u \int_0^\infty \rmd \mu \: \frac{e^{-a u^2 - b \mu}}{u - \zeta_\s + \left(2u^2 \zeta_{\kappa \s} + \mu  \zeta_{Bs} \right)}, \label{eq:iab_no_flr} \\
	\J_{a,b}^{(\s)}(\zeta_{\s}, \zeta_{\kappa \s}, \zeta_{Bs}) & = \frac{1}{\sqrt{\pi}} \int_{-\infty}^\infty \rmd u \int_0^\infty \rmd \mu \: \frac{u e^{-a u^2 - b \mu}}{u - \zeta_\s + \left(2u^2 \zeta_{\kappa \s} + \mu \zeta_{Bs} \right)}. \label{eq:jab_no_flr}
\end{align}
Furthermore, we consider the particular case in which the difference between the curvature and $\gradd \! B$ drifts, given by the right-hand side of \cref{eq:magnetic_drifts_difference}, is zero and so their associated drift frequencies can be taken to be equal, viz., 
\begin{align}
	\omega_{\kappa \s} = \omega_{\gradd \! B \s} \equiv \omega_{d\s} \quad \Rightarrow \quad \zeta_{\kappa \s} = \zeta_{\gradd \! B \s} \equiv \zeta_{d\s}.
	\label{eq:magnetic_drifts_equality}
\end{align}
Note that neither approximation should be interpreted as a consequence of some asymptotic ordering of the parameters describing our gyrokinetic system of equations. Instead, they should be viewed as formal approximations that allow us to obtain a solvable case of a more general one. Their relaxation is discussed in \cref{sec:from_DK_to_GK}. 

With these simplifications, we have reduced our problem to the evaluation of
\begin{align}
	\I_{a,b}(\zeta, \zeta_d) & = \frac{1}{\sqrt{\pi}} \int_{-\infty}^\infty \rmd u \int_0^\infty \rmd \mu \: \frac{e^{-a u^2 - b \mu}}{u - \zeta + \zeta_d\left(2u^2  + \mu  \right)}, \label{eq:iab_original} \\
	\J_{a,b}(\zeta, \zeta_d) & = \frac{1}{\sqrt{\pi}} \int_{-\infty}^\infty \rmd u \int_0^\infty \rmd \mu \: \frac{u e^{-a u^2 - b \mu}}{u - \zeta +\zeta_d \left(2u^2  + \mu  \right)}, \label{eq:jab_original}
\end{align}
where we have used \cref{eq:magnetic_drifts_equality} and have dropped the species index for the sake of compactness of notation.

\section{Previous solutions}
\label{sec:previous_solutions}
Before tackling the task of analytically integrating \cref{eq:iab_original} and \cref{eq:jab_original}, we shall briefly discuss some special cases in which these expressions are already known within the literature. A reader already familiar with these solutions may wish to skip ahead to \cref{sec:the_generalised_plasma_dispersion_function}, working backwards if further clarification is required. 

\subsection{The plasma dispersion function and Landau's solution}
\label{sec:plasma_dispersion_function}
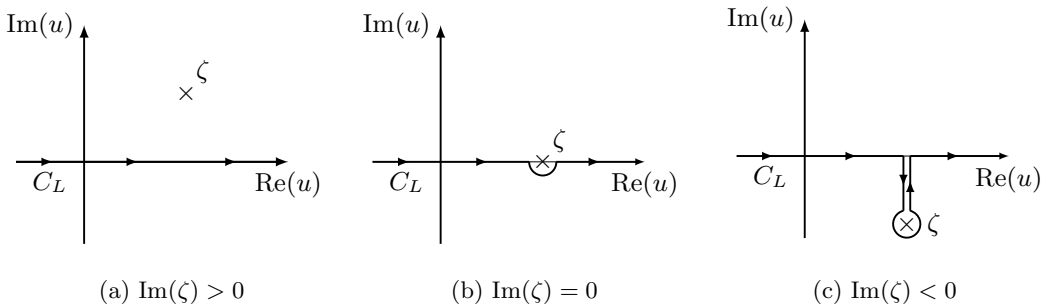
\begin{figure}
	\centering
	
	\begin{tabular}{ccc}
		\hspace*{-1mm}\scalebox{0.9}{\begin{tikzpicture}[scale=1, thick, every node/.style={scale=1.2}]
			\newcommand{\arrowIn}{
				\tikz \draw[-latex] (-1pt,0) -- (1pt,0);
			}

            \def\xaxislength{6}
            \def\negativexaxis{\xaxislength/6}
            \def\positivexaxis{\xaxislength/2}
            
            \def\yaxislength{6}
            \def\negativeyaxis{\yaxislength/5}
            \def\positiveyaxis{\yaxislength/3}
            
            \def\xoffset{0.25}
            \def\yoffset{2}
            \def\zoffset{0.1}

            \draw[-latex] (\xoffset - \negativexaxis,\yoffset,\zoffset) -- (\xoffset,\yoffset,\zoffset) -- (\xoffset + \positivexaxis,\yoffset,\zoffset);
            \draw (\xoffset + \positivexaxis,\yoffset,\zoffset) node[anchor=north] {$\Re(u)$};
            
            \draw[-latex] (\xoffset,\yoffset - \negativeyaxis,\zoffset) -- (\xoffset,\yoffset,\zoffset) -- (\xoffset,\yoffset + \positiveyaxis,\zoffset);
            \draw (\xoffset,\yoffset +\positiveyaxis,\zoffset) node[anchor=east] {$\Im(u)$};
            
            \def\contouroffset{0.1}
            
            \draw (\xoffset - \negativexaxis,\yoffset,\zoffset) 
            -- node[midway, allow upside down] {\arrowIn}
             node[midway, anchor = north] {$C_L$}
            (\xoffset,\yoffset,\zoffset) 
            -- node[midway, allow upside down] {\arrowIn}
            (\xoffset + \positivexaxis/2,\yoffset,\zoffset)
            -- node[midway, allow upside down] {\arrowIn}
            (\xoffset + \positivexaxis - \contouroffset,\yoffset,\zoffset);

            \draw (\xoffset +\positivexaxis/2 ,\yoffset + \yaxislength/6 ,\zoffset) node[anchor = center] {$\times$};
            \draw (\xoffset +\positivexaxis/2 ,\yoffset + \yaxislength/6 ,\zoffset) node[anchor = south west] {$\zeta$};

        \end{tikzpicture}} &
		\scalebox{0.9}{\begin{tikzpicture}[scale=1, thick, every node/.style={scale=1.2}]
			\newcommand{\arrowIn}{
				\tikz \draw[-latex] (-1pt,0) -- (1pt,0);
			}
			
            \def\xaxislength{6}
            \def\negativexaxis{\xaxislength/6}
            \def\positivexaxis{\xaxislength/2}
            
            \def\yaxislength{6}
            \def\negativeyaxis{\yaxislength/5}
            \def\positiveyaxis{\yaxislength/3}
            
            \def\xoffset{0.25}
            \def\yoffset{2}
            \def\zoffset{0.1}

            \draw[-latex, thin] (\xoffset - \negativexaxis,\yoffset,\zoffset) -- (\xoffset,\yoffset,\zoffset) -- (\xoffset + \positivexaxis,\yoffset,\zoffset);
            \draw (\xoffset + \positivexaxis,\yoffset,\zoffset) node[anchor=north] {$\Re(u)$};
            
            \draw[-latex] (\xoffset,\yoffset - \negativeyaxis,\zoffset) -- (\xoffset,\yoffset,\zoffset) -- (\xoffset,\yoffset + \positiveyaxis,\zoffset);
            \draw (\xoffset,\yoffset +\positiveyaxis,\zoffset) node[anchor=east] {$\Im(u)$};
            
            \def\lineseparation{0.05}
            \def\circleradius{0.2}
            \def\contouroffset{0.1}
            
            \draw (\xoffset - \negativexaxis,\yoffset,\zoffset) 
            -- node[midway, allow upside down] {\arrowIn}
             node[midway, anchor = north] {$C_L$}
            (\xoffset,\yoffset,\zoffset) 
            -- node[midway, allow upside down] {\arrowIn}
            (\xoffset + \positivexaxis/2 - \circleradius,\yoffset,\zoffset);
            
            \draw (\xoffset + \positivexaxis/2 + \circleradius ,\yoffset,\zoffset) arc (0 : -180 : \circleradius);
            
            \draw (\xoffset + \positivexaxis/2 + \circleradius,\yoffset,\zoffset)
            -- node[midway, allow upside down] {\arrowIn}
            (\xoffset + \positivexaxis - \contouroffset,\yoffset,\zoffset);

            
            \draw (\xoffset +\positivexaxis/2 ,\yoffset,\zoffset) node[anchor = center] {$\times$};
            \draw (\xoffset +\positivexaxis/2 ,\yoffset,\zoffset) node[anchor = south west] {$\zeta$};

        \end{tikzpicture}} &
		\hspace*{1mm}\scalebox{0.9}{\begin{tikzpicture}[scale=1, thick, every node/.style={scale=1.2}]
			\newcommand{\arrowIn}{
				\tikz \draw[-latex] (-1pt,0) -- (1pt,0);
			}
		
            \def\xaxislength{6}
            \def\negativexaxis{\xaxislength/6}
            \def\positivexaxis{\xaxislength/2}
            
            \def\yaxislength{6}
            \def\negativeyaxis{\yaxislength/5}
            \def\positiveyaxis{\yaxislength/3}
            
            \def\xoffset{0.25}
            \def\yoffset{2}
            \def\zoffset{0.1}

            \draw[-latex, thin] (\xoffset - \negativexaxis,\yoffset,\zoffset) -- (\xoffset,\yoffset,\zoffset) -- (\xoffset + \positivexaxis,\yoffset,\zoffset);
            \draw (\xoffset + \positivexaxis,\yoffset,\zoffset) node[anchor=north] {$\Re(u)$};
            
            \draw[-latex] (\xoffset,\yoffset - \negativeyaxis,\zoffset) -- (\xoffset,\yoffset,\zoffset) -- (\xoffset,\yoffset + \positiveyaxis,\zoffset);
            \draw (\xoffset,\yoffset +\positiveyaxis,\zoffset) node[anchor=east] {$\Im(u)$};
            
            \def\lineseparation{0.05}
            \def\circleradius{0.2}
            \def\circleangle{atan(\lineseparation/(\circleradius))}
            \def\poleoffset{0.2}
            \def\contouroffset{0.1}
            
            \draw (\xoffset - \negativexaxis,\yoffset,\zoffset) 
            --node[midway, allow upside down] {\arrowIn}
            node[midway, anchor = north] {$C_L$}
            (\xoffset,\yoffset,\zoffset) 
            -- node[midway, allow upside down] {\arrowIn}
            (\xoffset + \positivexaxis/2 - \lineseparation,\yoffset,\zoffset);
            
			\draw (\xoffset + \positivexaxis/2 - \lineseparation,\yoffset,\zoffset)
			-- node[midway, sloped, allow upside down] {\arrowIn}
			(\xoffset + \positivexaxis/2 - \lineseparation,\yoffset - \yaxislength/6 + \poleoffset,\zoffset);
			
			\draw (\xoffset + \positivexaxis/2 - \lineseparation , \yoffset - \yaxislength/6 + \poleoffset,\zoffset) arc (-270 + \circleangle : 90 - \circleangle : \circleradius);
			
			\draw[thick] (\xoffset + \positivexaxis/2 + \lineseparation,\yoffset - \yaxislength/6 + \poleoffset,\zoffset)
			-- node[midway, sloped, allow upside down] {\arrowIn}
			(\xoffset + \positivexaxis/2 + \lineseparation,\yoffset,\zoffset);
            
            \draw[thick]  (\xoffset + \positivexaxis/2 + \lineseparation,\yoffset,\zoffset)
            -- node[midway, allow upside down] {\arrowIn}
            (\xoffset + \positivexaxis - \contouroffset,\yoffset,\zoffset);

            \draw (\xoffset +\positivexaxis/2 ,\yoffset - \yaxislength/6 ,\zoffset) node[anchor = center] {$\times$};
            \draw (\xoffset +\positivexaxis/2 + \circleradius/2 + \lineseparation,\yoffset - \yaxislength/6 ,\zoffset) node[anchor = west] {$\zeta$};

        \end{tikzpicture}} \\\\
		(a) $\Im(\zeta) > 0$ &
		(b) $\Im(\zeta) = 0$ & 
		(c) $\Im(\zeta) < 0$
	\end{tabular}
	
	\caption{The Landau prescription for the contour of integration $C_L$ that gives the analytic continuation of \cref{eq:Z_function}. As the Laplace transform demands \(\Re(p) \geqslant \sigma > 0 \), the pole $u = \zeta$ is located in the upper-half plane [where \(\Im(\zeta) > 0\), see \cref{footnote:kpar}], above the contour of integration, as in panel (a). Therefore, the appropriate analytic continuation for \(\Re(p) \leqslant 0\) [i.e., \(\Im(\zeta) \leqslant 0\)] demands that the contour must be deformed so as to always remain below the pole, as in panels (b), (c). Cauchy's integral theorem ensures that we are free to deform the contour without changing the value of the integral, so long as it does not cross the pole. }
	\label{fig:landau_contour}
	\vspace*{-0.1725cm}
\end{figure}

In the absence of magnetic drifts (i.e., when $\zeta_d = 0$), \cref{eq:iab_original} and \cref{eq:jab_original} can straightforwardly be written in terms of the well-studied plasma dispersion function (\citealt{faddeeva54, fried61}):
\begin{align}
	 \Z(\zeta) \equiv \frac{1}{\sqrt{\pi}} \int_{-\infty}^\infty \rmd u \: \frac{e^{-u^2}}{u - \zeta},
	\label{eq:Z_function}
\end{align}
where the integral is defined for $\Im(\zeta)>0$ with the integration contour along the real $u$ axis, as in \cref{fig:landau_contour}(a). In particular, we have that
\begin{align}
	\left.\I_{a,b} \right|_{\zeta_d=0} = \frac{1}{b}\Z_a(\zeta), \quad \left.\J_{a,b}\right|_{\zeta_d=0} = \frac{1}{b}\left[\frac{1}{\sqrt{a}} + \zeta \Z_a(\zeta)\right],
	\label{eq:iab_jab_in_terms_of_Z_function}
\end{align}
where we have, for the sake of brevity, introduced the shorthand notation 
	\begin{align}
		Z_a(\zeta) \equiv Z(\sqrt{a} \zeta).
		\label{eq:Z_function_a}
\end{align}
The integral in \eqref{eq:Z_function} can be analytically continued to $\Im(\zeta) \leqslant 0$ by deforming the contour of integration in such a way as to always keep the pole above it, as shown in \cref{fig:landau_contour}(b), (c). This is known as the Landau prescription, and the resultant contour is the well-known Landau contour $C_L$ (\citealp{landau46}). 

The plasma dispersion function \cref{eq:Z_function} is ubiquitous in calculations of linear waves and instabilities in systems with a spatially uniform magnetic field; notable examples include the electron-temperature-gradient \citep[see, e.g.,][]{liu71, lee87} and ion-temperature-gradient (see, e.g., \citealt{rudakov61,coppi66,sauter90, brunner98, smolyakov2002}) instabilities, the latter of which we shall consider in \cref{sec:itg}. It is also worth noting that the Bessel functions can easily be incorporated into the integrals if \(\zeta_d=0\) because the resonant denominators are independent of \(\mu\). The resulting expressions involve modified Bessel functions and are well-known in the literature \citep[see, e.g.,][]{howes06}.

\subsection{Two-dimensional limit}
\label{sec:two_dimensional_limit}
In the two-dimensional limit, $\kpar \rightarrow 0$ with $\zeta \sim \zeta_d \rightarrow \infty$, it can be shown (via, e.g., a partial-fractions expansion of the integrand) that \cref{eq:iab_original} can be expressed exactly in terms of products of the plasma dispersion function \citep{biglari89}, viz., 
\begin{align}
	\I_{1,1} = - \frac{1}{2 \zeta_d } Z(\sqrt{\Omega})^2 + \order{\zeta_d^{-2}}, \quad  \J_{1,1} = \order{\zeta_d^{-2}}, \quad \Omega = \frac{\zeta}{2 \zeta_d},
	\label{eq:2d_limit}
\end{align}
with the integral for $\J_{a,b}$ vanishing to leading order because the integrand in \cref{eq:jab_original} is manifestly odd in \(x\) in this limit. The analytic continuation for \cref{eq:2d_limit} is significantly more subtle than in the case of the plasma dispersion function \cref{eq:Z_function}, owing to the presence of the branch point at $\zeta = 0$; we shall delay discussion of these subtleties until \cref{sec:analytic_continuation}. The solution \cref{eq:2d_limit} has been used extensively in the investigation of two-dimensional ITG instabilities (see, e.g., \citealt{similon84,biglari89,kuroda98,sugama99,ricci06,helander11,mishchenko18,zocco18}).

\subsection{Numerical methods}
\label{sec:numerical_methods}
Owing to their analytical complexity, previous literature has also been devoted to the numerical evaluation of \cref{eq:iab_original} and \cref{eq:jab_original} (see \citealt{beer96,gurcan14,gultekin18,gultekin20,parisi20}, and references contained therein). In many cases, this involves expressing these integrals in terms of one-dimensional integrals. For example, writing 
\begin{align}
	\frac{1}{u - \zeta + \zeta_d(2u^2 +\mu)} = i \int_{0}^{\sgn[\Im (\zeta)] \infty} \rmd \lambda \: e^{-i\lambda[u - \zeta +\zeta_d(2u^2 + \mu)]},
	\label{eq:lambda_trick}
\end{align}
allows the integration over $u$ and $\mu$ in \cref{eq:iab_original} and \cref{eq:jab_original} to be done analytically, leaving an integral over $\lambda$ that can be evaluated numerically (cf. \citealt{beer96,parisi20}). While this method is quite general --- in that it also allows the direct inclusion of the Bessel functions in \crefrange{eq:iab_flr}{eq:mab_flr} --- the numerical evaluation of the resultant expressions can often be slow, numerical errors may be difficult to quantify, and subtleties like multivaluedness and branch cuts easy to overlook. This motivates the goal of the present study, viz., to find expressions for these integrals in terms of known functions that can be better understood analytically and more readily computed numerically.

\section{The generalised plasma dispersion function}
\label{sec:the_generalised_plasma_dispersion_function}
In this section, we detail the method by which \cref{eq:iab_original} and \cref{eq:jab_original} can be expressed in terms of the plasma dispersion function \cref{eq:Z_function}, making the resultant expressions simpler to treat both analytically and numerically. When solving the integrals, we will assume that \(p\) remains within the region of analyticity $\Re(p) \geqslant \sigma > 0$, with $\sigma$ defined after \cref{eq:sigma_definition}. The analytic continuation will be performed only after obtaining expressions for \cref{eq:iab_original} and \cref{eq:jab_original} in terms of known functions. In the main text, we present the integration of \cref{eq:iab_original}; all other required expressions follow directly from this single integral, and have been relegated to appendices \ref{app:calculation_of_jab} and \ref{app:calculation_of_derivatives} due to their complexity. The remainder of this section proceeds as follows. \cref{sec:multivaluedness} discusses the multivalued nature of the integrand of \cref{eq:iab_original} before evaluating the integral over $u$ in terms of plasma dispersion function \cref{eq:Z_function}, allowing us, in \cref{sec:evaluation_of_iab}, to obtain a closed form expression for \cref{eq:iab_original} upon evaluating the remaining integral over $\mu$. In \cref{sec:derivatives}, we discuss how the \(\partial_a\) and \(\partial_b\) derivatives of \cref{eq:iab_original} and \cref{eq:jab_original} can be obtained, with detailed calculations relegated to \cref{app:calculation_of_derivatives}. Then, in \cref{sec:branches_disp}, we discuss some important properties of \cref{eq:iab_original} and \cref{eq:jab_original}.

\subsection{Multivaluedness}
\label{sec:multivaluedness}
To begin, it shall be useful to consider the integral over $u$ separately, and so we write \cref{eq:iab_original} as follows:
\begin{align}
	\I_{a,b} = \int_0^\infty \rmd \mu \: e^{-b\mu} \tilde{\I}_{a}, \quad \tilde{\I}_{a} = \frac{1}{\sqrt{\pi}} \int_{-\infty}^\infty \rmd u \frac{e^{-au^2}}{u - \zeta + \zeta_d(2u^2 + \mu)}.
		\label{eq:iab_tilde_definition}
\end{align}
Now, for each value of $\mu$, the denominator of $\tilde{\I}_{a}$ has two zeros at
\begin{equation}
	u=\frac{-1 \pm \sqrt{1 + 8\zeta_d(\zeta - \zeta_d\mu)}}{4\zeta_d}
	\label{eq:quadratic_roots}
\end{equation}
that produce poles on opposite sides of the integration contour along the real $u$ axis. Unsurprisingly, given the presence of square roots in \cref{eq:quadratic_roots}, $\tilde{\I}_{a}$ is a multivalued function. In particular, we shall find that \(\tilde{\I}_{a}\), and thus \(\Iab\), has two branches, just like the square root. To define these two branches, we need to choose a branch cut, which will allow us to `label' the two zeros in \cref{eq:quadratic_roots}. Note that this choice cannot (and does not) affect the time evolution of the potentials that results from the inverse Laplace transform of \cref{eq:vector_of_fields}. It turns out to be analytically convenient to consider the `principal' branch cut for the square-root function, for which \(\sqrt{z}\) is discontinuous across \(\Re(z) < 0\). We can then define the two branches of the square root, \(\sqrtp{z}\) and \(\sqrtm{z}\), where the principal branch satisfies \(\sqrtp{z} > 0\) for all positive real \(z\), and $\sgn[\Im (\sqrtp{z})] = \sgn[\Im(z)]$.

At this point, it is nontrivial to define the second branch of \(\I_{a,b}\). The choice of a branch for the square root does not determine the branch of the integral \cref{eq:iab_tilde_definition} but only the labels of the zeros in \cref{eq:quadratic_roots} --- observe that \cref{eq:iab_tilde_definition} makes no reference to any multivalued functions. Indeed, the function \(\I_{a,b}\) is defined as the integral in \cref{eq:iab_tilde_definition} \textit{only} for \(\Im(\zeta) > 0\); the multivaluedness becomes relevant after one considers the analytic continuation to \(\Im(\zeta) < 0\). To make this explicit, until we perform said continuation, we will make use of the labels \(\tilde{\I}_{a}^+\) and \(\I_{a,b}^+\) to indicate that our expressions only apply to this one branch.

Choosing to work with \(\sqrtp{}\), the zeros \cref{eq:quadratic_roots} can be written as
\begin{align}
	u = \mp \xpm, \quad \xpm \equiv \frac{\sqrtp{1 + 8\zeta_d(\zeta - \zeta_d\mu)} \pm 1}{4\zeta_d}.
	\label{eq:x_pm_definition}
\end{align}
Using a partial-fraction expansion of the integrand, it follows that 
\begin{align}
	\tilde{\I}_{a}^+ = \frac{1}{2\zeta_d(\xplus + \xminus)}\left( \frac{1}{\sqrt{\pi}}\int_{-\infty}^\infty \rmd u   \frac{e^{-au^2}}{u - \xminus} - \frac{1}{\sqrt{\pi}}\int_{-\infty}^\infty \rmd u \frac{e^{-au^2}}{u + \xplus }\right).
	\label{eq:iab_tilde_partial_fractions}
\end{align}
Now, given that $\Im(\zeta) > 0$, the sign of the imaginary part of \(\sqrtp{1 + 8\zeta_d(\zeta - \zeta_d\mu)}\) is determined by the sign of $\zeta_d$, viz., 
\begin{equation}
	\sgn\left[\Im{\sqrtp{1 + 8\zeta_d(\zeta - \zeta_d\mu)}}\right] = \sgn(\zeta_d),
\end{equation}
and so \cref{eq:x_pm_definition} implies that \(\Im(\xpm) > 0\). Therefore, the first integral in the brackets in \eqref{eq:iab_tilde_partial_fractions} is manifestly the plasma dispersion function, as the imaginary part of the pole at $u = \xminus$ has the correct sign for the definition \cref{eq:Z_function}, i.e., $\Im(\xminus) > 0$. The second integral has a pole at $u = -\xplus$ with the opposite sign of its imaginary part, i.e., $\Im(-\xplus) < 0$, meaning that it can also be turned into a plasma dispersion function under a straightforward change of variables $u \mapsto -u$ (this effectively flips the pole from being below the real \(u\) axis to being above it). Thus, it follows that \cref{eq:iab_tilde_partial_fractions} can be written as
\begin{align}
	\tilde{\I}_{a}^+ = \frac{1}{2\zeta_d} \frac{Z_a(\xplus) + Z_a(\xminus )}{\xplus + \xminus},
	\label{eq:iab_tilde_final}
\end{align}
where we have used the shorthand notation \cref{eq:Z_function_a} for \(\Z_a\). 

\subsection{Explicit evaluation of \(\Iab^+\)}
\label{sec:evaluation_of_iab}
Using \cref{eq:iab_tilde_final}, our expression for $\I_{a,b}^+$ thus becomes:
\begin{align}
	\I_{a,b}^+ = \frac{1}{2\zeta_d} \int_0^\infty \rmd \mu \: e^{-b \mu} \frac{Z_a(\xplus) + Z_a(\xminus )}{\xplus + \xminus}.
	\label{eq:iab_mu_integral}
\end{align}
Using \cref{eq:x_pm_definition}, together with the property \(\Z'(u) = -2 [1 + u\Z(u)]\), it can be deduced that
\begin{align}
	&\mu = \frac{\zeta}{\zeta_d} + \frac{1}{4 \zeta_d^2} - \left(\xplus^2 + \xminus^2 \right)
	\label{eq:mu_in_terms_of_xpm}
\end{align}
and
\begin{align}
	&\frac{e^{a \xpm^2}}{\xplus + \xminus} = \frac{1}{\sqrt{a}} \frac{\partial}{\partial \mu } \left[e^{a \xpm^2} Z_a(\xpm) \right].
	\label{eq:total_derivative_identity}
\end{align}
It is then a matter of straightforward algebra to show that \cref{eq:iab_mu_integral} can be rewritten as 
\begin{align}
	\I_{a,b}^+ = \frac{1}{2 \sqrt{a} \zeta_d } \int_0^\infty \rmd \mu \: e^{-(b-a) \mu} \frac{\partial}{\partial \mu} \left[ e^{-a \mu}   Z_a(\xplus) Z_a(\xminus) \right].
	\label{eq:iab_total_derivative}
\end{align}
Observe that \crefrange{eq:L_phiphi}{eq:L_bb} only reference $\I_{a,b}$, and its derivatives with respect to $a$ and $b$, evaluated at $a = b = 1$. We thus set $a=b$ in \cref{eq:iab_total_derivative} and, noting that \(Z_a(\xpm) \to 0\) as \(\mu \to \infty\) since \(\Im(\xpm) > 0\), we find
\begin{align}
	\I_{a,a}^+ = - \frac{1}{2 \sqrt{a} \zeta_d } Z_a(\zetaplus^+) Z_a(\zetaminus^+) ,
	\label{eq:Iabp}
\end{align}
where we have introduced
\begin{align}
	\zetapm^+ = \left. \xpm \right|_{\mu = 0} = \frac{\sqrtp{1 + 8\zeta_d\zeta} \pm 1}{4\zeta_d}.
	\label{eq:zetapm_definition}
\end{align}

From \cref{eq:Iabp} and \cref{eq:zetapm_definition}, it is clear that \(\I_{a,a}\) is a multivalued function with a branch point at \mbox{\(\zeta = -1/8\zeta_d\)}. Its second branch can be obtained by considering the \(\sqrtm{}\) branch of the square root in \cref{eq:zetapm_definition}. This means that both branches can be summarised by defining
\begin{equation}
	\zetapm^\branchsymb \equiv \frac{\sqrtbr{1 + 8\zeta_d\zeta} \pm 1}{4\zeta_d} = \frac{\lambda\sqrtp{1 + 8\zeta_d\zeta} \pm 1}{4\zeta_d}, 
	\label{eq:zetas_branches}
\end{equation}
where \(\branchsymb = \pm\) labels the branch. Therefore, \(\I_{a,a}\) can be written as
\begin{align}
	\I_{a,a}^{\branchsymb} = - \frac{1}{2 \sqrt{a} \zeta_d } Z_a(\zetaplus^\branchsymb) Z_a(\zetaminus^\branchsymb).
	\label{eq:genZ_function}
\end{align}

Equation \cref{eq:genZ_function} is the key result of this paper. We shall henceforth refer to it as the \textit{generalised plasma dispersion function}, in that it is the generalisation of the usual plasma dispersion function \cref{eq:Z_function} to include the resonances associated with the magnetic drifts arising in a non-uniform magnetic field. In \cref{sec:asymptotic_known}, we show that, in the appropriate limits, the generalised plasma dispersion function reduces to the already-known solutions discussed in \cref{sec:previous_solutions}. It is worth stressing that \cref{eq:genZ_function} is an exact result: no approximations have been made in deriving it from \cref{eq:iab_original}. Furthermore, the fact that \cref{eq:genZ_function} is composed of a product of plasma dispersion functions, for the evaluation of which there are numerous efficient algorithms, means that it is very fast to evaluate numerically. In \cref{sec:numerical_comparison}, we compare our expression for \(\Iab\) with the numerical solver by \citet{gurcan14}.

It can be shown, via a similar procedure to the one used to obtain \cref{eq:iab_total_derivative} (see \cref{app:calculation_of_jab}), that the related integral $\Jab^+$ \cref{eq:jab_original} can be expressed exactly in terms of $\Iabp$ and plasma dispersion functions as
\begin{align}
	\left( 1 - \frac{2b}{a} \right) \J_{a,b}^+ = \frac{1}{2 a \zeta_d }\left[Z_a(\zetaplus^+) - Z_a(\zetaminus^+) \right] + \frac{b}{2a \zeta_d} \I_{a,b}^+,
	\label{eq:jab_final}
\end{align}
and so
\begin{equation}
	\J_{a, a}^\branchsymb = -\frac{1}{2a\zeta_d} \left[Z_a(\zetaplus^\branchsymb) - Z_a(\zetaminus^\branchsymb) \right] - \frac{1}{2\zeta_d} \I_{a, a}^\branchsymb.
	\label{eq:jaa_general}
\end{equation}

It is crucial to realise that the \(\branchsymb=+\) branch of the functions \(\I_{a,a}^{\branchsymb}\) and \(\J_{a,a}^{\branchsymb}\) is the `more important' one, in the sense that it is the branch that is equal to the integrals \cref{eq:iab_original} and \cref{eq:jab_original} for \(\Im(\zeta) > 0\). Thus, it is also the branch that is used in the inverse Laplace transform over \(\CB\), as in \cref{eq:inverse_laplace_transform}. Therefore, we shall refer to the \(\branchsymb=+1\) branch as the `principal' branch of \(\I_{a,a}^{\branchsymb}\) and \(\J_{a,a}^{\branchsymb}\).

\subsection{Derivatives of the generalised plasma dispersion function}
\label{sec:derivatives}
In addition to \cref{eq:iab_original} and \cref{eq:jab_original}, the matrix elements \crefrange{eq:L_phiphi}{eq:L_bb} require the partial derivatives of these expressions with respect to $a$ and $b$. There are two factors that conspire to simplify the necessary calculations. First, we only need \(\I_{a,b}\), \(\J_{a,b}\), and their derivatives at \(a=b=1\). Secondly, the derivatives \(\partial_a\) and \(\partial_b\) often appear in the combination \(\partial_a + \partial_b\). Notice that, by the chain rule,
\begin{equation}
	(\partial_a + \partial_b)f_{a,b}\big\vert_{a=b=1} = \partial_a f_{a,a}\big\vert_{a=1}
\end{equation}
for any (appropriately smooth) function \(f\). Using this, we can rewrite \crefrange{eq:L_phiphi}{eq:L_bb} in a way that involves only \(\I_{a, a}\), \(\J_{a, a}\), \(\partial_a \I_{a,b}\vert_{a=b}\), \(\partial_b \I_{a,b}\vert_{a=b}\), \(\partial_b^2 \I_{a,b}\vert_{a=b}\), and \(\partial_b \J_{a,b}\vert_{a=b}\). For example,
\begin{equation}
	\text{L}_{\phi B} = \sum_\s \frac{q_\s n_{0\s}}{q_r n_{0r}} \left[\zeta_\s - \zeta_{*\s} + \eta_\s \zeta_{*\s} \left( \partial_a + \frac{3}{2} \right) \right] \left.\left(\partial_b \left.\I_{a,b}^{(\s)}\right|_{a=b}\right)\right|_{a=1}, \label{eq:L_phib_alternative}
\end{equation}
where we have also taken advantage of \cref{eq:integral_equalities}. Due to their unwieldy length, the calculations of the required derivatives of \(\I_{a,b}\) and \(\J_{a,b}\) are relegated to \cref{app:calculation_of_derivatives}.

\subsection{Branches of the dispersion function}
\label{sec:branches_disp}

Our choice of the principal branch cut for the square root gives the branches of the dispersion function \(D\) (see \cref{sec:gyrokinetic_linear_eigenvalue_problem}) several nice properties stemming from the relationship \(\sqrtp{z^*} = \sqrtp{z}^*\) for any complex \(z\). In \cref{app:branches}, we show that \(\I_{a, a}\) and \(\J_{a, a}\) satisfy
\begin{align}
	\I_{a, a}^\branchsymb(-\zeta^*, -\zeta_d) &= -\I_{a, a}^\branchsymb(\zeta, \zeta_d)^* \label{eq:iaa_minuszetaconj}, \\
	\J_{a, a}^\branchsymb(-\zeta^*, -\zeta_d) &= \J_{a, a}^\branchsymb(\zeta, \zeta_d)^* \label{eq:jaa_minuszetaconj}, 
\end{align}
and
\begin{align}
	\I_{a, a}^\branchsymb(\zeta^*, \zeta_d) &= \I_{a, a}^{-\branchsymb}(\zeta, \zeta_d)^* \label{eq:iaa_zetaconj}, \\
	\J_{a, a}^\branchsymb(\zeta^*, \zeta_d) &= \J_{a, a}^{-\branchsymb}(\zeta, \zeta_d)^* \label{eq:jaa_zetaconj}.
\end{align}
Relations \crefrange{eq:iaa_minuszetaconj}{eq:jaa_zetaconj} are also valid for the \(a\) and \(b\) derivatives of \(\I_{a,b}\) and \(\J_{a,b}\). Of course, the functions \(\Iab\) and \(\Jab\) are double-valued for each of the species \(\s\), and so the dispersion function \(D\) has \(2^{N}\) branches for a system with \(N\) species. Letting \(\vec{\branchsymb} \equiv (\branchsymb_1, \branchsymb_2, ..., \branchsymb_N)\) be the vector of choices of the branch for each species, we can prove that \(D\) satisfies (see \cref{app:branches})
\begin{align}
	D^{\vec{\branchsymb}}(p^*, -\vec{k}) &= D^{\vec{\branchsymb}}(p, \vec{k})^* \label{eq:D_minuszetaconj}, \\
	D^{\vec{\branchsymb}}(-p^*, \vec{k}) &= D^{-{\vec{\branchsymb}}}(p, \vec{k})^*. \label{eq:D_zetaconj} 
\end{align}
These imply two different pairings of roots of the dispersion relation \cref{eq:dispersion_relation}; see figures \ref{fig:unicurvy_principalbranch} and \ref{fig:unicurvy_rotatedbranch} in \cref{app:branches} for a visual illustration of \cref{eq:D_minuszetaconj} and \cref{eq:D_zetaconj}. Note that when using the superscript \(\vec{\branchsymb}\), we are referring to a \textit{particular} branch, while without it, \(D\) refers to all branches simultaneously.

Relation \cref{eq:D_minuszetaconj} implies that solutions to the dispersion relation, i.e., \(D=0\), come in pairs \((p, k_y) \leftrightarrow (p^*, -k_y)\), which is the condition for the fields \(\phi\), \(A_\parallel\), and \(\delta B_\parallel\) to remain real for all \(t\). Therefore, such a pairing is bound to exist for \textit{all} roots of \(D = 0\), i.e., when \textit{all} branches are considered. The choice of the principal branch of the square root makes this pairing also valid within each individual branch of \(D\), hence justifying our adoption of it in \cref{sec:multivaluedness}. In \cref{sec:analytic_continuation}, we shall see that there is a better choice of branch for the purposes of performing the inverse Laplace transform.

Additionally, \cref{eq:D_zetaconj} says that if \(p\) is a solution to \(D=0\) for a given poloidal wavenumber \(k_y\), then so is \(-p^*\) for the same \(k_y\) but for a different branch. At first glance, this might seem to imply that solutions to \cref{eq:dispersion_relation} always come in pairs, one stable and one unstable. While this is true if all branches of \(D\) are considered, the time evolution given by the inverse Laplace transform \cref{eq:inverse_laplace_transform} does not necessarily pick up contributions from all solutions to \(D=0\); one cannot mix-and-match roots from different branches at will. In \cref{sec:analytic_continuation}, we shall see that the roots of \cref{eq:dispersion_relation} picked up by \cref{eq:inverse_laplace_transform} depend on the choice of branch cut. However, only the principal branch, given by \(\vec{\branchsymb} = (+, +, ..., +)\), contributes linearly unstable solutions.

\section{Comparison with known results}
\label{sec:five}

\subsection{Asymptotic expansions of \(\Iab\)}
\label{sec:asymptotic_known}

\begin{figure}
	\centering \hspace*{-1cm}\includegraphics[scale=0.29]{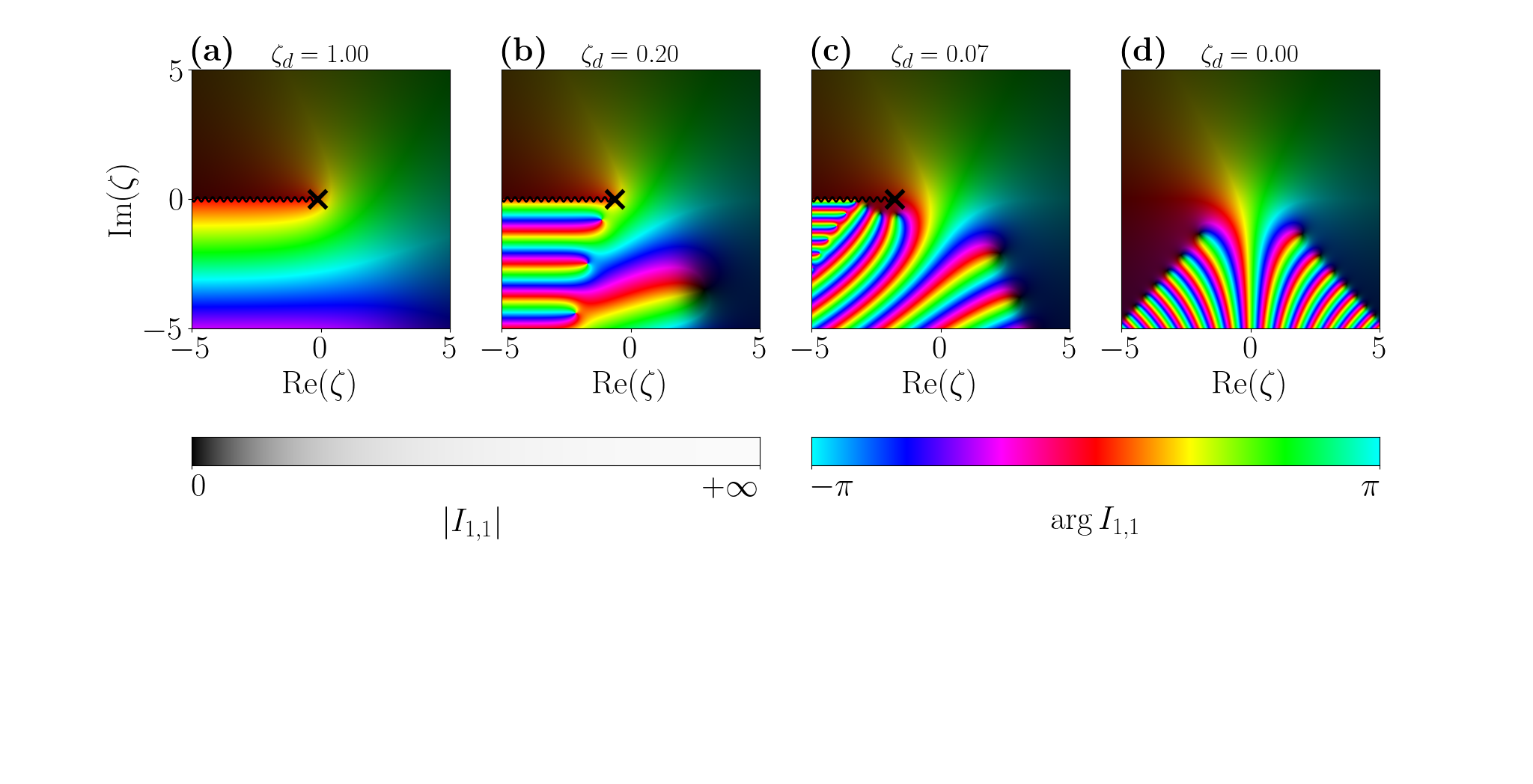}
	\vspace{-2.5cm}\caption{A plot of the principal branch \(\I^+_{1,1}(\zeta, \zeta_d)\) in the complex plane for decreasing values of \(\zeta_d\) (from left to right). The black cross denotes the branch point \(\zeta = -1/8\zeta_d\). Panel (d) shows \mbox{\(\I_{1,1}^+(\zeta, 0) = Z(\zeta)\)}. As \(\zeta_d \to 0^+\), the branch point, alongside the entire branch cut, is pushed towards \(\Re(\zeta) \to -\infty\). If \(\zeta_d\) were negative, the branch cut would instead join the branch point with \(\Re(\zeta)\to+\infty\), to which the branch cut would be pushed in the limit of \(\zeta_d\to0^-\).}
	\label{fig:I11_contours}
\end{figure}

Let us now show that \cref{eq:genZ_function} asymptotes to the known limits discussed in \cref{sec:previous_solutions}, as it should. First, in the limit of \(\zeta_d \to 0\), i.e., the limit of vanishing magnetic curvature, we employ the expansions
\begin{align}
	\zetaplus^+ &= \frac{1}{2\zeta_d} + \zeta - 2\zeta_d\zeta^2  + \order{\zeta_d^2} ,\\
	\zetaminus^+ &= \zeta - 2\zeta_d\zeta^2  + 8\zeta_d^2\zeta^3 + \order{\zeta_d^3} ,
	\label{eq:zetapm_expansion_small}
\end{align}
and the asymptotic form \(\Z(\zeta) \sim -\zeta^{-1}\) for finite \(\Im(\zeta)\) but \(|\Re(\zeta)| \to \infty\), to find
\begin{equation}
	\Z_a(\zetaplus^+) \sim \frac{-2\zeta_d}{\sqrt{a}},\quad \Z_a(\zetaminus^+) \sim \Z_a(\zeta). 
\end{equation}
Therefore, in the limit \(\zeta_d \to 0\), the principal branch satisfies
\begin{align}
	\I_{a,a}^+\sim\frac{1}{a}\Z_a(\zeta),
\end{align}
in agreement with \cref{eq:iab_jab_in_terms_of_Z_function}. This is visualised in \cref{fig:I11_contours}. One can perform an analogous calculation with $\J_{a,b}^+$ to obtain the second expression in \cref{eq:iab_jab_in_terms_of_Z_function}. Note that the second branch satisfies
\begin{equation}
	\I_{a,a}^- \sim -\frac{1}{a}\Z_a(-\zeta)
\end{equation}
in the limit \(\zeta_d \to 0\), which is not related to the correct expression for \(\I_{a, a}\) at zero magnetic curvature. The `connection' between the two branches, viz., the branch cut, is `sent to infinity' as \(\zeta_d \to 0\) (see  \cref{fig:I11_contours}), and so the second branch \(\I_{a, a}^-\) is `lost' in the limit of zero magnetic curvature. In this way, the dispersion function loses all but one of its branches and becomes single-valued.

\begin{figure}
	\centering \includegraphics[scale=0.27]{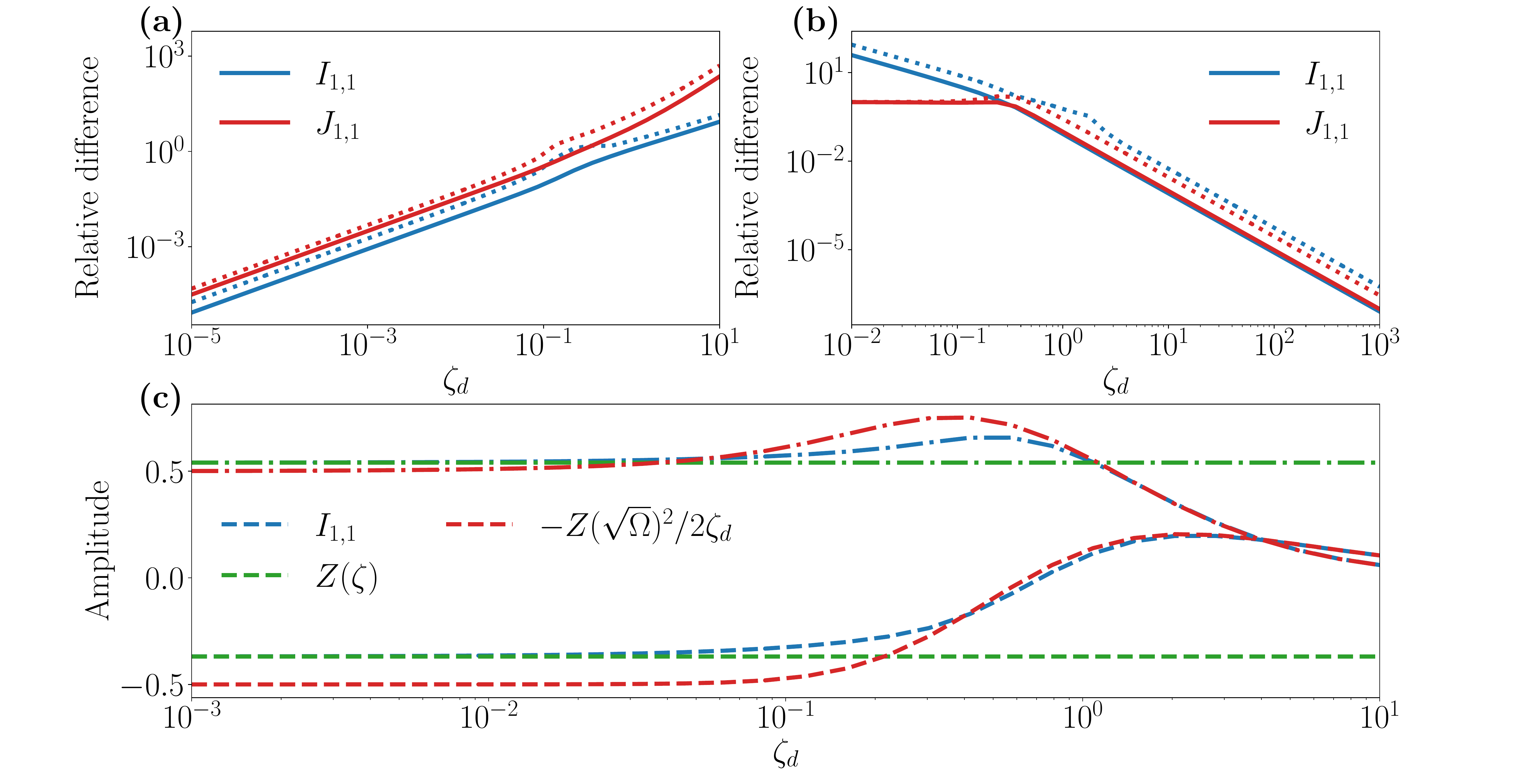}
	\caption{Plots of the asymptotic convergence of $\I_{a,b}$ and $\J_{a,b}$ to their known limits. Panels (a) and (b) demonstrate the convergence of \(\I_{1,1}^+\) and \(\J_{1,1}^+\), given by \cref{eq:genZ_function} and \cref{eq:jaa_general}, to their small- and large-\(\zeta_d\) limits, given by \cref{eq:iab_jab_in_terms_of_Z_function} and \cref{eq:2d_limit}, respectively. We define the relative difference of two functions \(f\) and \(g\) as \(|f-g|/\min\{|f|, |g|\}\). Note that this is ill-defined if one of the functions is identically zero, so for the \(\J_{1,1}\) comparison in panel (b), we plot simply \(|\J_{1,1}|\) because we expect to recover \(\J_{1,1} = 0\) in the 2D limit \cref{eq:2d_limit}. The solid and dotted lines in (a) and (b) show the average and maximum relative difference, respectively, as computed over a grid of \(32\times32\) points for \(\zeta\), equally spaced in \(\Re(\zeta) \in [1, 1]\), \(\Im(\zeta) \in [0, 1]\), for each value of \(\zeta_d\). Panel (c) demonstrates the convergence of the real (dashed) and imaginary (dash-dot) parts of \(\I_{1,1}^+\) to the small- and large-\(\zeta_d\) limits for a fixed \(\zeta = 1+i\), which are given by \(\Z(\zeta)\) and \(-\Z^2(\sqrt{\Omega})/\zeta_d\), with \(\Omega = \zeta/2\zeta_d\), respectively.}
	\label{fig:Z_comparison}
\end{figure}

On the other hand, the 2D limit can be found by taking the limit \(\zeta\sim\zeta_d\to\infty\), which is equivalent to dropping the \(u\) term from the denominators in \cref{eq:iab_original} and \cref{eq:jab_original}. In this case, 
\begin{align}
	\zetapm^+ = \sqrtp{\frac{\zeta}{2\zeta_d}} \pm \frac{1}{4\zeta_d} + O(\zeta_d^{-2}),
	\label{eq:zetapm_expansion_large}
\end{align}
and so one obtains \cref{eq:2d_limit}.

\Cref{fig:Z_comparison} compares the exact expressions \cref{eq:genZ_function} and \cref{eq:jaa_general} with their known asymptotic limits in the case of vanishing magnetic drifts \cref{eq:iab_jab_in_terms_of_Z_function} and 2D perturbations \cref{eq:2d_limit}, respectively. It is evident that, while these known asymptotic limits are obtained in the cases of small and large \(\zeta_d\), they are not a good approximation of \(\I_{a,b}\) and \(\J_{a,b}\) for \(\zeta_d\sim1\), as one would expect.

\subsection{Numerical comparison with \citet{gurcan14}}
\label{sec:numerical_comparison}

\begin{figure}
	\centering\includegraphics[scale=0.27]{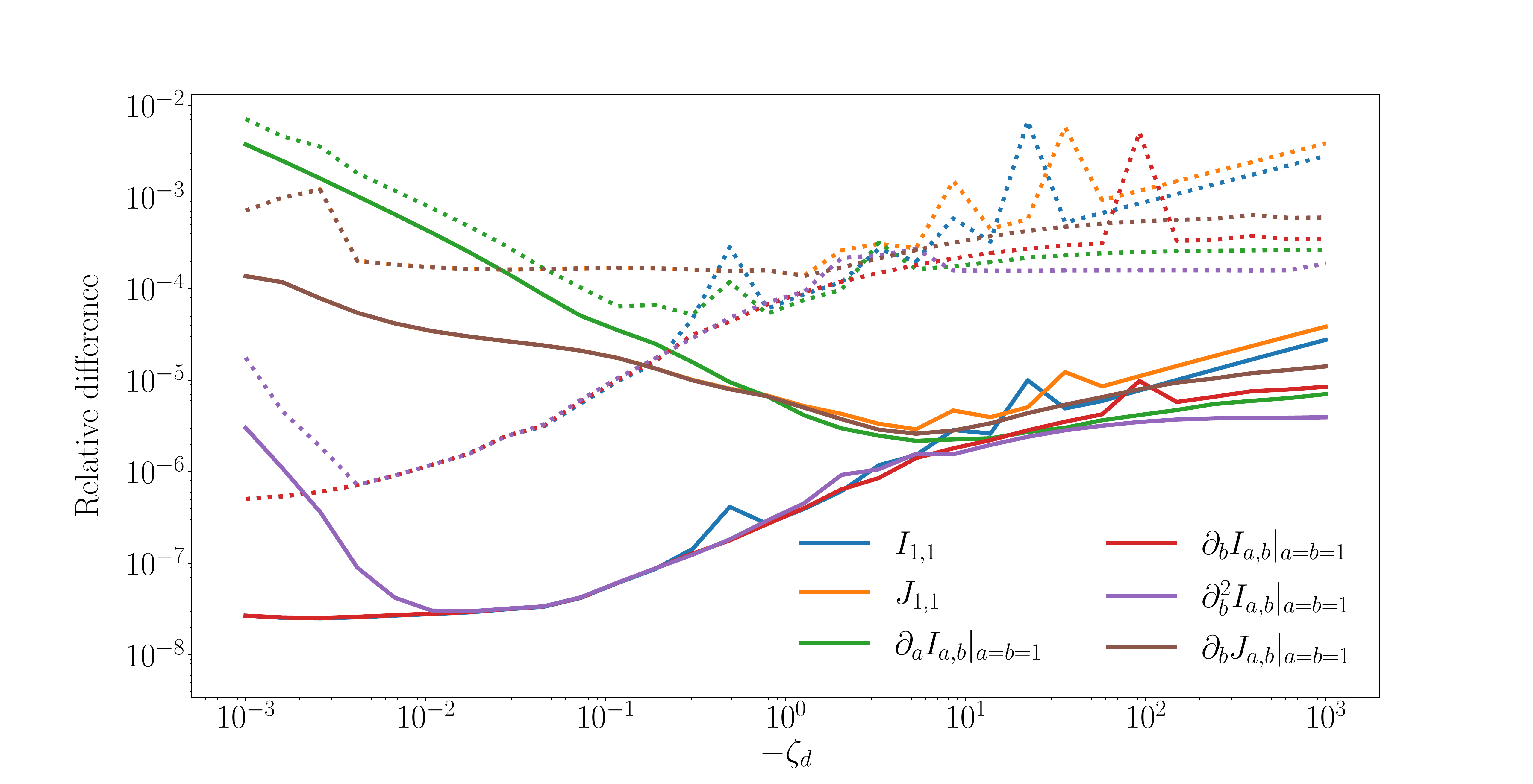}
	\caption{Mean (solid) and maximum (dotted) relative difference (defined as in \cref{fig:Z_comparison}) between expressions \cref{eq:genZ_function}, \cref{eq:jaa_general}, \cref{eq:diab_da}--\cref{eq:djab_db} and their equivalents derived from \cref{eq:gurcan_imn_equivalent}, computed via the code published at \href{https://github.com/gurcani/zpdgen}{https://github.com/gurcani/zpdgen}. For each \(\zeta_d\), we evaluated the respective functions at an equally spaced grid of \(32\times32\) points in the region \(\Re(\zeta)\in[-10, 10]\), \(\Im(\zeta)\in[0, 10]\).}
	\label{fig:gurcan_comparison}
\end{figure}

\citet{gurcan14} consider a very similar problem to the one on which this paper has focused but from a numerical perspective. In particular, they discuss the numerical integration of the function
\begin{equation}
	\mathcal{I}_{nm}(\zeta_\alpha, \zeta_\beta, b) = \frac{2}{\sqrt{\pi}}\int_0^\infty \rmd x_\perp\int_{-\infty}^\infty\rmd x_\parallel \: \frac{x_\perp^n x_\parallel^m J_0^2(\sqrt{2b}x_\perp)e^{-x_\parallel^2-x_\perp^2}}{x_\parallel^2 + x_\perp^2/2 + \zeta_\alpha - \zeta_\beta x_\parallel},
	\label{eq:gurcan_imn}
\end{equation}
defined for \(\Im(\zeta_\alpha) > 0\), real \(\zeta_\beta\) and \(b\), and \(n>1\). With a few algebraic manipulations, it can be shown that, for odd \(n\),
\begin{align}
	\mathcal{I}_{nm}\left(-\frac{\zeta}{2\zeta_d}, -\frac{1}{2\zeta_d}, 0\right) = -\frac{1}{\zeta_d}\left\{
	\begin{array}{cc}
		(-\partial_a)^{\frac{m}{2}}(-\partial_b)^{\frac{n-1}{2}} \I_{a,b}^+(\zeta, \zeta_d)\vert_{a=b=1}, &  \text{for \(m\) even},\\[4mm]
		(-\partial_a)^{\frac{m-1}{2}}(-\partial_b)^{\frac{n-1}{2}} \J_{a,b}^+(\zeta, \zeta_d)\vert_{a=b=1}, &  \text{for \(m\) odd}.
	\end{array}
	\right.
	\label{eq:gurcan_imn_equivalent}
\end{align}
The above expression is actually correct only for \(\zeta_d < 0\), otherwise \cref{eq:gurcan_imn} computes the second (\(\branchsymb = -\)) branch of \(\I_{a,b}\), \(\J_{a,b}\), and their derivatives. In the case \(\zeta_d < 0\), the requirement \(\Im(\zeta_\alpha) > 0\) implies that \(\Im(\zeta) > 0\). \Cref{fig:gurcan_comparison} shows a comparison between the values obtained via the results of this work [represented by \cref{eq:genZ_function}, \cref{eq:jaa_general}, \cref{eq:diab_da}--\cref{eq:djab_db}] and the \citet{gurcan14} result \cref{eq:gurcan_imn} in the region \(\Re(\zeta)\in[-10, 10]\), \(\Im(\zeta)\in[0, 10]\), \(\zeta_d \in [-10, -0.001]\). There is good agreement for all tested values of \(\zeta\) and \(\zeta_d\), with less than \(1\%\) relative difference in most cases. It is important to stress that as our solution uses only standard functions, e.g., the plasma dispersion function \(\Z\) and \(\sqrt{}\), for which there exist very efficient numerical algorithms. We found that even a na\"ive, unoptimised \texttt{Python} implementation took anywhere between \(20\) and \(80\) times less time to compute \(\I_{a,b}\), \(\J_{a,b}\), and their derivatives than the direct numerical integration of \cref{eq:gurcan_imn} implemented in Fortran at \href{https://github.com/gurcani/zpdgen}{https://github.com/gurcani/zpdgen}.

\section{Analytic continuation for the inverse Laplace transform}
\label{sec:analytic_continuation}
Together, the expressions \cref{eq:genZ_function} and \cref{eq:jaa_general} for $\I_{a,a}$ and $\J_{a,a}$, respectively, along with the derivatives \cref{eq:diab_da}--\cref{eq:djab_db}, allow us to calculate $\mathbf{L}$, and hence the Laplace-transformed fields \cref{eq:solving_for_fields}. Recall that in order to determine the evolution of the system as a function of time, we need to compute the inverse Laplace transform
\begin{align}
	\vec{\chi}_{\vec{k}}(t) = \frac{1}{2\pi i} \int_{\CB} \rmd p \: e^{pt} \hat{\vec{\chi}}_{\vec{k}}(p),
	\label{eq:fields_inverse_laplace}
\end{align}
where the contour of integration $\CB$ is once again as in \cref{fig:inverse_laplace_transform}, and we remind the reader that \(\hat{\vec{\chi}}_{\vec{k}}(p)\) is given by
\begin{align}
	\hat{\vec{\chi}}_{\vec{k}}(p)  = \frac{(\text{adj}\:\mathbf{L})	\vec{G}	}{\det\mathbf{L}},
	\label{eq:solving_for_fields_copy}
\end{align}
where the vector of initial conditions \(\vec{G}\) is given by \crefrange{eq:g_phi}{eq:g_B}.
 
\begin{figure}
	\centering
	
	\begin{tikzpicture}[thick]
	
	\tikzset{
		pics/.cd,
		vector out/.style={
			code={
				\draw[#1] (0,0)  circle (1) (45:1) -- (225:1) (135:1) -- (315:1);
			}
		}
	}

	\def\xaxislength{5}
	\def\yaxislength{5}
	\def\xoffset{0}
	\def\yoffset{0}
	\def\zoffset{0}
	\def\psone{1.2}
	\def\pstwo{-0.5}
	
	\draw[-latex] (-\xaxislength/2,\yoffset) -- (\xoffset,\yoffset) -- (\xoffset + \xaxislength/2,\yoffset);
	\draw (\xoffset + \xaxislength/2,\yoffset) node[anchor=north] {$\Re(p)$};
	
	\draw[-latex] (\xoffset,\yoffset - \yaxislength/2) -- (\xoffset,\yoffset) -- (\xoffset,\yoffset + \yaxislength/2);
	\draw (\xoffset,\yoffset +\yaxislength/2 ) node[anchor=east] {$\Im(p)$};

	\draw[decorate, decoration = zigzag, color=blue]  (\xoffset,\psone) -- (\xoffset,\yaxislength/2);
	
	\draw[decorate, decoration = zigzag]  (\xoffset - \xaxislength/2,\psone) -- (\xoffset,\psone);

	\draw (\xoffset, \psone) node[anchor = center, scale=3] {$\bcdot$};
	\draw (\xoffset, \psone) node[anchor = west] {$p_{\s_1}$};

	\draw[decorate, decoration = zigzag, color=blue]  (\xoffset,\pstwo) -- (\xoffset,-\yaxislength/2);
	
	\draw[decorate, decoration = zigzag]  (\xoffset - \xaxislength/2,\pstwo) -- (\xoffset,\pstwo);
	
	\draw (\xoffset, \pstwo) node[anchor = center, scale=3] {$\bcdot$};
	\draw (\xoffset, \pstwo) node[anchor = west] {$p_{\s_2}$};

\end{tikzpicture}
	
	\caption{This diagram shows the `principal' (in blue) and `dispersion' (in black) branch cuts for a plasma with one negatively and one positively charged species, labelled as \(\s_1\) and \(\s_2\), respectively.}
	\label{fig:moving_branch_cuts}
\end{figure}
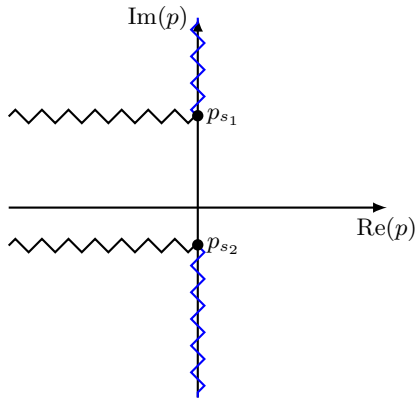

The results of \cref{sec:evaluation_of_iab} show that the entries of \(\Lmat\) have branch points at \(\zeta_\s = -1/8\zeta_{d\s}\), or equivalently, at \(p = p_\s\), where
\begin{equation}
	p_\s \equiv \frac{i \kpar^2 \vths^2}{8 \omega_{d\s}},
	\label{eq:ps_def}
\end{equation}
but are otherwise free of poles since, apart from the square roots and the associated branch cuts, they are composed of entire functions. Recall that we have defined the branches of the dispersion function \(D\) using the principal branch of the square root in \cref{eq:x_pm_definition}. Therefore, the relevant branch \(D^{\vec{\branchsymb}}\) that enters the inverse Laplace transform is the principal branch given by \(\vec{\branchsymb} = (+, ..., +)\). This has branch cuts that connect the branch points \(\zeta_\s = -1/8\zeta_{d\s}\) to \(\zeta_\s \to -\sgn(\zeta_{d\s}) \infty\), or, equivalently, \(p = p_\s\) to \mbox{\(p \to i\sgn(\zeta_{d\s})\infty\)}. While this choice of the principal branch and branch cuts was convenient for obtaining the closed forms of $\I_{a,b}$, $\J_{a,b}$, and \(D\), and their properties, it is not necessarily the best one for performing the inverse Laplace transform \cref{eq:fields_inverse_laplace}. Instead, we would like to rotate the branch cuts by \(\sgn(\zeta_{d\s}) \pi/2\) around \(p_\s\), so that they are parallel to the real \(p\) axis, as shown in \cref{fig:moving_branch_cuts}. Let us call the branch \(\mainD\) of the dispersion function obtained this way the `dispersion' branch. Crucially, the rotation of the branch cuts does not disturb the values of the dispersion function at \(\Re(p) > 0\). Therefore, \(\mainD(p) = D^{(+,...+)}(p)\) for \(\Re(p) > 0\). This ensures that the `unphysical' unstable zeros of the other branches of the dispersion function, which are a consequence of \cref{eq:D_zetaconj}, do not contribute to the solution (see also discussion in \cref{sec:branches_disp}); the only unstable solutions that are picked up by the inverse Laplace transform are those of the principal branch.

\begin{figure}
	\centering
	
	\scalebox{0.9}{\begin{tikzpicture}[scale=1, thick, every node/.style={scale=1.2}]
			\newcommand{\arrowIn}{
				\tikz \draw[-latex] (-1pt,0) -- (1pt,0);
			}
		
            \def\xaxislength{10}
            \def\yaxislength{7.5}
            \def\xoffset{0.25}
            \def\yoffset{2}
            \def\zoffset{0.1}
            
            \draw[-latex] (\xoffset - \xaxislength/3,\yoffset,\zoffset) -- (\xoffset,\yoffset,\zoffset) -- (\xoffset + \xaxislength/2,\yoffset,\zoffset);
            \draw (\xoffset + \xaxislength/2,\yoffset,\zoffset) node[anchor=north] {$\Re(p)$};
            
            \draw[-latex] (\xoffset,\yoffset - \yaxislength/2,\zoffset) -- (\xoffset,\yoffset,\zoffset) -- (\xoffset,\yoffset + \yaxislength/2,\zoffset);
            \draw (\xoffset,\yoffset +\yaxislength/2 ,\zoffset) node[anchor=east] {$\Im(p)$};
            
            \def\sigmaoffset{\xaxislength/4}
            \draw (\xoffset + \sigmaoffset,\yoffset,\zoffset) node[anchor= north west, opacity = 0.5] {$\sigma$};
            \draw[dashed] (\xoffset + \sigmaoffset,\yoffset - \yaxislength/2,\zoffset) -- (\xoffset + \sigmaoffset,\yoffset + \yaxislength/2,\zoffset);

            \draw (\xoffset + \sigmaoffset/2,\yoffset + \yaxislength/3 ,\zoffset) node[anchor = center] {$\times$};
            \draw (\xoffset - \sigmaoffset/2,\yoffset + \yaxislength/5 ,\zoffset) node[anchor = center] {$\times$};
            \draw (\xoffset + \sigmaoffset/4,\yoffset - \yaxislength/8 ,\zoffset) node[anchor = center] {$\times$};
            \draw (\xoffset + 2*\sigmaoffset/4,\yoffset - 7*\yaxislength/16,\zoffset) node[anchor = center] {$\times$};
            
			\draw[decorate, decoration = zigzag]  (\xoffset - \xaxislength/3,\yoffset - 9*\yaxislength/32,\zoffset) -- (\xoffset,\yoffset - 9*\yaxislength/32,\zoffset);
            
            \draw (\xoffset + \sigmaoffset,\yoffset + \yaxislength/6,\zoffset) node[anchor = north west
            , opacity = 0.5] {$\CB$};
            \draw (\xoffset - \xaxislength/4,\yoffset + \yaxislength/6,\zoffset) node[anchor = north west] {$\CInvL$};
            \draw (\xoffset - \xaxislength/4,\yoffset,\zoffset) node[anchor = north west] {$\rho$};

        	\def\lineseparation{0.05}
        	\def\poleoffset{0.2}
        	\def\contourstyle{solid}
        	
        	\def\circleradius{0.2}
        	\def\circleangle{atan(\lineseparation/(\circleradius))}
        	
        	\def\branchoffset{0.25}
        	\def\branchangle{atan(\branchoffset/\circleradius)}
        	
        	\def\arrowshift{0.2}
        	
        	\draw[\contourstyle] (\xoffset - \xaxislength/4,\yoffset + \yaxislength/2,\zoffset) 
        	-- node[midway, sloped, allow upside down, rotate = 180] {\arrowIn}
        	(\xoffset - \xaxislength/4,\yoffset + \yaxislength/3 + \lineseparation,\zoffset);
        	
        	\draw[\contourstyle] (\xoffset - \xaxislength/4,\yoffset + \yaxislength/3 - \lineseparation,\zoffset)
        	-- node[midway, sloped, allow upside down, rotate = 180] {\arrowIn}
        	(\xoffset - \xaxislength/4,\yoffset + \yaxislength/5 + \lineseparation,\zoffset);
        	
        	\draw[\contourstyle] (\xoffset - \xaxislength/4,\yoffset + \yaxislength/5 - \lineseparation,\zoffset)
        	-- node[midway, sloped, allow upside down, rotate = 180] {\arrowIn}
        	(\xoffset - \xaxislength/4,\yoffset - \yaxislength/8 + \lineseparation,\zoffset);
        	
        	\draw[\contourstyle] (\xoffset - \xaxislength/4,\yoffset - \yaxislength/8 - \lineseparation,\zoffset)
        	-- node[midway, sloped, allow upside down, rotate = 180] {\arrowIn}
        	(\xoffset - \xaxislength/4,\yoffset - 9*\yaxislength/32 + \branchoffset,\zoffset);
        	
        	\draw[\contourstyle] 	(\xoffset - \xaxislength/4,\yoffset - 9*\yaxislength/32 - \branchoffset,\zoffset)
        	-- node[midway, sloped, allow upside down, rotate = 180] {\arrowIn}
        	(\xoffset - \xaxislength/4,\yoffset - 7*\yaxislength/16 + \lineseparation,\zoffset);
        	
        	\draw[\contourstyle] (\xoffset - \xaxislength/4,\yoffset - 7*\yaxislength/16 - \lineseparation,\zoffset)
        	-- node[midway, sloped, allow upside down, rotate = 180] {\arrowIn}
        	(\xoffset - \xaxislength/4,\yoffset - \yaxislength/2,\zoffset);

        	\draw[\contourstyle] (\xoffset + \sigmaoffset/2 - \poleoffset,\yoffset + \yaxislength/3 + \lineseparation ,\zoffset)
        	-- node[midway, sloped, allow upside down] {\arrowIn}
        	(\xoffset - \xaxislength/4,\yoffset + \yaxislength/3 + \lineseparation,\zoffset);
        	
        	
        	\draw (\xoffset + \sigmaoffset/2 - \poleoffset, \yoffset + \yaxislength/3 - \lineseparation,\zoffset) arc (-180 + \circleangle : 180 - \circleangle : \circleradius);
        	
        	\draw[\contourstyle] (\xoffset - \xaxislength/4,\yoffset + \yaxislength/3 - \lineseparation,\zoffset)
        	-- node[midway, sloped, allow upside down] {\arrowIn}
        	(\xoffset + \sigmaoffset/2 - \poleoffset,\yoffset + \yaxislength/3 - \lineseparation ,\zoffset);

        	\draw[\contourstyle] (\xoffset - \sigmaoffset/2 - \poleoffset, \yoffset + \yaxislength/5 + \lineseparation,\zoffset)
        	-- node[midway, sloped, allow upside down] {\arrowIn}
        	(\xoffset - \xaxislength/4,\yoffset + \yaxislength/5 + \lineseparation,\zoffset);
        	
        	\draw (\xoffset - \sigmaoffset/2 - \poleoffset, \yoffset + \yaxislength/5 - \lineseparation,\zoffset) arc (-180 + \circleangle : 180 - \circleangle : \circleradius);
        	
        	\draw[\contourstyle] (\xoffset - \xaxislength/4,\yoffset + \yaxislength/5 - \lineseparation,\zoffset)
        	-- node[midway, sloped, allow upside down] {\arrowIn}
        	(\xoffset - \sigmaoffset/2 - \poleoffset, \yoffset + \yaxislength/5 - \lineseparation,\zoffset);
        	
        	\draw[\contourstyle]  (\xoffset + \sigmaoffset/4 - \poleoffset,\yoffset - \yaxislength/8 + \lineseparation,\zoffset)
        	-- node[midway, sloped, allow upside down] {\arrowIn}
        	(\xoffset - \xaxislength/4, \yoffset - \yaxislength/8 + \lineseparation,\zoffset);
        	
        	\draw 	(\xoffset + \sigmaoffset/4 - \poleoffset,\yoffset - \yaxislength/8 - \lineseparation,\zoffset) arc (-180 + \circleangle : 180 - \circleangle : \circleradius);
        	
        	\draw[\contourstyle]  (\xoffset - \xaxislength/4, \yoffset - \yaxislength/8 - \lineseparation,\zoffset)
        	-- node[midway, sloped, allow upside down] {\arrowIn}
        	(\xoffset + \sigmaoffset/4 - \poleoffset,\yoffset - \yaxislength/8 - \lineseparation,\zoffset);
        	
        	\draw[\contourstyle] (\xoffset,\yoffset - 9*\yaxislength/32 + \branchoffset,\zoffset)
        	-- node[midway, sloped, allow upside down] {\arrowIn}
        	(\xoffset - \xaxislength/4,\yoffset - 9*\yaxislength/32 + \branchoffset,\zoffset);
        	
        	\draw (\xoffset,\yoffset - 9*\yaxislength/32 - \branchoffset,\zoffset) arc (-90 : 90 : \branchoffset);
        	
        	\draw[\contourstyle] (\xoffset - \xaxislength/4,\yoffset - 9*\yaxislength/32 - \branchoffset,\zoffset)
        	-- node[midway, sloped, allow upside down] {\arrowIn}
        	(\xoffset,\yoffset - 9*\yaxislength/32 - \branchoffset,\zoffset);
        	
        	\draw[\contourstyle] (\xoffset + 2*\sigmaoffset/4 - \poleoffset,\yoffset - 7*\yaxislength/16 + \lineseparation,\zoffset)
        	-- node[midway, sloped, allow upside down] {\arrowIn} 
        	(\xoffset - \xaxislength/4,\yoffset - 7*\yaxislength/16 + \lineseparation,\zoffset);
        	
        	\draw 		(\xoffset + 2*\sigmaoffset/4 - \poleoffset,\yoffset - 7*\yaxislength/16 - \lineseparation,\zoffset) arc (-180 + \circleangle : 180 - \circleangle : \circleradius);
        	
        	\draw[\contourstyle] (\xoffset - \xaxislength/4,\yoffset - 7*\yaxislength/16 - \lineseparation,\zoffset)
        	-- node[midway, sloped, allow upside down] {\arrowIn}
            (\xoffset + 2*\sigmaoffset/4 - \poleoffset,\yoffset - 7*\yaxislength/16 - \lineseparation,\zoffset);

        \end{tikzpicture}}
	
	\caption{Same as in \cref{fig:inverse_laplace_transform}, except that the contour associated with the inverse Laplace transformation \cref{eq:fields_inverse_laplace} has now been shifted to \(\Re(p) = \rho\), deforming it such that it does not cross any of the poles or the branch cut. We denote this new contour \(\CInvL\). The original contour is shown by the vertical dashed line. The integrals along \(\CB\) and \(\CInvL\) are equal by Cauchy's integral theorem.}
	\label{fig:inverse_laplace_transform_continuation}
\end{figure}
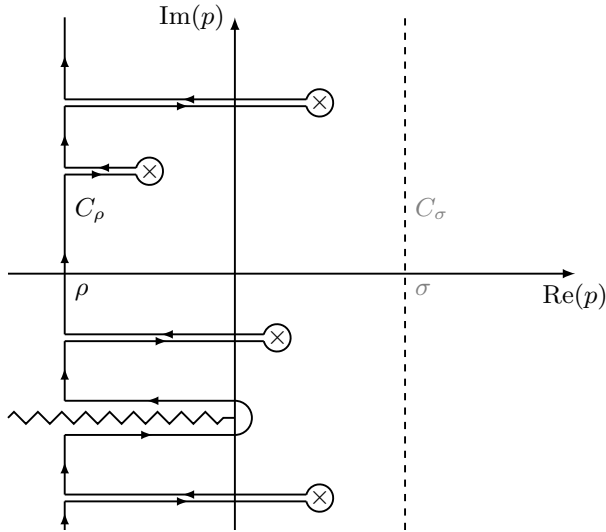

With this choice for the branch cuts of the dispersion function, we are ready to perform the inverse Laplace transform \cref{eq:fields_inverse_laplace}. This is done in the usual way, viz., by pushing the integration contour \(\CB\) towards \(\Re(p) \to -\infty\), with the proviso that it must be deformed so as not to cross any singularities, e.g., poles or branch cuts. Pushing the contour to the vertical line at \(\Re(p) = \rho\), we find the new integration contour \(\CInvL\) (see \cref{fig:inverse_laplace_transform_continuation}). Since there are no singularities between \(\CB\) and \(\CInvL\), Cauchy's integral theorem ensures that the integrals over these two contours are equal. Taking the limit of \(\rho \to -\infty\), it is evident that the contributions arising from the vertical segments of $\CInvL$ are exponentially small\footnote{They are exponentially small at any \(t > 0\) because the integrand of the inverse Laplace transformation \cref{eq:inverse_laplace_transform} contains a factor \(e^{\rho t}\).}, while those arising from the integration along the horizontal segments leading towards and away from the poles cancel, leaving the contributions from the poles. The integration around the branch cuts is more subtle and will be discussed shortly.

There are several singularities present in \cref{eq:solving_for_fields_copy}, and hence in the integrand in \cref{eq:fields_inverse_laplace}. The first is the so-called `ballistic response' associated with the initial conditions contained within $\vec{G}$, arising from simple poles located along $\Re(p) = 0$, viz., 
\begin{align}
	\lim_{\rho \rightarrow -\infty} \frac{1}{2 \pi i} \int_{\CInvL} \rmd p \: e^{pt} \frac{\gsk}{p + i \kpar \vpar + i \omega_{D\s}} = \gsk e^{-i(\kpar \vpar + \omega_{D\s})t},
	\label{eq:ballistic_response}
\end{align}
where we have assumed that $\gsk$ is a smooth function. Plugging this into \crefrange{eq:g_phi}{eq:g_B}, we find that the contribution to \(\vec{\chi}_{\vec{k}}(t)\) due to the ballistic response can be written as
\begin{align}
	\vec{\chi}_{\vec{k}0}(t) =& \sum_\s  \int \rmd^3 \vec{v} \ \mathbf{L}^{-1}(-i\kpar\vpar -i\omega_{D\s}) \gsk e^{-i(\kpar \vpar + \omega_{D\s})t} \nonumber \\ &\left( \frac{q_\s n_{0\s}}{q_r n_{0r}} \frac{1}{n_{0\s}} J_0 (b_\s), \frac{q_\s n_{0\s} \vths}{q_r n_{0r} \vthr } \frac{1}{n_{0\s}}\frac{\vpar}{\vths} J_0 (b_\s), - \frac{\beta_\s}{2}\frac{1}{n_{0\s}}\frac{\vperp^2}{\vths^2} \frac{2 J_1(b_\s)}{b_\s} \right)^T.
\end{align}	
There is a wealth of interesting physics that can arise from the ballistic response, see, e.g., \citet{ewart22} and references therein, in the context of the Vlasov-Poisson system. However, this is not the focus of the present work and so will not be discussed further.

Another source of non-analyticity are the solutions to the dispersion relation $\mainD = 0$, should any of these exist. The contributions to \cref{eq:fields_inverse_laplace} arising from the zeros \(p = p_j\) of \(\mainD\) can be written as
\begin{equation}
	\label{eq:pole_contributions}
	\sum_j \Res[\hat{\vec{\chi}}_{\vec{k}}(p), p_j] e^{p_j t}.
\end{equation}
It is evident that unlike the ballistic response, whose time dependence is an oscillating exponential, the terms \cref{eq:pole_contributions} can, in general, be exponentially decaying (i.e., stable) for \(\Re(p_j) < 0\) or growing (i.e., unstable) for \(\Re(p_j) > 0\). 

Finally, singularities may arise from the functions \crefrange{eq:iab_flr}{eq:mab_flr} that are contained within both \(\text{adj}\:\mathbf{L}\) and \(\det\mathbf{L}\). As discussed above, these functions are free of poles, but are multivalued. Deforming the integration contour \(\CInvL\) around their branch cuts (see \cref{fig:inverse_laplace_transform_continuation}) gives a nontrivial contribution to \cref{eq:fields_inverse_laplace}. Letting \(\vec{B}_\s(t)\) be the contribution from the integral around the branch cut due to a given species \(\s\), we can finally write the full solution for \(\vec{\chi}_{\vec{k}}(t)\) as
\begin{equation}
	\vec{\chi}_{\vec{k}}(t) = \vec{\chi}_{\vec{k}0}(t) + \sum_j \Res[\hat{\vec{\chi}}_{\vec{k}}(p), p_j] e^{p_j t} + \sum_\s \vec{B}_\s(t).
	\label{eq:chi_time_evolution}
\end{equation}

In \cref{app:integral_around_the_branch_cut}, we show that, in the long-time limit \(t \to \infty\), the branch-cut contribution \(\vec{B}_\s\) for each species is dominated by that arising from the branch point itself, and exhibits an algebraic decay \(\propto t^{-3/2}\). The same algebraic decay was found by \cite{kim94, kuroda98} in their treatment of the toroidal ITG mode. Such a `continuum mode' \citep{kuroda98,sugama99} is a direct consequence of the multivaluedness of \cref{eq:iab_original} and \cref{eq:jab_original}, in that such multivaluedness gives rise to a branch point and to the resulting discontinuity. This behaviour is qualitatively different from that of a plasma in a straight magnetic field, whose dispersion function is single-valued, meaning that there are no branch cuts and hence no continuum modes. Note that nonexponentially decaying solutions to similar initial-value problems can also be found in other contexts; see, e.g., \citet{taylor65, sedlacek95}.

Equation \cref{eq:chi_time_evolution} is our final expression for the time evolution of \(\vec{\chi}_{\vec{k}}(t)\). Depending on whether there are any unstable solutions, we find that either: (i) there are solutions to \(\mainD(p) = 0\) for \(\Re(p) > 0\). In that case, the long-time solution is dominated by the solution with largest \(\Re(p)\); or (ii) there are no solutions to \(\mainD(p) = 0\) for \(\Re(p) > 0\). In that case, the long-time solution is dominated by the ballistic response \cref{eq:ballistic_response} and by waves with frequencies \(\omega = ip_\s = -\kpar^2\vths^2/8\omega_{d\s}\) that exhibit a nonexponential decay \(\propto t^{-3/2}\).

\section{From drift kinetics to gyrokinetics}
\label{sec:from_DK_to_GK}

The analytical forms of the integrals derived in \cref{sec:the_generalised_plasma_dispersion_function} are not without their limitations: in their derivation, we assumed both the drift-kinetic limit and the case of equal magnetic drifts (see \cref{sec:low_beta_drift_kinetic_limit}). We will now devote some space to a brief discussion of how one can relax these assumptions. 

\subsection{Bessel functions}
\label{sec:bessel}

The drift-kinetic assumption is perhaps the more egregious approximation, especially given that the presence of finite-Larmor-radius effects, or otherwise, can have a nontrivial impact on the plasma dynamics \citep[see, e.g.,][and references therein]{smolyakov2002,parisi20,parisi22}. Thankfully, however, it can be relaxed if one is willing to pay the price of complicated analytical expressions. Noting that $2 J_0(b_\s) J_1(b_\s) = - \partial J_0^2(b_\s)/\partial b_\s$, it is clear that the Bessel functions $J_0$ and $J_1$ always appear quadratically in \cref{eq:iab_flr}--\cref{eq:mab_flr}, for which there are known, rapidly converging Taylor series \citep{neumann1871, watson66}:
\begin{equation}
	J_n^2(b_\s) = \sum_{m=0}^\infty \frac{(-1)^m (2n+2m)!}{m!(2n+m)![(n+m)!]^2}\left(\frac{b_\s}{2}\right)^{2n+2m}.
	\label{eq:j_taylor_series}
\end{equation}
Using this expansion in \cref{eq:iab_flr}--\cref{eq:mab_flr}, one can, in principle, compute each of the resulting integrals analytically, and thus obtain an absolutely convergent series for the resulting gyrokinetic dispersion relation. This is done by noticing that their argument $b_\s$ only appears quadratically as $b_\s^2 = \mu \kperp^2 \rho_\s^2$, and thus the additional factors of $\mu$ can be handled by partial differentiation with respect to $b$ before setting $a = b$ in \cref{eq:iab_flr}--\cref{eq:mab_flr}. For example, \cref{eq:iab_flr} would give
\begin{align}
	\Ical_{a,b}^{(\s)}(\zeta_\s, \zeta_{d\s}, \zeta_{d\s}) & = \sum_{m=0}^\infty \frac{(2m)!}{(m!)^4}\left(\frac{\kperp \rho_\s}{2}\right)^{2m} \partial_b^m \I_{a,b}(\zeta_\s, \zeta_{d\s}),
\end{align}
with the other required integrals, viz., \(\Jcal^{(\s)}_{a,b}\) and \(\partial_a \Ical^{(\s)}_{a,b}\), satisfying similar expressions. We remind the reader that \(\Ical_{a,b}^{(\s)}\) refers to the FLR-containing integral \cref{eq:iab_flr}, while \(\I_{a,b}\) is the integral \cref{eq:iab_original} on which we have focused throughout most of this paper. Doing this calculation by hand seems rather daunting given the complicated expressions even for the low-order derivatives \(\partial_b^2 \I_{a,b}\) and \(\partial_b \J_{a,b}\) [see \cref{eq:d2iab_db2} and \cref{eq:djab_db}, respectively]. In practice, however, only a few terms would be needed due to the rapid convergence of the Taylor series \cref{eq:j_taylor_series}. Those wishing to compute these terms to an arbitrary order may want to do so by using symbolic libraries (e.g., those in \texttt{Wolfram Mathematica}) in order to calculate the derivatives analytically, which can then be imported into an associated numerical solver. An alternative approach would be to implement a recursive scheme to calculate numerically the \(m^\text{th}\)-order derivatives from the \((m-1)^\text{th}\) ones.

\subsection{General magnetic drifts}

Our second approximation was to neglect the difference between the curvature and $\gradd \! B$ drifts, taking their associated drift frequencies to be equal, i.e., $\zeta_{\kappa \s} = \zeta_{\gradd \! B \s}$, as in \cref{eq:magnetic_drifts_equality}. While this approximation is relatively well-satisfied in the context of magnetic-confinement fusion, there are certainly other systems in which it is not, e.g., space and astrophysical plasmas. By a simple change of variables to \(\mu' = \zeta_{Bs}\mu/\zeta_{\kappa \s}\) in \cref{eq:iab_no_flr}, we find
\begin{align}
	\I_{a,b}^{(\s)}(\zeta_\s, \zeta_{\kappa \s}, \zeta_{Bs}) &= \frac{1}{\sqrt{\pi}} \int_{-\infty}^\infty \rmd u \int_0^\infty \rmd \mu \: \frac{e^{-a u^2 - b \mu}}{u - \zeta_\s + \left(2u^2 \zeta_{\kappa \s} + \mu  \zeta_{Bs} \right)} \nonumber \\
	&= \frac{\zeta_{\kappa \s}}{\sqrt{\pi}\zeta_{Bs}} \int_{-\infty}^\infty \rmd u \int_0^\infty \rmd \mu' \: \frac{e^{-a u^2 - (b\zeta_{\kappa \s}/\zeta_{Bs})\mu'}}{u - \zeta_\s + \zeta_{\kappa \s}\left(2u^2  + \mu' \right)}.
\end{align}
Therefore, the integral \cref{eq:iab_no_flr} that enters \crefrange{eq:L_phiphi}{eq:L_bb} at \(a=b\) can be found in terms of the known integrals \cref{eq:iab_original} via
\begin{equation}
	\I_{a,a}^{(\s)}(\zeta_\s, \zeta_{\kappa \s}, \zeta_{Bs}) = \frac{\zeta_{\kappa \s}}{\zeta_{Bs}} \I_{a,b}(\zeta_\s, \zeta_{\kappa \s}),
\end{equation}
where now \(b = a \zeta_{\kappa \s}/\zeta_{Bs}\). Finally, to find \(\I_{a,b}\) for \(a\neq b\), one can Taylor expand
\begin{equation}
	\I_{a,b} = \sum_{m=0}^\infty \frac{(b-a)^m}{m!} \left.\partial_b^m \I_{a,b}\right|_{a=b}.
	\label{eq:general_ab_Taylor}
\end{equation}
Fortunately, just as in \cref{sec:bessel}, one can find closed, albeit complicated, analytical expressions for \(\partial_b^m \I_{a,b}|_{a=b}\) for any \(m\). The expansion should converge for arbitrary positive \(a\) and \(b\) since \cref{eq:general_ab_Taylor} is equivalent to expanding \(e^{-(b-a)\mu}\) in \cref{eq:iab_total_derivative} using its absolutely convergent Taylor series. 

\section{Electrostatic ITG: a detailed example}
\label{sec:itg}

To illustrate the results of \cref{sec:the_generalised_plasma_dispersion_function}, we provide an explicit calculation of the dispersion relation in the simple case of an electrostatic, ion-scale, temperature-gradient-driven instability, and compare the solution with well-known kinetic and fluid limits. 

In particular, we consider a two-species plasma of ions and electrons of comparable temperatures, \(T_{0i}\sim T_{0e}\). Since we want to consider electrostatic physics, we assume
\begin{equation}
	\beta_s \sim (\kperp d_s)^{-2} \ll 1,
\end{equation}
where \(d_s = \sqrt{m_sc^2 / 4\pi n_{0s} q_s^2}\) is the skin depth. Therefore, to lowest order, \(\Apark\) and \(\dBpark\) do not contribute to 
\cref{eq:fourier_transformed_gyrokinetic_potential}, and \cref{eq:eigenvalue_problem} simplifies to
\begin{align}
	\text{L}_{\phi \phi} \frac{q_r \phikhat}{T_{0r}} + \text{G}_\phi = 0.
	\label{eq:eigenvalue_problem_itg}
\end{align}
This implies that the dispersion relation is given simply by \(\text{L}_{\phi \phi} = 0\). Furthermore, we consider the frequencies of the perturbations to be comparable to the parallel streaming and drift frequencies of the ions, as well as the magnetic-drift frequency, viz.,
\begin{equation}
	\label{eq:itg_ordering}
	p\sim\kpar\vthi\sim\omega_{di}\sim\omega_{*i}\sim\eta_i \omega_{*i}.
\end{equation}
The relevant equilibrium length scales in our problem are thus the ion-density and ion-temperature gradients \(L_{n_i}^{-1}\) and \(L_{T_i}^{-1}\), respectively [see \cref{eq:equilibrium_gradients}], and the gradient of the magnetic field \(L_B^{-1} \equiv -\partial \ln B_0 / \partial x\). In the small-mass-ratio limit, \(m_e/m_i \ll 1\), \cref{eq:itg_ordering} implies
\begin{equation}
	\kpar \vthe \sim \sqrt{\frac{m_i}{m_e}} \kpar \vthi \gg p,
\end{equation}
i.e., the electrons stream quickly along the fields lines. Thus, \(\zeta_e \sim \zeta_{*e} \ll \zeta_i \sim \zeta_{*i}\), and the electron contributions to \(\text{L}_{\phi \phi}\) can be ignored. Choosing \(q_r = q_i = Ze\), \(T_{0r} = T_{0i}\), and \(n_{0r} = n_{0i}\), the expression \cref{eq:L_phiphi} simplifies to
\begin{equation}
	-\text{L}_{\phi\phi} = 1 + \tau + \left[ \zeta_i - \zeta_{*i} + \eta_i \zeta_{*i} \left(\partial_a + \frac{3}{2}\right) \right] \left. \I_{a, a}^{(i)} \right|_{a=1},
\end{equation}
where \(\tau \equiv T_{0i}/ZT_{0e}\) is the temperature ratio. To avoid carrying around an extra minus sign, we shall define \(D \equiv -\text{L}_{\phi \phi}\), the object whose zeros we shall be interested in. Using \cref{eq:genZ_function}, we obtain the principal branch of the ITG dispersion relation
\begin{equation}
	\label{eq:cITG}
	D = 1 + \tau - \frac{\zeta - \zeta_*}{2 \zeta_d} \Z_+\Z_- + \frac{\eta \zeta_*}{2\zeta_d} \left[ \left(\zetaplus\Z_- + \zetaminus\Z_+\right) + \left(\frac{\zeta}{\zeta_d} + \frac{1}{4\zeta_d^2} - 1\right) \Z_+\Z_- \right] = 0,
\end{equation}
where we have dropped the \(i\) subscripts, \(\zeta_\pm\) are given by \cref{eq:zetapm_definition}, and we are using the shorthand notation \(\Z_\pm \equiv \Z(\zeta_\pm)\). Note that the principal branch (i.e., \(\branchsymb=+\)) is implicitly used everywhere, but we have dropped the associated superscripts to reduce the notational clutter.

\begin{figure}
	\centering\includegraphics[scale=0.27]{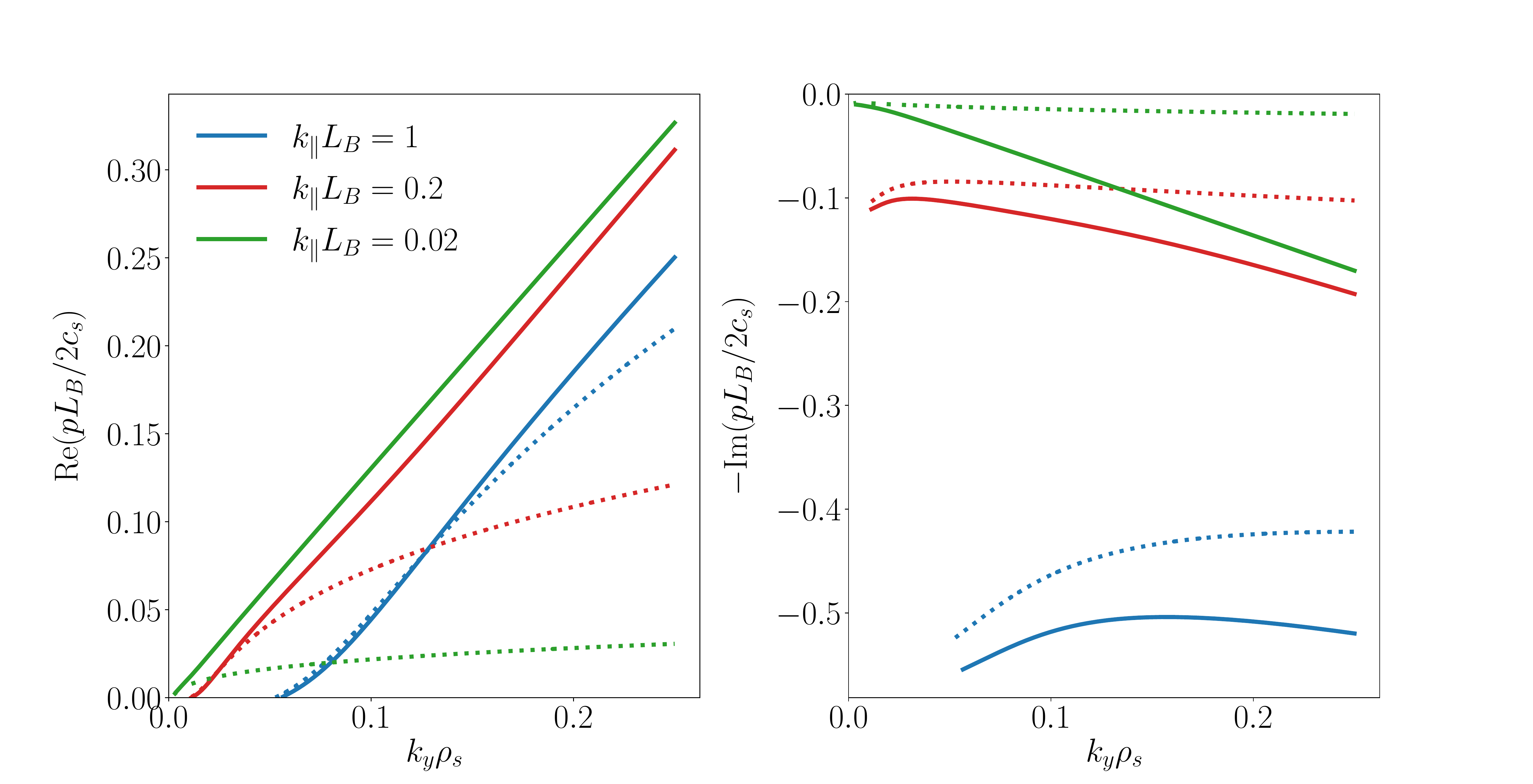}
	\caption{ A comparison between the growth rate and frequency of the most unstable solution to the kinetic dispersion relation with magnetic effects \cref{eq:cITG} and the slab dispersion relation \cref{eq:sITG}, represented by the solid and dotted lines, respectively. Here, \(\rho_s = \rho_i / \sqrt{2\tau}\) is the ion sound radius, and we have set \(\tau=0.1\) and \(\tau L_B / 2L_{T_{i}} = 2\).}
	\label{fig:cITG_vs_sITG}
\end{figure}
\begin{figure}
	\centering\includegraphics[scale=0.27]{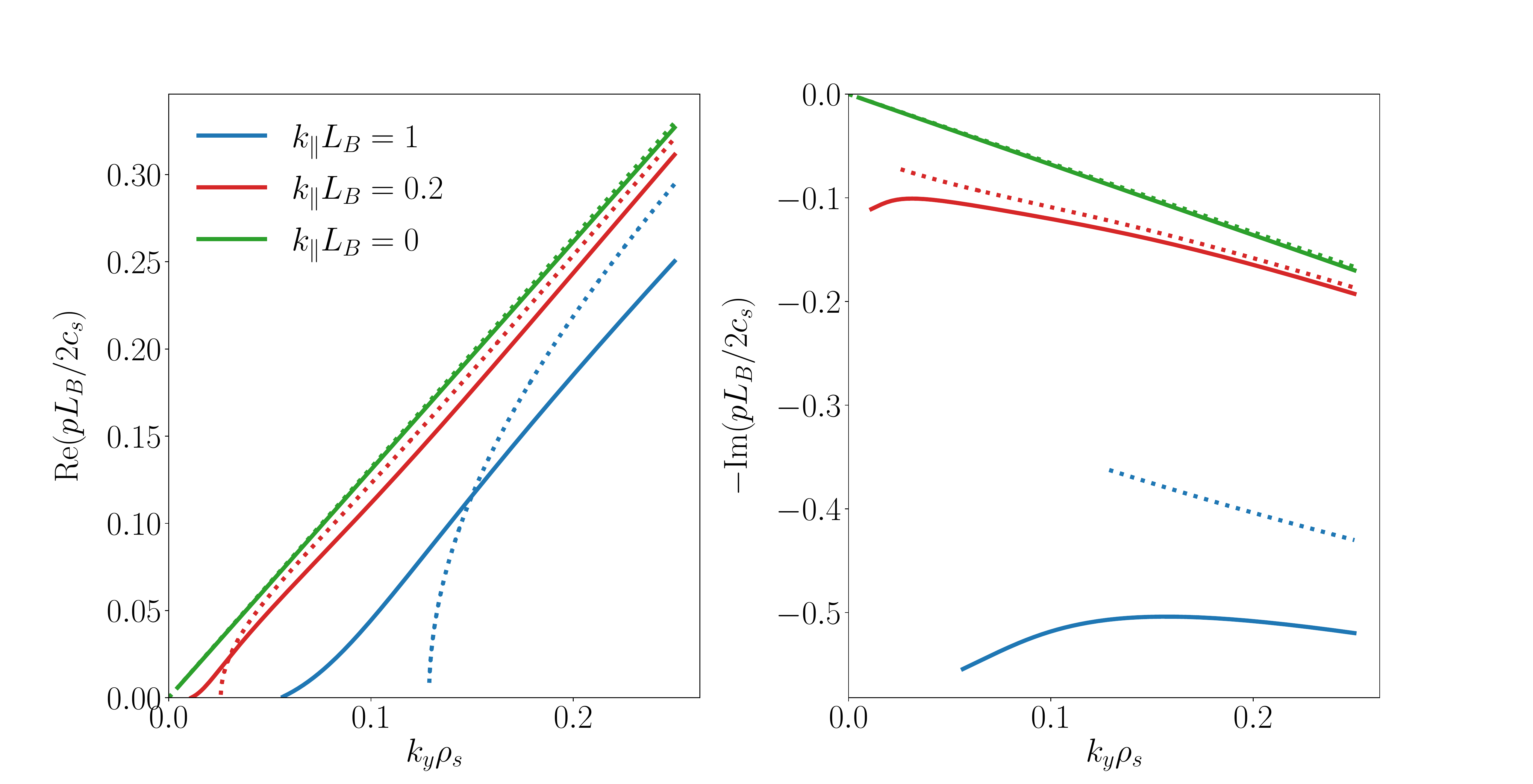}
	\caption{A comparison between the growth rate and frequency of the most unstable solution to the kinetic dispersion relation \cref{eq:cITG} and that obtained from the fluid equations \crefrange{eq:fluid_phi}{eq:fluid_T}, represented by the solid and dotted lines, respectively. The parameters used are the same as in \cref{fig:cITG_vs_sITG}.}
	\label{fig:cITG_vs_fluid}
\end{figure}

We can use \cref{eq:zetapm_expansion_small} and \cref{eq:zetapm_expansion_large} to verify that \cref{eq:cITG} converges to the correct limits in the case of: vanishingly small magnetic gradients (i.e., \(\zeta_d \to 0\)) 
\begin{align}
	&D_\text{slab} = 1 + \tau + (\zeta - \zeta_*)\Z(\zeta) + \eta \zeta_* \left[\zeta + \zeta^2\Z(\zeta) - \frac{1}{2}\Z(\zeta)\right] = 0;
	\label{eq:sITG} 
\end{align}
and of 2D perturbations (i.e., \(\zeta\sim\zeta_*\sim\zeta_d\to\infty\))
\begin{align}
	&D_\text{2D} = 1 + \tau - (\Omega - \Omega_*)\Z(\sqrt{\Omega})^2 + \eta\Omega_*\left[2 \sqrt{\Omega} \Z(\sqrt{\Omega}) + \left(2\Omega - 1\right)\Z(\sqrt{\Omega})^2\right]= 0, \label{eq:cITG_2D} 
\end{align}
where \(\Omega = \zeta/2\zeta_d = ip/2\omega_d\) and \(\Omega_* = \zeta_* / 2\zeta_d = \omega_*/2\omega_d\). Note that \cref{eq:cITG_2D} agrees with the expressions obtained by \citet{biglari89, zocco18} in a similar limit to \cref{eq:itg_ordering}. 

In \cref{fig:cITG_vs_sITG}, we compare the solutions to \cref{eq:cITG} and \cref{eq:sITG} for the case of zero density gradient, viz., \(\omega_{*} = 0\), but nonzero temperature gradient, so \(\eta\omega_{*} \propto L_{T_{i}}^{-1} \neq 0\). The growth rates agree well only at simultaneously large perpendicular and small parallel wavelengths; this is to be expected given that the slab dispersion relation \cref{eq:sITG} does not capture the effect of magnetic drifts, which are most important at large parallel wavelengths. There is poorer agreement between the frequencies of the two dispersion relations.

\begin{figure}
	\centering\includegraphics[scale=0.27]{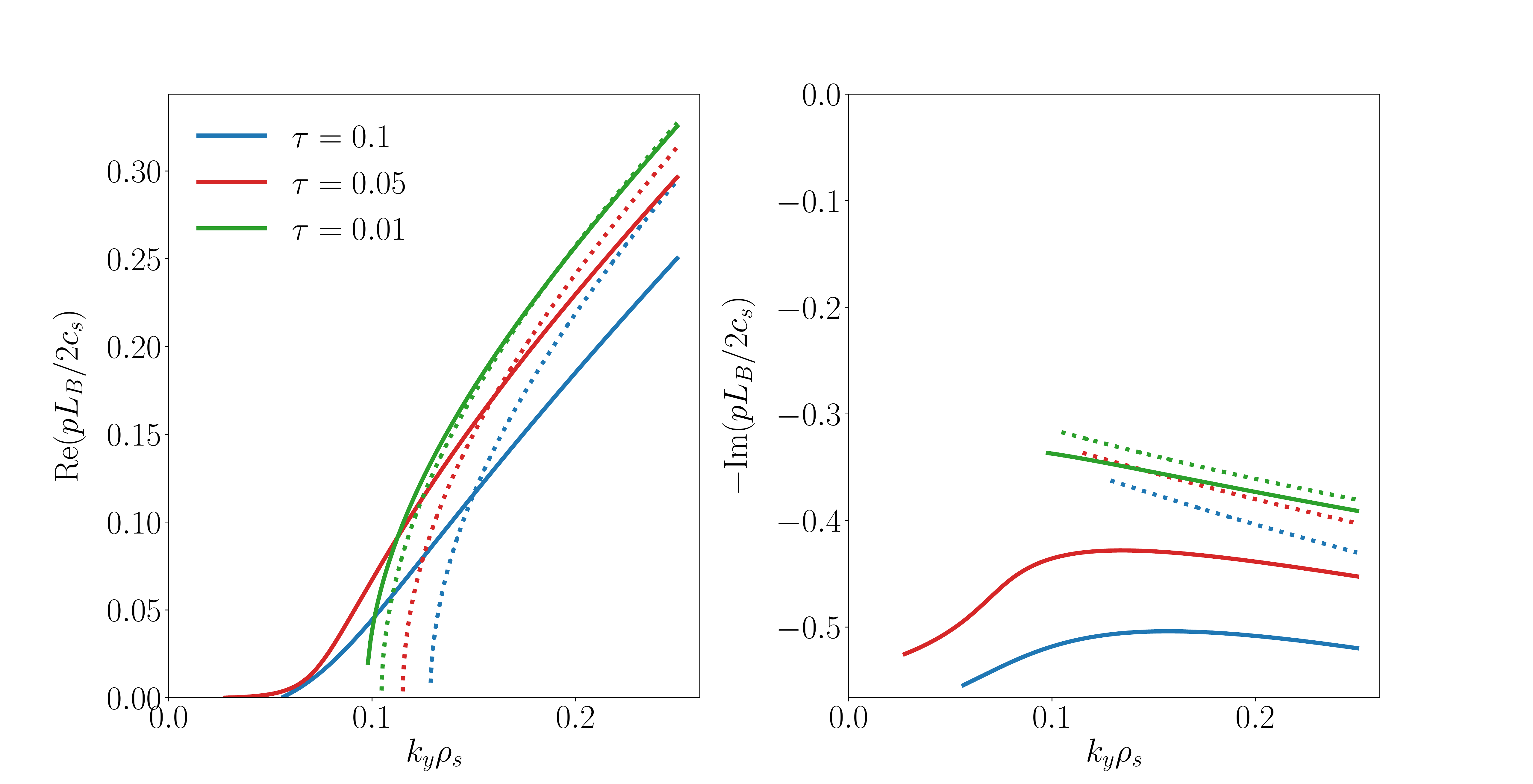}
	\caption{A comparison between the growth rate and frequency of the most unstable solution to the kinetic dispersion relation \cref{eq:cITG} and that of the fluid equations \crefrange{eq:fluid_phi}{eq:fluid_T}, represented by the solid and dotted lines. Here we have set \(\kpar L_B = 1\) and \(\tau L_B / 2L_{T_{i}} = 2\).}
	\label{fig:cITG_vs_fluid_tau}
\end{figure}

We can also compare the solutions to \cref{eq:cITG} with those obtained from a simple three-field fluid model of the ITG instability in a slab with magnetic curvature. The model consists of the following equations:
\begin{align}
	&\tau\ptf{\varphi} + \pzf{\upar} - \frac{\rho_i \vthi}{L_B} \pyf{}\left[(1+\tau)\varphi + \frac{\delta T_i}{T_{0i}}\right] = 0, \label{eq:fluid_phi}\\
	&\ptf{\upar} + \frac{\vthi^2}{2}\pzf{}\left[(1+\tau)\varphi + \frac{\delta T_i}{T_{0i}}\right] - \frac{2\rho_i\vthi}{L_B} \pyf{\upar} = 0, \label{eq:fluid_u}
	\\
	&\ptf{}\frac{\delta T_i}{T_{0i}} + \frac{2}{3}\pzf{\upar} + \frac{\rho_i \vthi}{2 L_{T_i}} \pyf{\varphi} - \frac{2}{3}\frac{\rho_i\vthi}{L_B} \pyf{}\left[(1+\tau)\varphi + \frac{7}{2}\frac{\delta T_i}{T_{0i}}\right] = 0, 
	\label{eq:fluid_T}
\end{align}
where \(\varphi \equiv Ze\phi/T_{0i}\), \(\upar\), and \(\delta T_i/T_{0i}\) are the perturbed electrostatic potential, ion parallel flow, and ion temperature, respectively. These equations can be derived by substituting a perturbed Maxwellian for \(h_i\) in the ion gyrokinetic equation and taking the three relevant velocity moments (cf. \citealt{newton10} or the cold-ion fluid model in \citealt{ivanov22} but with additional \(\tau \sim 1\) terms). \Cref{fig:cITG_vs_fluid} shows a comparison between the kinetic and fluid growth rates at fixed value of \(\tau\) and varying \(\kpar L_B\). We see that the fluid approximation is decent for small \(\kpar L_B\), but fails for larger ones because of its lack of kinetic effects. Making the ions cold, i.e., lowering \(\tau\), improves the accuracy of the fluid approximation, as in \cref{fig:cITG_vs_fluid_tau}.

\section{Summary and discussion}
\label{sec:summary} 
We have considered the problem of local linear gyrokinetics in a curved magnetic field, expressing the associated dispersion relation in terms of velocity-space integrals featuring resonances arising both from parallel streaming and from magnetic drifts (\cref{sec:gyrokinetic_linear_theory}). Previously, exact solutions for these integrals were known either in the absence of magnetic drifts --- leading to the well-known plasma dispersion function $\Z (\zeta)$ --- or in the two-dimensional limit (\cref{sec:previous_solutions}). In the case of drift kinetics (i.e., no finite-Larmor-radius effects) and equal magnetic drifts, we showed that these resonances can in fact be handled simultaneously without any additional approximations or expansions, and that the integrals can be expressed exactly in terms of a generalised plasma dispersion function consisting of products of $\Z$ functions, and its derivatives (\cref{sec:the_generalised_plasma_dispersion_function}). Since there exist known algorithms for the computation of the $\Z$ function, the resulting expressions are efficient to evaluate numerically, and can easily be handled analytically through known asymptotic expansions. Solutions to the exact dispersion relation for the electrostatic ITG instability, derived using this method, were then compared with approximate solutions in the previously known limits, showing poor agreement for the majority of parameters and wavenumbers considered (\cref{sec:itg}). This demonstrates that, in order to properly capture the growth rate and frequency of kinetic instabilities in the presence of a curved magnetic field, one must simultaneously resolve the resonances associated with parallel streaming and magnetic drifts, for which this paper provides the first known exact analytical solution. 

In \cref{sec:from_DK_to_GK}, we discussed how the assumptions of no finite-Larmor-radius effects and equal magnetic drifts can be relaxed using absolutely convergent Taylor-series expansions, and thus solve the more general linear gyrokinetic system. This results in expressions that naturally capture the multivaluedness of the underlying dispersion relation and handle the integration of resonant denominators exactly.

An immediate practical application of this work would be to use the derived analytical expressions to implement an efficient and accurate solver for drift-kinetic/gyrokinetic instabilities in the local limit considered in this paper. Such a solver could be used to benchmark both reduced models and gyrokinetic solvers. It could also be exploited to explore the equilibrium parameter space in search of new instabilities or to investigate the properties of subdominant ones, i.e., those whose growth rate is smaller than the largest growth rate in the system; this is typically difficult to do in most gyrokinetic solvers. Such subdominant instabilities have been proposed as one of the possible explanations for the lack of saturation observed in certain electromagnetic gyrokinetic simulations. With this in mind, we consider the implementation of such a gyrokinetic dispersion-relation solver to be a natural extension of this work that will produce a useful practical tool in the study of gyrokinetic instabilities and turbulence.

\section*{Funding}
This work has been carried out within the framework of the EUROfusion Consortium and has received funding from the Euratom research and training programme 2014–2018 and 2019–2020 under Grant Agreement No. 633053, and from the UKRI Energy Programme (EP/T012250/1). The views and opinions expressed herein do not necessarily reflect those of the European Commission. This work was supported by the Engineering and Physical Sciences Research Council (EPSRC) [EP/R034737/1]. TA was supported by a UK EPSRC studentship.

\section*{Declaration of interests}
The authors report no conflict of interest.


\begin{appendix}
	
\section{Calculation of $\J_{a,b}$}
\label{app:calculation_of_jab}
In this appendix, we derive the expression \cref{eq:jab_final} for $\J_{a,b}$. The calculation proceeds in a similar way to that of \(\I_{a,b}\) in \cref{sec:evaluation_of_iab}. Starting from \cref{eq:jab_original}, we consider the integral over $u$ separately, and so write 
	\begin{align}
		\J_{a,b} = \int_0^\infty \rmd \mu \: e^{-b\mu} \tilde{\J}_{a}, \quad \tilde{\J}_{a} = \frac{1}{\sqrt{\pi}} \int_{-\infty}^\infty \rmd u \frac{u e^{-au^2}}{u - \zeta + \zeta_d(2u^2 + \mu)}.
		\label{eq:jab_tilde_definition}
	\end{align}
Defining $\xpm$ as in \cref{eq:x_pm_definition} and making the same choice for the branch cut and square-root branch, a partial-fractions expansion of the integrand yields
\begin{align}
	\tilde{\J}_{a}^+ = \frac{1}{2\zeta_d(\xplus + \xminus)}\left( \frac{\xminus}{\sqrt{\pi}}\int_{-\infty}^\infty \rmd u   \frac{ e^{-au^2}}{u - \xminus} + \frac{\xplus}{\sqrt{\pi}}\int_{-\infty}^\infty \rmd u \frac{ e^{-au^2}}{u + \xplus }\right).
	\label{eq:jab_tilde_partial_fractions}
\end{align}
As previously, the sign of the imaginary part of $\xpm$ is always positive. Therefore, the first integral in the brackets in \cref{eq:jab_tilde_partial_fractions} is manifestly the plasma dispersion function, while the second can be turned into a plasma dispersion function under the change of variables $u \mapsto -u$. Thus, it follows that \cref{eq:jab_tilde_partial_fractions} can be written as 
\begin{align}
	\tilde{\J}_{a}^+ = -\frac{1}{2\zeta_d} \frac{\xplus Z_a(\xplus) - \xminus Z_a( \xminus )}{\xplus + \xminus}.
	\label{eq:jab_tilde_final}
\end{align}
Using the property (confirmed by direct calculation)
\begin{align}
	\xpm Z_a(\xpm) = - \frac{1}{\sqrt{a}} + \frac{\xplus + \xminus}{a} \frac{\partial Z_a(\xpm)}{\partial \mu},
	\label{eq:mu_derivative_identity}
\end{align}
we can write \(\tilde{\J}_a^+\) as
\begin{align}
	\tilde{\J}_{a}^+ = -\frac{1}{2a\zeta_d} \frac{\partial}{\partial \mu} \left[Z_a(\xplus) - Z_a (\xminus)\right].
	\label{eq:jab_tilde_asmuderivative}
\end{align}
Alternatively, using
\begin{equation}
	\xpm = \frac{1}{2} \left(\xplus + \xminus\right) \pm \frac{1}{4\zeta_d},
\end{equation}
we can also write
\begin{align}
	\tilde{\J}_{a}^+ = -\frac{1}{4\zeta_d} \left[Z_a(\xplus) - Z_a (\xminus)\right] - \frac{1}{4\zeta_d}\tilde{\I}_a^+.
\end{align}
Therefore, by substitution into the first expression in \cref{eq:jab_tilde_definition}, we find
\begin{align}
	\J_{a,b}^+ & = -\frac{1}{4\zeta_d} \int_0^\infty \rmd \mu \: e^{-b\mu} \left[Z_a(\xplus) - Z_a (\xminus)\right] - \frac{1}{4\zeta_d} \int_0^\infty \rmd \mu \: e^{-b\mu} \tilde{\I}_a^+ \nonumber \\
	& = -\frac{1}{4b\zeta_d}\left[Z_a(\zetaplus) - Z_a (\zetaminus)\right] + \frac{a}{2b}\J_{a,b}^+ - \frac{1}{4\zeta_d}I_{a, b}^+,
	\label{eq:jab_final_rearrange}
\end{align}
where, in going from the first line to the second, we have integrated by parts in the first integral and have used \cref{eq:jab_tilde_asmuderivative}. Equation \cref{eq:jab_final_rearrange} can be straightforwardly rearranged to yield \cref{eq:jab_final}.

\section{Calculation of derivatives of $\I_{a,b}$}
\label{app:calculation_of_derivatives}
Even though we are unable to evaluate \cref{eq:iab_mu_integral} exactly in the case where $a$ and $b$ are distinct, we are still able to find its derivatives with respect to $a$ and $b$ at \(a=b=1\), a task to which this appendix is devoted.

\subsection{Derivatives for general $a$ and $b$}
\label{app:derivatives_for_general_a_and_b}
To avoid clutter, we shall suppress the \(\branchsymb=+\) indices until \cref{app:derivatives_a_equal_b}. For this subsection, assume that all expressions \(\I_{a,b}\), \(\J_{a,b}\), \(\Q_{a,b}\), and \(\zetapm\) come with a \(\branchsymb=+\).

Using \cref{eq:x_pm_definition}, we can show that
\begin{align}
	\frac{\partial Z_a(\xpm)}{\partial a} = - \frac{\xpm(\xplus+ \xminus)}{a} \frac{\partial Z_a(\xpm)}{\partial \mu },
	\label{eq:derivative_identity}
\end{align}
and so the derivative of \cref{eq:iab_mu_integral} with respect to $a$ becomes
\begin{align}
	& \partial_a I_{a,b}  = - \frac{1}{2 a \zeta_d} \int_0^\infty \rmd \mu \: e^{- b \mu} \left[ \xplus \frac{\partial Z_a(\xplus)}{\partial \mu} + \xminus \frac{\partial Z_a(\xminus)}{\partial \mu} \right] \nonumber\\
	& = - \frac{1}{2a \zeta_d} \int_0^\infty \rmd \mu \: e^{- b \mu} \left\{ \frac{1}{2\zeta_d} \frac{\partial}{\partial \mu} \left[Z_a (\xplus) - Z_a ( \xminus) \right] + \left[ \xminus \frac{\partial Z_a( \xplus)}{\partial \mu} + \xplus \frac{\partial Z_a( \xminus)}{\partial \mu} \right]\right\} \nonumber \\
	& = \frac{1}{2 a \zeta_d} \left[\zetaminus Z_a( \zetaplus) + \zetaplus Z_a( \zetaminus) \right] + \frac{1}{2 \zeta_d }\J_{a,b} - \frac{1}{2a} \I_{a,b} - \frac{b}{a} \Q_{a,b}, \label{eq:iab_derivative_a_Q} 
\end{align}
where we have defined the integral
\begin{align}
	\Q_{a,b} = \frac{1}{2 \zeta_d} \int_0^\infty \rmd \mu \: e^{- b \mu} \left[ \xminus Z_a( \xplus) + \xplus Z_a( \xminus) \right].
	\label{eq:Q_integral_definition}
\end{align}
In going from the first line of \cref{eq:iab_derivative_a_Q} to the second, we made use of the fact, obvious from the definition \cref{eq:x_pm_definition}, that
\begin{align}
	\xpm = \xmp \pm \frac{1}{2 \zeta_d},
	\label{eq:xpm_relationships}
\end{align}
while going from the second to the third, we have recognised the first expression in the curly brackets as \cref{eq:jab_tilde_asmuderivative} and integrated by parts the second. 

Similarly, taking a derivative of \cref{eq:iab_mu_integral} with respect to $b$, and making use of \cref{eq:mu_in_terms_of_xpm} and \cref{eq:zetapm_definition}, we have that
\begin{align}
	\partial_b \I_{a,b} = - (\zetaplus^2 + \zetaminus^2) \I_{a,b} + \frac{1}{2\zeta_d} \int_0^\infty \rmd \mu \: e^{- b \mu} \frac{\xplus^2 +\xminus^2}{\xplus + \xminus} \left[ Z_a( \xplus) + Z_a( \xminus) \right].
	\label{eq:iab_b_derivative_xpm}
\end{align}
Since 
\begin{align}
	\left( \xplus^2 + \xminus^2 \right) \left[ Z_a ( \xplus) + Z_a ( \xminus) \right] &= \left( \xplus + \xminus \right) \left[ \xminus Z_a ( \xplus) + \xplus Z_a ( \xminus) \right] \nonumber \\
	& \quad + \left( \xplus -\xminus \right) \left[\xplus Z_a ( \xplus) - \xminus Z_a ( \xminus) \right],
\end{align}
\cref{eq:iab_b_derivative_xpm} becomes 
\begin{align}
	\partial_b \I_{a,b} = - (\zetaplus^2 + \zetaminus^2) \I_{a,b} - \frac{1}{2\zeta_d} \J_{a,b} + \Q_{a,b},
	\label{eq:iab_derivative_b_Q}
\end{align}
where we have made use of \cref{eq:jab_tilde_asmuderivative} again. It is clear from \cref{eq:iab_derivative_a_Q} and \cref{eq:iab_derivative_b_Q} that we need to find $\Q_{a,b}$ in order to obtain expressions for $\partial_a \I_{a,b}$ and $\partial_b \I_{a,b}$. Though it is possible to do so via direct manipulation of the integrand of \cref{eq:Q_integral_definition}, we prefer an alternative approach. Using 
\begin{align}
	\frac{u}{u - \zeta + \zeta_d (2 u^2 + \mu)} = 1 + \frac{\zeta - \zeta_d (2u^2 +\mu)}{u - \zeta + \zeta_d (2u^2 + \mu) }
	\label{eq:jab_integrand_partial_fractions}
\end{align}
in \eqref{eq:jab_original} gives
\begin{align}
	\J_{a,b} = \frac{1}{\sqrt{a}b} + \zeta \I_{a,b} + \zeta_d(2\partial_a + \partial_b) \I_{a,b}.
	\label{eq:jab_in_terms_of_derivatives}
\end{align}
Substituting \cref{eq:iab_derivative_a_Q} and \cref{eq:iab_derivative_b_Q} into \cref{eq:jab_in_terms_of_derivatives}, and rearranging, we obtain the following expression for $\Q_{a,b}$ in terms of $\I_{a,b}$ and $\J_{a,b}$:
\begin{align}
	\left(1 - \frac{2b}{a} \right)\Q_{a,b} & = - \frac{1}{\sqrt{a} b \zeta_d } - \frac{1}{a \zeta_d}  \left[ \zetaminus Z_a( \zetaplus) + \zetaplus Z_a( \zetaminus) \right] + \frac{1}{2\zeta_d} \J_{a,b} + \left(\frac{1}{a} + \frac{1}{4\zeta_d^2} \right) \I_{a,b}.
	\label{eq:Q_explicit_expression}
\end{align}
In a similar way, taking a \(\partial_b\) derivative of \cref{eq:jab_final}, we find
\begin{align}
	\left(1 - \frac{2b}{a} \right) \partial_b \J_{a,b} & = \frac{2}{a} \left( \J_{a,b} + \frac{1}{4\zeta_d} \I_{a,b} \right) + \frac{b}{2 a \zeta_d } \partial_b \I_{a,b}. \label{eq:jab_derivative_b}
\end{align}

\subsection{Derivatives at \(a=b\)}
\label{app:derivatives_a_equal_b}

Finally, using \cref{eq:jab_final}, \cref{eq:iab_derivative_a_Q}, \cref{eq:iab_derivative_b_Q}, and \cref{eq:Q_explicit_expression}, setting \(a=b\), and simultaneously expressing both branches using \cref{eq:zetas_branches}, we obtain
\begin{align}
	\left.\partial_a\I_{a,b}^\branchsymb\right|_{a=b} = & -\frac{1}{a^{3/2}\zeta_d} + \left(\frac{1}{2a} - \frac{1}{4\zeta_d^2}\right)\I_{a,a}^\branchsymb - \frac{1}{2a\zeta_d^2}\left[Z_a( \zetaplus^\branchsymb) - Z_a ( \zetaminus^\branchsymb)\right] \nonumber \\
	& -\frac{1}{2a\zeta_d}\left[\zetaminus^\branchsymb Z_a( \zetaplus^\branchsymb) + \zetaplus^\branchsymb Z_a ( \zetaminus^\branchsymb)\right],
	\label{eq:diab_da}
\end{align}
\begin{align}
	\left.\partial_b \I_{a,b}^\branchsymb\right|_{a=b} = &\frac{1}{a^{3/2}\zeta_d} - \left(\frac{1}{a} + \frac{\zeta}{\zeta_d}\right)\I_{a,a}^\branchsymb + 
     \frac{1}{a\zeta_d}\left[\zetaplus^\branchsymb\Z_a(\zetaplus^\branchsymb) + \zetaminus^\branchsymb\Z_a(\zetaminus^\branchsymb)\right],
	\label{eq:diab_db}
\end{align}
\begin{align}
	\left.\partial_b^2 \I_{a,b}^\branchsymb\right|_{a=b} = &-\frac{1}{a^{5/2}\zeta_d} -\frac{2}{a}\Q_{a, a}^\branchsymb - \left(\frac{1}{a} + \frac{\zeta}{2\zeta_d} + \frac{1}{2\zeta_d^2}\right)\partial_b \left.\I_{a,b}^\branchsymb\right|_{a=b} - \frac{1}{\zeta_d}\partial_b \left.\J_{a,b}^\branchsymb\right|_{a=b},
	\label{eq:d2iab_db2}
\end{align}
\begin{align}
	\left.\partial_b \J_{a,b}^\branchsymb\right|_{a=b} =& -\frac{1}{2a^{3/2}\zeta_d^2} + \frac{1}{2\zeta_d}\left(\frac{2}{a} + \frac{\zeta}{\zeta_d}\right)\I_{a,a}^\branchsymb - \frac{1}{2a\zeta_d^2}\left[\zetaplus^\branchsymb\Z_a(\zetaplus^\branchsymb) + \zetaminus^\branchsymb\Z_a(\zetaminus^\branchsymb)\right] \nonumber \\ & + \frac{1}{a^2\zeta_d}\left[\Z_a(\zetaplus^\branchsymb) -\Z_a(\zetaminus^\branchsymb)\right],
	\label{eq:djab_db}
\end{align}
\begin{align}
	\Q_{a, a}^\branchsymb = \frac{1}{a^{3/2}\zeta_d} - \left(\frac{1}{a} + \frac{1}{4\zeta_d^2}\right)\I_{a,a}^\branchsymb + \frac{1}{a\zeta_d}\left[\zetaminus^\branchsymb\Z_a(\zetaplus^\branchsymb) + \zetaplus^\branchsymb\Z_a(\zetaminus^\branchsymb)\right] - \frac{1}{2\zeta_d}\J_{a, a}^\branchsymb.
	\label{eq:qaa}
\end{align}

\section{Properties of the branches of the dispersion function}
\label{app:branches}
The main convenience of choosing the branch cut along the negative real line in \cref{sec:multivaluedness} is the relationship \(\sqrtp{z^*} = \sqrtp{z}^*\) for any \(z \in \mathbb{C}\). It is then easy to see that the expressions \cref{eq:zetas_branches} satisfy
\begin{align}
	\zetapm^\branchsymb(-\zeta^*, -\zeta_d) &= -\zetapm^\branchsymb(\zeta, \zeta_d)^* \label{eq:zetapm_minuszetaconj} \\
	\zetapm^\branchsymb(\zeta^*, \zeta_d) &= \zetapm^\branchsymb(\zeta, \zeta_d)^*, \label{eq:zetapm_zetaconj}
\end{align}
and that \cref{eq:zetas_branches} implies
\begin{equation}
	\zetapm^{-\branchsymb} = -\zeta_\mp^\branchsymb.
\end{equation}
Additionally, it is straightforward to show that the \(\Z\) function satisfies
\begin{equation}
	\Z(\zeta^*) = -\Z(-\zeta)^* \label{eq:z_zetaconj}.
\end{equation}
Then, using \cref{eq:genZ_function} and \crefrange{eq:zetapm_minuszetaconj}{eq:z_zetaconj}, we have
\begin{align}
	\I_{a,a}^\branchsymb(-\zeta^*, -\zeta_d) &= -\frac{1}{2\sqrt{a}(-\zeta_d)}\Z_a(-\zetaplus^{\branchsymb*})\Z_a(-\zetaminus^{\branchsymb*}) \nonumber \\
	&=\frac{1}{2\sqrt{a}\zeta_d}[-\Z_a(\zetaplus^\branchsymb)]^*[-\Z_a(\zetaminus^\branchsymb)]^* \nonumber \\
	&= -\I_{a,a}^\branchsymb(\zeta, \zeta_d)^*, \label{eq:iaa_minuszetaconj_app}
\end{align}
and
\begin{align}
	\I_{a,a}^\branchsymb(\zeta^*, \zeta_d) &= -\frac{1}{2\sqrt{a}\zeta_d}\Z_a(\zetaplus^{\branchsymb*})\Z_a(\zetaminus^{\branchsymb*}) \nonumber \\
	&=-\frac{1}{2\sqrt{a}\zeta_d}[-\Z_a(-\zetaplus^\branchsymb)]^*[-\Z_a(-\zetaminus^\branchsymb)]^* \nonumber \\
	&=-\frac{1}{2\sqrt{a}\zeta_d}\Z_a(\zetaminus^{-\branchsymb})^*\Z_a(\zetaplus^{-\branchsymb})^* \nonumber \\
	&= \I_{a,a}^{-\branchsymb}(\zeta, \zeta_d)^*. \label{eq:iaa_zetaconj_app}
\end{align} 
 Similarly, using \cref{eq:jaa_general}, we find
\begin{align}
	\J_{a, a}^\branchsymb(-\zeta^*, -\zeta_d) &= \J_{a, a}^\branchsymb(\zeta, \zeta_d)^* \label{eq:jaa_minuszetaconj_app} \\
	\J_{a, a}^\branchsymb(\zeta^*, \zeta_d) &=\J_{a, a}^{-\branchsymb}(\zeta, \zeta_d)^* \label{eq:jaa_zetaconj_app}.
\end{align}
The derivatives of \(\I_{a,b}\), given by \crefrange{eq:diab_da}{eq:d2iab_db2}, and \(\partial_b\J_{a,b}\vert_{a=b}\), given by \cref{eq:djab_db}, can also be shown to have the properties \crefrange{eq:iaa_minuszetaconj_app}{eq:iaa_zetaconj_app} and \crefrange{eq:jaa_minuszetaconj_app}{eq:jaa_zetaconj_app}, respectively.

\begin{figure}
	\centering \hspace*{-2cm}\includegraphics[scale=0.27]{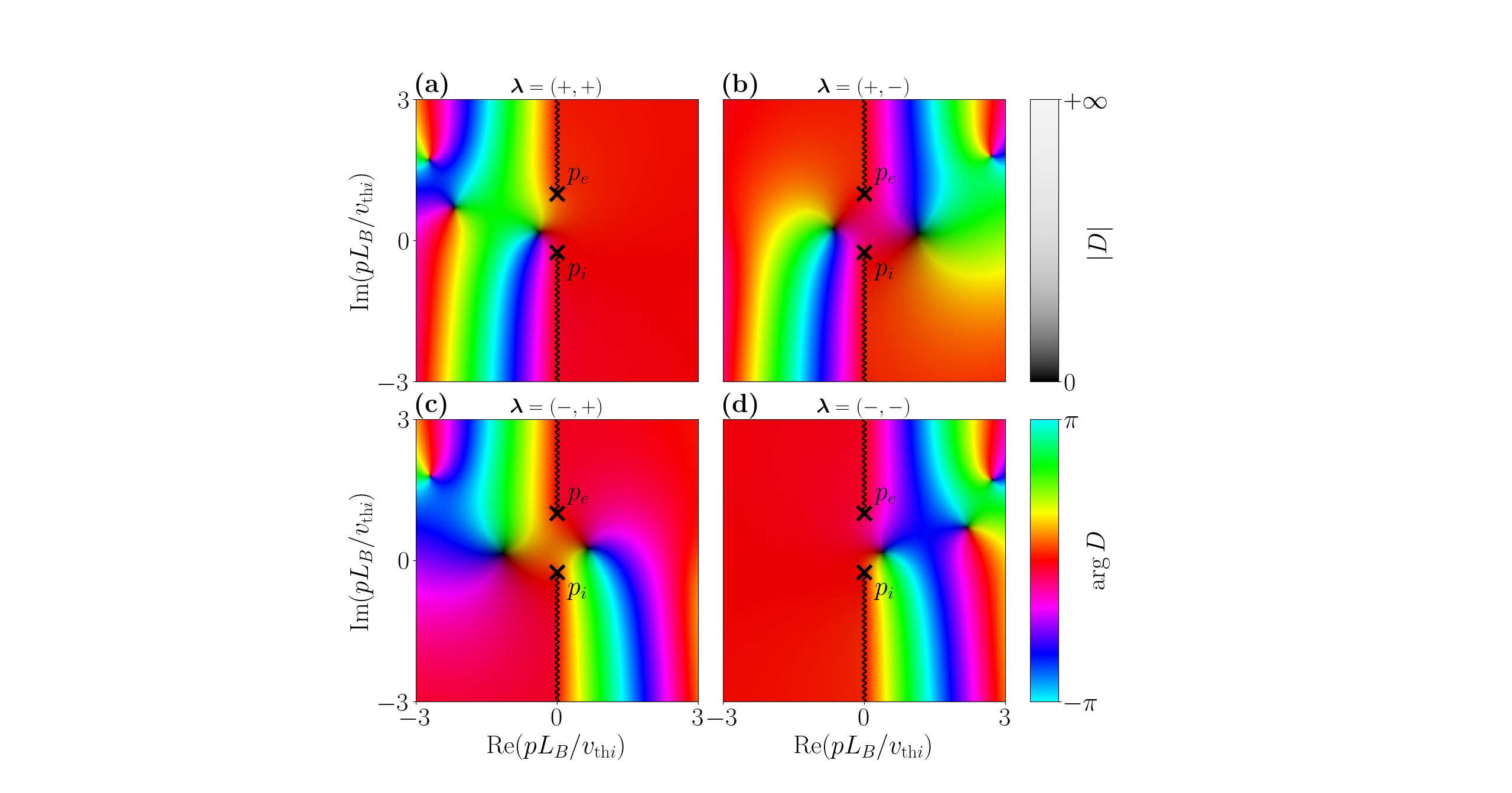}
	\caption{A plot in the complex plane of the dispersion function \(D(p)=\text{L}_{\phi\phi}\) for an electrostatic, two-species plasma composed of ion and electrons, for the following parameters: \(m_i / m_e = 2\), \(q_i = -q_e=e\), \(T_{0i} = T_{0e}\), \(k_y \rho_i = 1\), \(\kpar L_B = 1\), and \(L_{T_i} = L_B\). The panels show the four branches of \(D\), labelled by \(\vec{\branchsymb} = (\branchsymb_i, \branchsymb_e)\) as shown (see \cref{sec:branches_disp}). Here we are using the principal branch cut for the square root. The colour brightness shows the magnitude \(|D|\), while its hue shows the phase \(\arg D\). The relation \cref{eq:D_zetaconj_appendix}, \(D^{\vec{\branchsymb}}(-p^*, \vec{k}) = D^{-\vec{\branchsymb}}(p, \vec{k})^*\), is evident in the pairs (a),(d) and (b),(c): flipping the sign of \(\vec{\branchsymb}\) corresponds to mirroring the real part of \(p\) and taking the complex conjugate of \(D\) (note the change in colour). Furthermore, crossing the electron branch cut flips the sign of \(\branchsymb_e\) and so corresponds to jumping horizontally between the panels; crossing the ion branch cut corresponds to jumping vertically between them.}
	\label{fig:unicurvy_principalbranch}
\end{figure}

\begin{figure}
	\centering \hspace*{-2cm}\includegraphics[scale=0.27]{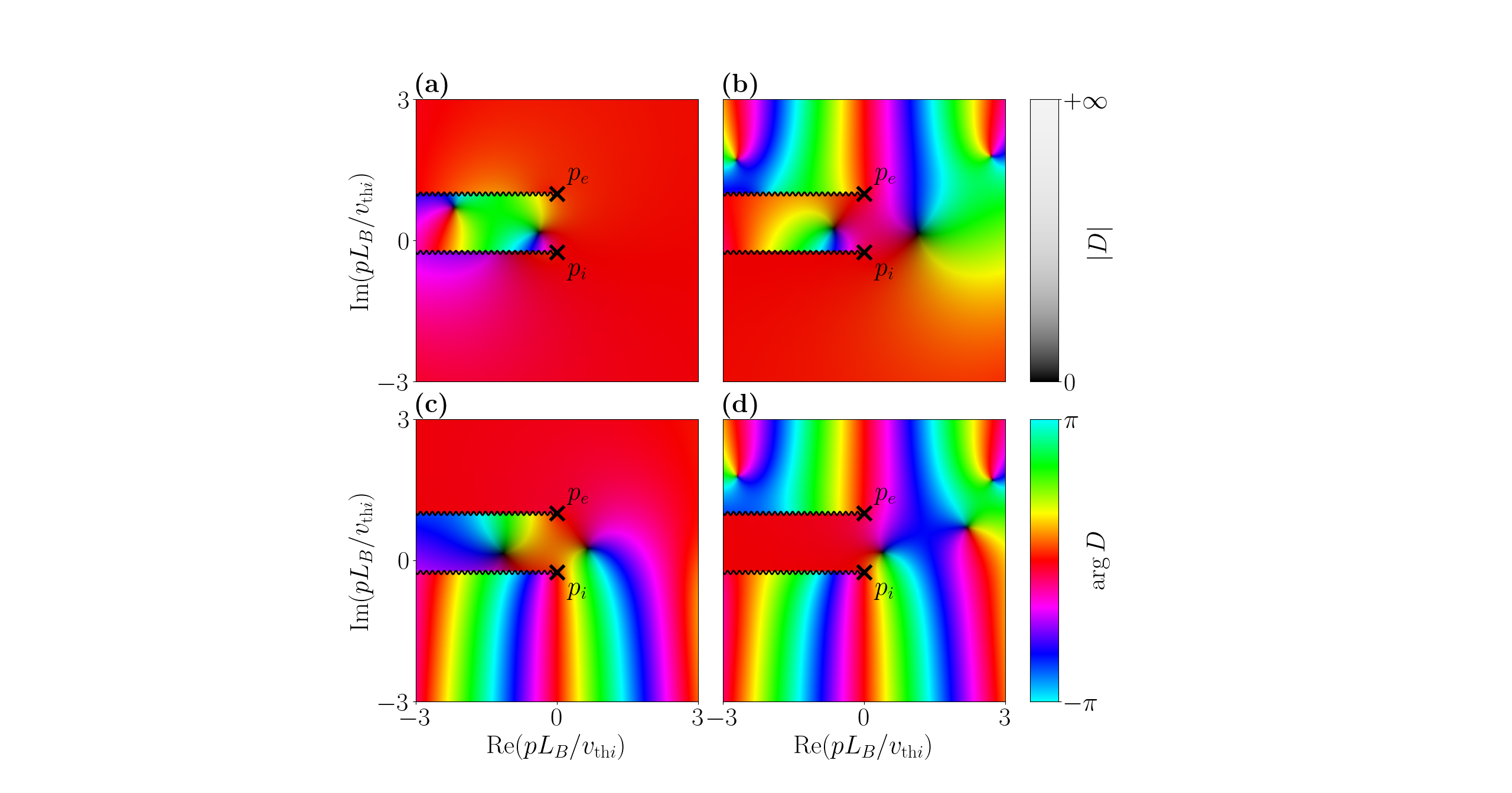}
	\caption{The same as \cref{fig:unicurvy_principalbranch} but with the branch cuts rotated to point towards \(\text{Re}(p) \to -\infty\). As previously, crossing the electron branch cut flips the sign of \(\branchsymb_e\) and so corresponds to jumping horizontally between the panels, while crossing the ion branch cut corresponds to jumping vertically between the panels. For practical purposes, we are only interested in the `dispersion' branch \(\mainD\) (see discussion in \cref{sec:analytic_continuation}) shown in (a) as it is that one that enters the inverse Laplace transform. }
	\label{fig:unicurvy_rotatedbranch}
\end{figure}

Recall that the frequencies, which enter the dispersion matrix elements \crefrange{eq:L_phiphi}{eq:L_bb}, are functions of \(\zeta_s \propto p\), \(\zeta_{*\s} \propto k_y\), and \(\zeta_{d\s} \propto k_y\) [see \cref{eq:drift_frequency}, \cref{eq:magnetic_drift_frequencies}, \cref{eq:normalised_frequencies}, and \cref{eq:magnetic_drifts_equality}]. It is then evident that \(p \mapsto p^*\) maps \(\zeta_s \mapsto -\zeta_s^*\), \(p \mapsto -p^*\) maps \(\zeta_s \mapsto \zeta_s^*\), and the inversion \(\vec{k} \mapsto -\vec{k}\) results in \(\zeta_{*\s} \mapsto -\zeta_{*\s}\) and \(\zeta_{d\s} \mapsto -\zeta_{d\s}\) (recall that the sign of the parallel wavenumber \(\kpar\) does not enter the normalised frequencies, as we noted in \cref{footnote:kpar}). Combining this with \crefrange{eq:iaa_minuszetaconj_app}{eq:jaa_zetaconj_app}, it is then straightforward to show that the dispersion matrix \(\mathbf{L}\) and its elements \crefrange{eq:L_phiphi}{eq:L_bb} satisfy
\begin{equation}
	\begin{pmatrix}
		\text{L}^{\vec{\branchsymb}}_{\phi \phi} & \text{L}^{\vec{\branchsymb}}_{\phi A} & \text{L}^{\vec{\branchsymb}}_{\phi B}\\
		\text{L}^{\vec{\branchsymb}}_{A \phi} & \text{L}^{\vec{\branchsymb}}_{A A} & \text{L}^{\vec{\branchsymb}}_{A B}\\
		\text{L}^{\vec{\branchsymb}}_{B \phi} & \text{L}^{\vec{\branchsymb}}_{B A} & \text{L}^{\vec{\branchsymb}}_{B B}\\
	\end{pmatrix}
	(p^*, -\vec{k}) = 
	\begin{pmatrix}
		\text{L}^{\vec{\branchsymb}}_{\phi \phi} & -\text{L}^{\vec{\branchsymb}}_{\phi A} & \text{L}^{\vec{\branchsymb}}_{\phi B}\\
		-\text{L}^{\vec{\branchsymb}}_{A \phi} & \text{L}^{\vec{\branchsymb}}_{A A} & -\text{L}^{\vec{\branchsymb}}_{A B}\\
		\text{L}^{\vec{\branchsymb}}_{B \phi} & -\text{L}^{\vec{\branchsymb}}_{B A} & \text{L}^{\vec{\branchsymb}}_{B B}\\
	\end{pmatrix}^*
	(p, \vec{k}),
	\label{eq:stuff1}
\end{equation}
and
\begin{equation}
	\mathbf{L}^{\vec{\branchsymb}}(-p^*, \vec{k}) = \mathbf{L}^{-\vec{\branchsymb}}(p, \vec{k})^*
	\label{eq:stuff2}
\end{equation}
where the vector \(\vec{\branchsymb} = (\branchsymb_1, \branchsymb_2, ..., \branchsymb_N)\) labels the branches the double-valued functions that constitute \(\mathbf{L}\), for each of the \(N\) particle species. Therefore, for the dispersion function \(D=\det\mathbf{L}\), we have, from \cref{eq:stuff1},
\begin{align}
	D^{\vec{\branchsymb}}(p^*, -\vec{k}) &= D^{\vec{\branchsymb}}(p, \vec{k})^* ,
\end{align}
and, from \cref{eq:stuff2},
\begin{align}
	D^{\vec{\branchsymb}}(-p^*, \vec{k}) &= D^{-\vec{\branchsymb}}(p, \vec{k})^* 
	\label{eq:D_zetaconj_appendix}.
\end{align}
Figures \ref{fig:unicurvy_principalbranch} and \ref{fig:unicurvy_rotatedbranch} show an example of the four branches of the dispersion function in the case of a two-species plasma. In particular, the property \cref{eq:D_zetaconj_appendix} is illustrated clearly in \cref{fig:unicurvy_principalbranch}.

\section{Integral around the branch cut}
\label{app:integral_around_the_branch_cut}

This appendix is devoted to calculating the asymptotic contribution to \cref{eq:fields_inverse_laplace} in the limit \(t\to\infty\) arising from the integral around one of the branch cuts of \(\hat{\vec{\chi}}_{\vec{k}}(p)\). Similar calculations already exist in the literature \citep[e.g.,][]{kim94, kuroda98}; we are including one here for completeness.

\begin{figure}
	\centering
	
	\scalebox{0.9}{\begin{tikzpicture}[scale=1, thick, every node/.style={scale=1.2}]
			\newcommand{\arrowIn}{
				\tikz \draw[-latex] (-1pt,0) -- (1pt,0);
			}
		
            \def\yaxislength{5}
            \def\xaxislength{8}
            \def\xoffset{0}
            \def\yoffset{0}
            \def\zoffset{0}
            
            \coordinate (origin) at (\xoffset,\yoffset,\zoffset);

            \coordinate (xaxishalf) at (\xaxislength/2, 0, 0);
            \coordinate (xaxisyshift) at (0, \yaxislength/2, 0);
            \coordinate (xaxisneg) at ($ (origin) - (xaxishalf) - (xaxisyshift)$);
            \coordinate (xaxispos) at ($ (origin) + 0.6*(xaxishalf) - (xaxisyshift)$);
            
            \draw[-latex] (xaxisneg) 
            --
            (xaxispos);
            
            \draw (xaxispos) node[anchor=west] {$\Re(p)$};
            
            \coordinate (yaxishalf) at (0, \yaxislength/2, 0);
            \coordinate (yaxisneg) at ($ (origin) - (yaxishalf)$);
            \coordinate (yaxispos) at ($ (origin) + (yaxishalf)$);
            
            \draw[-latex] (yaxisneg) 
            -- 
            (origin)
            -- 
            (yaxispos);
            
            \draw (yaxispos) node[anchor=east] {$\Im(p)$};
            
            \def\branchlength{4}
            
            \coordinate (branchpoint) at ($(origin) + (0, {0 * \yaxislength}, 0)$);
            
            \draw[decorate, decoration = zigzag]  (branchpoint) -- ($(branchpoint) - (\branchlength, 0, 0)$);
            \draw (branchpoint) node[anchor = west] {$p_\s$};
            
            \def\contourradius{1.0}
            
            \coordinate (cplus) at ($ (branchpoint) + (0, \contourradius, 0) $);
            \coordinate (cminus) at ($ (branchpoint) - (0, \contourradius, 0) $);
            
            \draw ($(cminus) - (\branchlength, 0, 0)$)
            -- node[midway, sloped, allow upside down] {\arrowIn}
            node[midway, anchor = north] {$C_-$}
            (cminus);
            
            \draw ($(cminus) - (\branchlength, 0, 0)$) node[anchor = east] {$\Im(p) = \Im(p_\s) - \varepsilon$};
            
            \draw (cminus) arc (-90 : 90 : \contourradius)
            [arrow inside = {end=latex, opt={black,scale=1}} {0.52}]
            node[midway, anchor = west] {$C_\varepsilon$};
            
            \draw (cplus)
            -- node[midway, sloped, allow upside down] {\arrowIn}
            node[midway, anchor = south] {$C_+$}
            ($(cplus) - (\branchlength, 0, 0)$);
            
            \draw ($(cplus) - (\branchlength, 0, 0)$) node[anchor = east] {$\Im(p) = \Im(p_\s) + \varepsilon$};

        \end{tikzpicture}}
	
	\caption{The contour of integration $C_\text{br}$ around the branch cut --- chosen to be parallel to the real $p$ axis --- with the latter indicated by the zigzag line. $C_\pm$ are the horizontal, semi-infinite segments along $\Im(p) = \Im(p_\s) \pm \varepsilon$ that connect the vertical contour at $\Re(p) \to -\infty$ (see \cref{fig:inverse_laplace_transform_continuation}) to the semi-circular arc $C_\varepsilon$ around the branch point.}
	\label{fig:branch_cut_integration}
\end{figure}
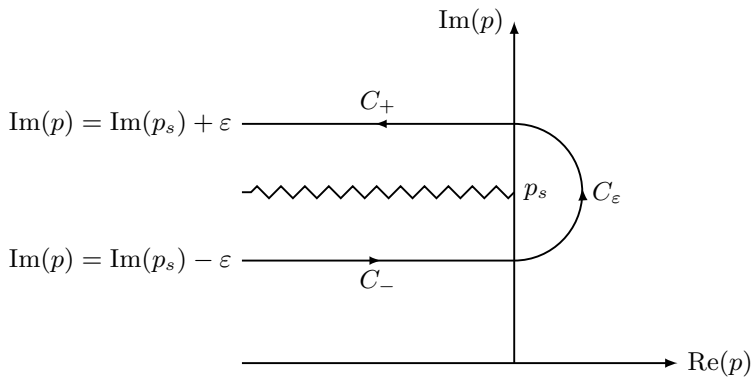

Recall that there is one branch cut for each particle species, associated with the branch point \(p_\s\) \cref{eq:ps_def}. We choose the branch cut to be parallel to the real $p$ axis and denote the contour around this branch cut $C_\text{br}$, as in \cref{fig:branch_cut_integration}. $C_\text{br}$ consists of a semi-circular arc $C_\varepsilon$ of radius \(\varepsilon\) around the branch point, where we choose $\varepsilon \sim t^{-2}$, and two horizontal, semi-infinite segments $C_\pm$ along $\Im(p) = \Im(p_\s) \pm \varepsilon$, viz.,
\begin{align}
	 \int_{C_\text{br}} \rmd p \: e^{pt} \hat{\vec{\chi}}_{\vec{k}}(p) = \int_{C_- + C_\varepsilon + C_+} \rmd p \: e^{pt} \hat{\vec{\chi}}_{\vec{k}}(p).
	\label{eq:branch_cut_integral}
\end{align}
Let us calculate each of the contributions to \cref{eq:branch_cut_integral} in turn. 

For \(C_\varepsilon\), we change variables to $p = p_\s + \varepsilon e^{i \theta}$ for $\theta \in [-\pi/2, \pi/2]$. It straightforwardly follows that, since  \(\varepsilon \sim t^{-2}\),
\begin{align}
	\left|\int_{C_\varepsilon} \rmd p \: e^{pt} \hat{\vec{\chi}}_{\vec{k}}(p) \right| \leqslant   \left|\hat{\vec{\chi}}_{\vec{k}}(p_\s) \right| \left[1 + O(\varepsilon)\right]\varepsilon \int_{-\pi/2}^{\pi/2} \rmd \theta \: e^{\varepsilon t \cos \theta} = O(t^{-2}).
	\label{eq:branch_cut_integral_circle}
\end{align}

Turning our attention to $C_\pm$, we set \(p = p_\s + \xi \pm i\varepsilon\), respectively, and find
\begin{align}
	\int_{C_\pm} \rmd p \: e^{pt} \hat{\vec{\chi}}_{\vec{k}}(p) & = \mp \int_{-\infty}^0 \rmd \xi \: e^{\xi t} e^{(p_\s \pm i \varepsilon)t} \hat{\vec{\chi}}_{\vec{k}}(p + p_\s \pm i \varepsilon) \nonumber \\
	& = \mp  \left[\int_{-\infty}^{-\delta } + \int_{-\delta }^{0}  \right]\rmd \xi \: e^{\xi t} e^{(p_\s \pm i \varepsilon)t}\hat{\vec{\chi}}_{\vec{k}}(p + p_\s \pm i \varepsilon),
	\label{eq:branch_cut_integral_pm_splitting}
\end{align}
where we have split the integration interval using some positive real $\delta \ll 1$. The first integral in the square brackets of \cref{eq:branch_cut_integral_pm_splitting} is bounded by an exponential, viz., 
\begin{align}
	\left| \int_{-\infty}^{-\delta } \rmd \xi \: e^{\xi t} e^{(p_\s \pm i \varepsilon)t} \hat{\vec{\chi}}_{\vec{k}}(\xi + p_\s \pm i \varepsilon)\right| \leqslant e^{-\delta  t} \int_{-\infty}^{-\delta } \rmd \xi \: |\hat{\vec{\chi}}_{\vec{k}}(\xi + p_\s \pm i \varepsilon)| = O(e^{-\delta  t}),
	\label{eq:branch_cut_integral_pm_exp_bound}
\end{align}
and so it is exponentially small in the limit of $t \rightarrow \infty$. For the second integral, we know that \(|\xi|\leqslant\delta\ll1\), and so it is natural to Taylor-expand the integrand. Note that the function \(\hat{\vec{\chi}}_{\vec{k}}\) contains both parts that are discontinuous across the branch cut (related to species \(\s\)), as well as some that are continuous (related to species other than \(\s\)). The discontinuity is due to the square-root terms in \(\I_{a,b}\) and \(\J_{a,b}\), manifest in the expression for \(\zetapm\) \cref{eq:zetas_branches}. These square roots appear only as arguments of analytic functions. Therefore, the discontinuity of \(\hat{\vec{\chi}}_{\vec{k}}\) across the branch cut can be made explicit by writing
\begin{equation}
	\hat{\vec{\chi}}_{\vec{k}} = \hat{\vec{\chi}}_{\vec{k}}(p, \sqrtp{p-p_\s}),
	\label{eq:chi_twoargs}
\end{equation}
where \(\hat{\vec{\chi}}_{\vec{k}}\) is an analytic function of both of its arguments.\footnote{The principal branch is the appropriate one for \(\sqrt{p-p_\s}\) only after performing the rotation of the branch cuts to align them in the horizontal direction in the \(p\) complex plane, see \cref{sec:analytic_continuation}.} Noting that
\begin{align}
	&\sqrtp{p-p_\s} = \sqrtp{\xi} + O(\varepsilon), \\
	&\sqrtp{p-p_\s} = -\sqrtp{\xi} + O(\varepsilon),
\end{align}
for \(p = p_\s + \xi \pm i\varepsilon\), respectively, we find
\begin{align}
	\int_{-\delta }^{0}\rmd \xi \: e^{\xi t} e^{(p_\s \pm i \varepsilon)t}\hat{\vec{\chi}}_{\vec{k}}(\xi + p_\s \pm i \varepsilon) \approx \int_{-\delta }^{0}\rmd \xi \: e^{\xi t} e^{p_\s t} \left[ \hat{\vec{\chi}}_{\vec{k}}(p_\s) \pm \sqrtp{\xi } \frac{\partial \hat{\vec{\chi}}_{\vec{k}}(p_\s)}{\partial \sqrtp{p-p_\s}}  \right],
	\label{eq:Cpm_integrals_expanded}
\end{align}
where we have ignored terms \(O(\delta)\) or \(O(\varepsilon)\) in the square brackets and \(\partial \hat{\vec{\chi}}_{\vec{k}}(p_\s) / \partial \sqrtp{p-p_\s}\) denotes the partial derivative of \(\hat{\vec{\chi}}_{\vec{k}}\) with respect to its second parameter in \cref{eq:chi_twoargs} evaluated at \(p=p_\s\), i.e., at \(\xi = 0\). Using \cref{eq:Cpm_integrals_expanded}, we then find
\begin{align}
	\left[\int_{C_-} + \int_{C_+}\right] \rmd p \: e^{pt} \hat{\vec{\chi}}_{\vec{k}}(p) \sim 2 e^{p_\s t} \int_{-\delta }^{0}\rmd \xi \: e^{\xi t} \sqrtp{\xi} \frac{\partial \hat{\vec{\chi}}_{\vec{k}}(p_\s)}{\partial \sqrtp{p-p_\s}} \quad \text{as \(t \to \infty\)}.
\end{align}
Using
\begin{align}
	 \int_{-\delta }^0 \rmd \xi \: e^{\xi t} \sqrt{\xi} = t^{-3/2} \int_{- t \delta}^0 \rmd \eta \: e^{\eta} \sqrt{\eta} \sim t^{-3/2} \frac{i \sqrt{\pi}}{2} \quad \text{as \(t \to \infty\)}, 
	\label{eq:asymptotic_integral}
\end{align}
we finally arrive at
\begin{align}
	\int_{C_\text{br}} \rmd p \: e^{pt} \hat{\vec{\chi}}_{\vec{k}}(p) \sim t^{-3/2} e^{ip_\s t} \sqrt{\pi} \frac{\partial \hat{\vec{\chi}}_{\vec{k}}(p_\s)}{\partial \sqrtp{p-p_\s}} \quad \text{as \(t \to \infty\)}, 
	\label{eq:branch_cut_integral_evaluated}
\end{align}
which is the required result.

\end{appendix}

\bibliography{bibliography.bib}{}

\begin{thebibliography}{53}
\expandafter\ifx\csname natexlab\endcsname\relax\def\natexlab#1{#1}\fi
\def\au#1{#1} \def\ed#1{#1} \def\yr#1{#1}\def\at#1{#1}\def\jt#1{\textit{#1}}
  \def\bt#1{#1}\def\bvol#1{\textbf{#1}} \def\vol#1{#1} \def\pg#1{#1}
  \def\publ#1{#1}\def\arxiv#1{#1}\def\org#1{#1}\def\st#1{\textit{#1}}

\bibitem[{Abel} {\em et~al.\/}(2013){Abel}, {Plunk}, {Wang}, {Barnes},
  {Cowley}, {Dorland} \& {Schekochihin}]{abel13}
{\sc \au{{Abel}, I.~G.}, \au{{Plunk}, G.~G.}, \au{{Wang}, E.}, \au{{Barnes},
  M.}, \au{{Cowley}, S.~C.}, \au{{Dorland}, W.} \& \au{{Schekochihin}, A.~A.}}
  \yr{2013}  \at{{Multiscale gyrokinetics for rotating tokamak plasmas:
  fluctuations, transport and energy flows}}.  \jt{Rep.\ Prog.\ Phys.}
  \bvol{76},  \pg{116201}.

\bibitem[{Abramowitz} \& {Stegun}(1972)]{abramowitz72}
{\sc \au{{Abramowitz}, M.} \& \au{{Stegun}, I.~A.}} \yr{1972} {\em {Handbook of
  Mathematical Functions}\/}.

\bibitem[{Adkins} {\em et~al.\/}(2022){Adkins}, {Schekochihin}, {Ivanov} \&
  {Roach}]{adkins22}
{\sc \au{{Adkins}, T.}, \au{{Schekochihin}, A.~A.}, \au{{Ivanov}, P.~G.} \&
  \au{{Roach}, C.~M.}} \yr{2022}  \at{Electromagnetic instabilities and plasma
  turbulence driven by electron-temperature gradient}.  \jt{J.\ Plasma Phys.}
  \bvol{88},  \pg{905880410}.

\bibitem[Beer {\em et~al.\/}(1995)Beer, Cowley \& Hammett]{beer95}
{\sc \au{Beer, M.~A.}, \au{Cowley, S.~C.} \& \au{Hammett, G.~W.}} \yr{1995}
  \at{Field‐aligned coordinates for nonlinear simulations of tokamak
  turbulence}.  \jt{Phys.\ Plasmas}  \bvol{2},  \pg{2687}.

\bibitem[{Beer} \& {Hammett}(1996)]{beer96}
{\sc \au{{Beer}, M.~A.} \& \au{{Hammett}, G.~W.}} \yr{1996}  \at{{Toroidal
  gyrofluid equations for simulations of tokamak turbulence}}.  \jt{Phys.\
  Plasmas}  \bvol{3},  \pg{4046}.

\bibitem[{Biglari} {\em et~al.\/}(1989){Biglari}, {Diamond} \&
  {Rosenbluth}]{biglari89}
{\sc \au{{Biglari}, H.}, \au{{Diamond}, P.~H.} \& \au{{Rosenbluth}, M.~N.}}
  \yr{1989}  \at{{Toroidal ion-pressure-gradient-driven drift instabilities and
  transport revisited}}.  \jt{Physics of Fluids B}  \bvol{1},  \pg{109}.

\bibitem[Brunner \& Vaclavik(1998)]{brunner98}
{\sc \au{Brunner, S.} \& \au{Vaclavik, J.}} \yr{1998}  \at{Global approach to
  the spectral problem of microinstabilities in a cylindrical plasma using a
  gyrokinetic model}.  \jt{Phys. Plasmas}  \bvol{5},  \pg{365}.

\bibitem[Catto(2019)]{catto2019}
{\sc \au{Catto, P.~J.}} \yr{2019}  \at{Practical gyrokinetics}.  \jt{{J.~Plasma
  Phys.}}  \bvol{85},  \pg{925850301}.

\bibitem[Coppi {\em et~al.\/}(1966)Coppi, Furth, Rosenbluth \&
  Sagdeev]{coppi66}
{\sc \au{Coppi, B.}, \au{Furth, H.~P.}, \au{Rosenbluth, M.~N.} \& \au{Sagdeev,
  R.~Z.}} \yr{1966}  \at{Drift instability due to impurity ions}.  \jt{Phys.
  Rev. Lett.}  \bvol{17},  \pg{377}.

\bibitem[{Coppi} {\em et~al.\/}(1967){Coppi}, {Rosenbluth} \&
  {Sagdeev}]{coppi67}
{\sc \au{{Coppi}, B.}, \au{{Rosenbluth}, M.~N.} \& \au{{Sagdeev}, R.~Z.}}
  \yr{1967}  \at{{Instabilities due to temperature gradients in complex
  magnetic field configurations}}.  \jt{Phys.\ Fluids}  \bvol{10},  \pg{582}.

\bibitem[{Cowley} {\em et~al.\/}(1991){Cowley}, {Kulsrud} \& {Sudan}]{cowley91}
{\sc \au{{Cowley}, S.~C.}, \au{{Kulsrud}, R.~M.} \& \au{{Sudan}, R.}} \yr{1991}
   \at{{Considerations of ion-temperature-gradient-driven turbulence}}.
  \jt{Phys.\ Fluids B}  \bvol{3},  \pg{2767}.

\bibitem[Ewart {\em et~al.\/}(2022)Ewart, Brown, Adkins \&
  Schekochihin]{ewart22}
{\sc \au{Ewart, R.~J.}, \au{Brown, A.}, \au{Adkins, T.} \& \au{Schekochihin,
  A.~A.}} \yr{2022}  \at{Collisionless relaxation of a lynden-bell plasma}.
  \jt{J.\ Plasma Phys.}  \bvol{88},  \pg{925880501}.

\bibitem[{Faddeeva} \& {Terent'ev}(1954)]{faddeeva54}
{\sc \au{{Faddeeva}, V.~N.} \& \au{{Terent'ev}, N.~M.}} \yr{1954} {\em Tables
  of Values of the Function $w(z)=\exp(-z^2)(1+2i/\sqrt{\pi}\int_0^z \exp(t^2)
  \rmd t)$ for Complex Argument\/}.  \publ{Moscow: Gostekhizdat, English
  translation: New York: Pergamon Press, 1961}.

\bibitem[{Fried} \& {Conte}(1961)]{fried61}
{\sc \au{{Fried}, B.~D.} \& \au{{Conte}, S.~D.}} \yr{1961} {\em {The Plasma
  Dispersion Function}\/}.  \publ{New York: Academic Press}.

\bibitem[{Frieman} \& {Chen}(1982)]{frieman82}
{\sc \au{{Frieman}, E.~A.} \& \au{{Chen}, L.}} \yr{1982}  \at{{Nonlinear
  gyrokinetic equations for low-frequency electromagnetic waves in general
  plasma equilibria}}.  \jt{Phys.\ Fluids}  \bvol{25},  \pg{502}.

\bibitem[{G{\"u}ltekin} \& {G{\"u}rcan}(2018)]{gultekin18}
{\sc \au{{G{\"u}ltekin}, {\"O}.} \& \au{{G{\"u}rcan}, {\"O}.~D.}} \yr{2018}
  \at{{Stable and unstable roots of ion temperature gradient driven mode using
  curvature modified plasma dispersion functions}}.  \jt{Plasma Phys.\
  Control.\ Fusion}  \bvol{60},  \pg{025021}.

\bibitem[{G{\"u}ltekin} \& {G{\"u}rcan}(2020)]{gultekin20}
{\sc \au{{G{\"u}ltekin}, {\"O}.} \& \au{{G{\"u}rcan}, {\"O}.~D.}} \yr{2020}
  \at{{Generalized curvature modified plasma dispersion functions and Dupree
  renormalization of toroidal ITG}}.  \jt{Plasma Phys.\ Control.\ Fusion}
  \bvol{62},  \pg{025018}.

\bibitem[{G{\"u}rcan}(2014)]{gurcan14}
{\sc \au{{G{\"u}rcan}, {\"O}.~D.}} \yr{2014}  \at{{Numerical computation of the
  modified plasma dispersion function with curvature}}.  \jt{J.\ Comp.\ Phys.}
  \bvol{269},  \pg{156}.

\bibitem[Guzdar {\em et~al.\/}(1983)Guzdar, Chen, Tang \& Rutherford]{guzdar83}
{\sc \au{Guzdar, P.~N.}, \au{Chen, L.}, \au{Tang, W.~M.} \& \au{Rutherford,
  P.~H.}} \yr{1983}  \at{Ion‐temperature‐gradient instability in toroidal
  plasmas}.  \jt{Phys. Fluids}  \bvol{26},  \pg{673}.

\bibitem[{Helander} {\em et~al.\/}(2011){Helander}, {Mishchenko}, {Kleiber} \&
  {Xanthopoulos}]{helander11}
{\sc \au{{Helander}, P.}, \au{{Mishchenko}, A.}, \au{{Kleiber}, R.} \&
  \au{{Xanthopoulos}, P.}} \yr{2011}  \at{{Oscillations of zonal flows in
  stellarators}}.  \jt{Plasma Phys.\ Control.\ Fusion}  \bvol{53},
  \pg{054006}.

\bibitem[{Howes} {\em et~al.\/}(2006){Howes}, {Cowley}, {Dorland}, {Hammett},
  {Quataert} \& {Schekochihin}]{howes06}
{\sc \au{{Howes}, G.~G.}, \au{{Cowley}, S.~C.}, \au{{Dorland}, W.},
  \au{{Hammett}, G.~W.}, \au{{Quataert}, E.} \& \au{{Schekochihin}, A.~A.}}
  \yr{2006}  \at{{Astrophysical gyrokinetics: basic equations and linear
  theory}}.  \jt{Astrophys.~J.}  \bvol{651},  \pg{590}.

\bibitem[Hugill(1983)]{hugill83}
{\sc \au{Hugill, J.}} \yr{1983}  \at{Transport in tokamaks -- a review of
  experiment}.  \jt{Nucl. Fusion}  \bvol{23},  \pg{331}.

\bibitem[Ivanov {\em et~al.\/}(2022)Ivanov, Schekochihin \& Dorland]{ivanov22}
{\sc \au{Ivanov, P.~G.}, \au{Schekochihin, A.~A.} \& \au{Dorland, W.}}
  \yr{2022}  \at{Dimits transition in three-dimensional
  ion-temperature-gradient turbulence}.  \jt{J.\ Plasma Phys.}  \bvol{88},
  \pg{905880506}.

\bibitem[{Ivanov} {\em et~al.\/}(2020){Ivanov}, {Schekochihin}, {Dorland},
  {Field} \& {Parra}]{ivanov20}
{\sc \au{{Ivanov}, P.~G.}, \au{{Schekochihin}, A.~A.}, \au{{Dorland}, W.},
  \au{{Field}, A.~R.} \& \au{{Parra}, F.~I.}} \yr{2020}  \at{{Zonally dominated
  dynamics and Dimits threshold in curvature-driven ITG turbulence}}.  \jt{J.\
  Plasma Phys.}  \bvol{86},  \pg{855860502}.

\bibitem[{Kim} {\em et~al.\/}(1994){Kim}, {Kishimoto}, {Horton} \&
  {Tajima}]{kim94}
{\sc \au{{Kim}, J.~Y.}, \au{{Kishimoto}, Y.}, \au{{Horton}, W.} \&
  \au{{Tajima}, T.}} \yr{1994}  \at{{Kinetic resonance damping rate of the
  toroidal ion temperature gradient mode}}.  \jt{Phys.\ Plasmas}  \bvol{1},
  \pg{927}.

\bibitem[{Kotschenreuther} {\em et~al.\/}(1995){Kotschenreuther}, {Dorland},
  {Beer} \& {Hammett}]{kotschenreuther95}
{\sc \au{{Kotschenreuther}, M.}, \au{{Dorland}, W.}, \au{{Beer}, M.~A.} \&
  \au{{Hammett}, G.~W.}} \yr{1995}  \at{{Quantitative predictions of tokamak
  energy confinement from first-principles simulations with kinetic effects}}.
  \jt{Phys.\ Plasmas}  \bvol{2},  \pg{2381}.

\bibitem[{Kuroda} {\em et~al.\/}(1998){Kuroda}, {Sugama}, {Kanno}, {Okamoto} \&
  {Horton}]{kuroda98}
{\sc \au{{Kuroda}, T.}, \au{{Sugama}, H.}, \au{{Kanno}, R.}, \au{{Okamoto}, M.}
  \& \au{{Horton}, W.}} \yr{1998}  \at{{Initial Value Problem of the Toroidal
  Ion Temperature Gradient Mode}}.  \jt{J. Phys. Soc. Japan}  \bvol{67},
  \pg{3787}.

\bibitem[{Landau}(1946)]{landau46}
{\sc \au{{Landau}, L.}} \yr{1946}  \at{{On the vibration of the electronic
  plasma}}.  \jt{Zh.\ Eksp.\ Teor.\ Fiz.}  \bvol{16},  \pg{574}.

\bibitem[Lee {\em et~al.\/}(1987)Lee, Dong, Guzdar \& Liu]{lee87}
{\sc \au{Lee, Y.~C.}, \au{Dong, J.~Q.}, \au{Guzdar, P.~N.} \& \au{Liu, C.~S.}}
  \yr{1987}  \at{Collisionless electron temperature gradient instability}.
  \jt{Phys.\ Fluids}  \bvol{30},  \pg{1331}.

\bibitem[Liewer(1985)]{liewer85}
{\sc \au{Liewer, P.~C.}} \yr{1985}  \at{Measurements of microturbulence in
  tokamaks and comparisons with theories of turbulence and anomalous
  transport}.  \jt{Nucl. Fusion}  \bvol{25},  \pg{543}.

\bibitem[Liu(1971)]{liu71}
{\sc \au{Liu, C.~S.}} \yr{1971}  \at{Instabilities in a magnetoplasma with skin
  current}.  \jt{Phys. Rev. Lett.}  \bvol{27},  \pg{1637--1640}.

\bibitem[{Mishchenko} {\em et~al.\/}(2018){Mishchenko}, {Plunk} \&
  {Helander}]{mishchenko18}
{\sc \au{{Mishchenko}, A.}, \au{{Plunk}, G.~G.} \& \au{{Helander}, P.}}
  \yr{2018}  \at{{Electrostatic stability of electron-positron plasmas in
  dipole geometry}}.  \jt{J.\ Plasma Phys.}  \bvol{84},  \pg{905840201}.

\bibitem[{Neumann}(1871)]{neumann1871}
{\sc \au{{Neumann}, C.}} \yr{1871}  \at{{Ueber die Entwickelung einer Function
  nach Quadraten und Produkten der Fourier-Bessel'schen Functionen}}.
  \jt{Math. Ann.}  \bvol{3},  \pg{581}.

\bibitem[{Newton} {\em et~al.\/}(2010){Newton}, {Cowley} \&
  {Loureiro}]{newton10}
{\sc \au{{Newton}, S.~L.}, \au{{Cowley}, S.~C.} \& \au{{Loureiro}, N.~F.}}
  \yr{2010}  \at{{Understanding the effect of sheared flow on
  microinstabilities}}.  \jt{Plasma Phys.\ Control.\ Fusion}  \bvol{52},
  \pg{125001}.

\bibitem[Ongena {\em et~al.\/}(2016)Ongena, Koch, Wolf \& Zohm]{ongena2016}
{\sc \au{Ongena, J.}, \au{Koch, R.}, \au{Wolf, R.} \& \au{Zohm, H.}} \yr{2016}
  \at{Magnetic-confinement fusion}.  \jt{Nature Phys.}  \bvol{12},  \pg{398}.

\bibitem[Parisi {\em et~al.\/}(2020)Parisi, Parra, Roach, Giroud, Dorland,
  Hatch, Barnes, Hillesheim, Aiba, Ball, Ivanov \& contributors]{parisi20}
{\sc \au{Parisi, J.~F.}, \au{Parra, F.~I.}, \au{Roach, C.~M.}, \au{Giroud, C.},
  \au{Dorland, W.}, \au{Hatch, D.~R.}, \au{Barnes, M.}, \au{Hillesheim, J.~C.},
  \au{Aiba, N.}, \au{Ball, J.}, \au{Ivanov, P.~G.} \& \au{contributors, JET}}
  \yr{2020}  \at{Toroidal and slab {ETG} instability dominance in the linear
  spectrum of {JET}-{ILW} pedestals}.  \jt{Nucl.\ Fusion}  \bvol{60},
  \pg{126045}.

\bibitem[{Parisi} {\em et~al.\/}(2022){Parisi}, {Parra}, {Roach}, {Hardman},
  {Schekochihin}, {Abel}, {Aiba}, {Ball}, {Barnes}, {Chapman-Oplopoiou},
  {Dickinson}, {Dorland}, {Giroud}, {Hatch}, {Hillesheim}, {Ruiz Ruiz},
  {Saarelma}, {St-Onge} \& Contributors]{parisi22}
{\sc \au{{Parisi}, J.~F.}, \au{{Parra}, F.~I.}, \au{{Roach}, C.~M.},
  \au{{Hardman}, M.~R.}, \au{{Schekochihin}, A.~A.}, \au{{Abel}, I.~G.},
  \au{{Aiba}, N.}, \au{{Ball}, J.}, \au{{Barnes}, M.}, \au{{Chapman-Oplopoiou},
  B.}, \au{{Dickinson}, D.}, \au{{Dorland}, W.}, \au{{Giroud}, C.},
  \au{{Hatch}, D.~R.}, \au{{Hillesheim}, J.~C.}, \au{{Ruiz Ruiz}, J.},
  \au{{Saarelma}, S.}, \au{{St-Onge}, D.} \& \au{Contributors, JET}} \yr{2022}
  \at{Three-dimensional inhomogeneity of electron-temperature-gradient
  turbulence in the edge of tokamak plasmas}.  \jt{Nucl.\ Fusion}  \bvol{62},
  \pg{086045}.

\bibitem[Pogutse(1968)]{pogutse68}
{\sc \au{Pogutse, O.~P.}} \yr{1968}  \at{Magnetic drift instability in a
  collisionless plasma}.  \jt{Plasma Physics}  \bvol{10},  \pg{649}.

\bibitem[{Ricci} {\em et~al.\/}(2006){Ricci}, {Rogers}, {Dorland} \&
  {Barnes}]{ricci06}
{\sc \au{{Ricci}, P.}, \au{{Rogers}, B.~N.}, \au{{Dorland}, W.} \&
  \au{{Barnes}, M.}} \yr{2006}  \at{{Gyrokinetic linear theory of the entropy
  mode in a Z pinch}}.  \jt{Physics of Plasmas}  \bvol{13},  \pg{062102}.

\bibitem[{Rudakov} \& {Sagdeev}(1961)]{rudakov61}
{\sc \au{{Rudakov}, L.~I.} \& \au{{Sagdeev}, R.~Z.}} \yr{1961}  \at{{On the
  instability of inhomogeneous rarefied plasma in a strong magnetic field}}.
  \jt{Dokl.\ Acad.\ Nauk SSSR}  \bvol{138},  \pg{581}.

\bibitem[Sauter {\em et~al.\/}(1990)Sauter, Vaclavik \& Skiff]{sauter90}
{\sc \au{Sauter, O.}, \au{Vaclavik, J.} \& \au{Skiff, F.}} \yr{1990}  \at{A
  nonlocal analysis of electrostatic waves in hot inhomogeneous bounded
  plasmas}.  \jt{Phys. Fluids B: Plasma Physics}  \bvol{2},  \pg{475}.

\bibitem[Sedl\`aček(1995)]{sedlacek95}
{\sc \au{Sedl\`aček, Z.}} \yr{1995}  \at{Continuum damping in plasma physics}.
   \jt{AIP Conference Proceedings}  \bvol{345},  \pg{119}.

\bibitem[{Similon} {\em et~al.\/}(1984){Similon}, {Sedlak}, {Stotler}, {Berk},
  {Horton} \& {Choi}]{similon84}
{\sc \au{{Similon}, P.}, \au{{Sedlak}, J.~E.}, \au{{Stotler}, D.}, \au{{Berk},
  H.~L.}, \au{{Horton}, W.} \& \au{{Choi}, D.}} \yr{1984}  \at{{Guiding-Center
  Dispersion Function}}.  \jt{J.\ Comp.\ Phys.}  \bvol{54},  \pg{260}.

\bibitem[Smolyakov {\em et~al.\/}(2002)Smolyakov, Yagi \&
  Kishimoto]{smolyakov2002}
{\sc \au{Smolyakov, A.~I.}, \au{Yagi, M.} \& \au{Kishimoto, Y.}} \yr{2002}
  \at{Short wavelength temperature gradient driven modes in tokamak plasmas}.
  \jt{Phys. Rev. Lett.}  \bvol{89},  \pg{125005}.

\bibitem[{Sugama}(1999)]{sugama99}
{\sc \au{{Sugama}, H.}} \yr{1999}  \at{Damping of toroidal ion temperature
  gradient modes}.  \jt{Phys.\ Plasmas}  \bvol{6},  \pg{3527}.

\bibitem[{Sugama} {\em et~al.\/}(1996){Sugama}, {Okamoto}, {Horton} \&
  {Wakatani}]{sugama96}
{\sc \au{{Sugama}, H.}, \au{{Okamoto}, M.}, \au{{Horton}, W.} \&
  \au{{Wakatani}, M.}} \yr{1996}  \at{{Transport processes and entropy
  production in toroidal plasmas with gyrokinetic electromagnetic turbulence}}.
   \jt{Phys.\ Plasmas}  \bvol{3},  \pg{2379}.

\bibitem[Taylor(1965)]{taylor65}
{\sc \au{Taylor, E.~C.}} \yr{1965}  \at{Landau solution of the plasma
  oscillation problem}.  \jt{Phys.\ Fluids}  \bvol{8},  \pg{2250}.

\bibitem[{Terry} {\em et~al.\/}(1982){Terry}, {Anderson} \& {Horton}]{terry82}
{\sc \au{{Terry}, P.}, \au{{Anderson}, W.} \& \au{{Horton}, W.}} \yr{1982}
  \at{Kinetic effects on the toroidal ion pressure gradient drift mode}.
  \jt{Nucl.\ Fusion}  \bvol{22},  \pg{487}.

\bibitem[{Waltz}(1988)]{waltz88}
{\sc \au{{Waltz}, R.~E.}} \yr{1988}  \at{{Three-dimensional global numerical
  simulation of ion temperature gradient mode turbulence}}.  \jt{Phys.\ Fluids}
   \bvol{31},  \pg{1962}.

\bibitem[{Watson}(1966)]{watson66}
{\sc \au{{Watson}, G.~N.}} \yr{1966} {\em {A Treatise on the theory of Bessel
  functions}\/}, 2nd edn.  \publ{Cambridge university press}.

\bibitem[Wootton {\em et~al.\/}(1990)Wootton, Carreras, Matsumoto, McGuire,
  Peebles, Ritz, Terry \& Zweben]{wootton90}
{\sc \au{Wootton, A.~J.}, \au{Carreras, B.~A.}, \au{Matsumoto, H.},
  \au{McGuire, K.}, \au{Peebles, W.~A.}, \au{Ritz, C.~P.}, \au{Terry, P.~W.} \&
  \au{Zweben, S.~J.}} \yr{1990}  \at{Fluctuations and anomalous transport in
  tokamaks}.  \jt{Phys. Fluids B}  \bvol{2},  \pg{2879}.

\bibitem[Xanthopoulos {\em et~al.\/}(2007)Xanthopoulos, Merz, G\"orler \&
  Jenko]{xanthopoulos2007}
{\sc \au{Xanthopoulos, P.}, \au{Merz, F.}, \au{G\"orler, T.} \& \au{Jenko, F.}}
  \yr{2007}  \at{Nonlinear gyrokinetic simulations of ion-temperature-gradient
  turbulence for the optimized wendelstein 7-x stellarator}.  \jt{Phys. Rev.
  Lett.}  \bvol{99},  \pg{035002}.

\bibitem[{Zocco} {\em et~al.\/}(2018){Zocco}, {Xanthopoulos}, {Doerk}, {Connor}
  \& {Helander}]{zocco18}
{\sc \au{{Zocco}, A.}, \au{{Xanthopoulos}, P.}, \au{{Doerk}, H.}, \au{{Connor},
  J.~W.} \& \au{{Helander}, P.}} \yr{2018}  \at{{Threshold for the
  destabilisation of the ion-temperature-gradient mode in magnetically confined
  toroidal plasmas}}.  \jt{J.\ Plasma Phys.}  \bvol{84},  \pg{715840101}.

\end{thebibliography}
\bibliographystyle{jpp}

\end{document}